\documentclass[12pt]{article}
\usepackage[reqno]{amsmath}
\usepackage{bbm}
\usepackage{epsfig}
\usepackage{array}
\usepackage{float}
\usepackage{dsfont}
\usepackage{amstext}
\usepackage{rotating}

\usepackage{a4}

\usepackage{a4wide}
\def\ra{\rightarrow}
\def\be{\begin{equation}}
\def\ee{\end{equation}}
\def\gs{\mathrel{
   \rlap{\raise 0.511ex \hbox{$>$}}{\lower 0.511ex \hbox{$\sim$}}}}
\def\ls{\mathrel{
   \rlap{\raise 0.511ex \hbox{$<$}}{\lower 0.511ex \hbox{$\sim$}}}}

\newcommand{\ba}{\begin{array}{c}}
\newcommand{\baz}{\begin{array}{cc}}
\newcommand{\bad}{\begin{array}{ccc}}
\newcommand{\bav}{\begin{array}{cccc}}
\newcommand{\bea}{\begin{equation} \begin{array}{c}}
\newcommand{\eea}{ \end{array} \end{equation}}
\newcommand{\ea}{\end{array}}
\newcommand{\D}{\displaystyle}
\newcommand{\dms}{\mbox{$\Delta m^2_{\odot}$}}
\newcommand{\dma}{\mbox{$\Delta m^2_{\rm A}$}}

\newcommand{\sss}{\sin^2 \theta_{12}}
\newcommand{\sch}{\sin^2 \theta_{13}}

\newcommand{\Slash}[1]{\mbox{\scriptsize $#1\hspace{-.5em}/$}}

\begin{document}

\title{
\hfill {\small hep--ph/0612047} 
\vskip 0.4cm
\bf 
Neutrino Mixing and Neutrino Telescopes
}
\author{
Werner Rodejohann\thanks{email: \tt werner.rodejohann@mpi-hd.mpg.de} 
\\\\
{\normalsize \it Max--Planck--Institut f\"ur Kernphysik,}\\
{\normalsize \it  Postfach 103980, D--69029 Heidelberg, Germany}
}
\date{}
\maketitle
\thispagestyle{empty}
\vspace{-0.8cm}
\begin{abstract}
\noindent  
Measuring flux ratios of high energy neutrinos 
is an alternative method to determine the neutrino 
mixing angles and the $CP$ phase. 
We conduct a systematic analysis of the neutrino 
mixing probabilities and of various flux ratios measurable 
at neutrino telescopes. 
The considered cases are neutrinos 
from pion, neutron and muon-damped sources. 
The flux ratios involve measurements 
with and without taking advantage of the Glashow resonance. 
Explicit formulae in case of 
$\mu$--$\tau$ symmetry ($\theta_{13}=0$ and $\theta_{23}=\pi/4$) 
and its special case tri-bimaximal mixing ($\sin^2 \theta_{12} = 1/3$) are 
obtained, and the leading corrections due to non-zero $\theta_{13}$ and 
non-maximal $\theta_{23}$ are given. The first order correction 
is universal as it appears in basically all ratios. 
We study in detail its dependence on $\theta_{13}$, 
$\theta_{23}$ and the $CP$ phase $\delta$, finding 
that the dependence on $\theta_{23}$ 
is strongest. 
The flavor compositions for the considered 
neutrino sources are evaluated in terms of this correction. 
A measurement of a flux ratio is a clean measurement of 
the universal correction (and therefore of $\theta_{13}$, 
$\theta_{23}$ and $\delta$) if the zeroth order ratio does not depend on 
$\theta_{12}$. This favors pion sources over the other cases, which in 
turn are good candidates to probe $\theta_{12}$.
The only situations in which the universal correction does not appear 
are certain ratios in case of a neutron and muon-damped 
source, which depend mainly on $\theta_{12}$ and receive only 
quadratic corrections from the other parameters. 
We further show that there are only two independent 
neutrino mixing probabilities, give the allowed ranges of 
the considered flux ratios and of 
all probabilities, and show that none of the latter can be zero or one. 
Finally, we analyze situations in which 
$\theta_{13}$ is sizable and $\theta_{23}$ is close to $\pi/4$, and in which 
$\theta_{13}$ is close to zero and $\theta_{23}- \pi/4$ is sizable.

\end{abstract}

\newpage

\section{\label{sec:intro}Introduction}

A major part of the ongoing activities in astroparticle physics is related to 
neutrino mixing and oscillations \cite{reviews}. 
The current information on the involved 
mixing parameters is already impressive, and in the next decades 
enormous improvement on their precision is expected \cite{10years}. 
While most efforts towards this goal have been put into 
terrestrial experiments with man-made 
neutrino sources, recently more attention has been paid to the possibility 
that high energy ($\gs$ TeV) 
neutrinos \cite{UHEs} might also provide valuable 
information on the neutrino mixing 
parameters\footnote{Note that cosmic rays have already turned out to 
be very useful for neutrino physics because 
the zenith angle dependent deficit of muon neutrinos, which are created 
by cosmic rays in the 
Earth's atmosphere, has provided the first compelling evidence for neutrino 
oscillations.} [4--16]. As the neutrino mass-squared differences do not 
play a role for high energy neutrinos, we are left in this framework 
with four observables, namely three mixing angles and 
one $CP$ phase.\\

In this paper we will try to systematically study the situations in which 
these four relevant neutrino parameters can be probed 
through measurements of neutrino flux ratios at neutrino telescopes. 
We consider neutrinos from 
three possible sources: pion decays, 
neutron decays and muon-damped sources 
(generated in environments in which muons, 
but not pions, loose energy \cite{energy,010}). 
Measurements of high energy neutrino flux ratios will eventually take 
place in the km$^3$ scale neutrino telescope IceCube \cite{IceCube}. 
Another option is a km$^3$ experiment in the Mediterranean, 
as investigated by the KM3Net network \cite{km3}, 
which coordinates the potential joining 
of the ANTARES \cite{ANTARES}, NESTOR \cite{NESTOR} and 
NEMO \cite{NEMO} projects. These will 
follow the smaller scale AMANDA \cite{Amanda} and Lake Baikal \cite{Baikal} 
experiments. The Cherenkov light of muon neutrinos leaves characteristic tracks 
in neutrino telescopes, which can be used to distinguish them from electron 
and tau neutrinos and to extract the ratio of muon to 
electron plus tau neutrinos from the ratio of tracks to showers \cite{R}.  
Electron and tau neutrinos are harder to disentangle, though there are 
characteristic differences \cite{R,tauvse,tauvse1}. At neutrino energies 
close to $6.3 \cdot 10^{6}$ GeV the Glashow resonance enhances 
the process $\bar \nu_e \, e^- \ra W^-$, which could be used to identify 
anti-electron neutrinos \cite{GR0,wrong}. However, our approach here 
focusses on the general properties of the 
mixing probabilities and their dependence on the 
neutrino mixing parameters. We will therefore consider several possible 
neutrino flux ratios and analyze them in terms of their dependence 
on those, mostly ignoring for the time being experimental aspects. 

In what regards our current knowledge of the neutrino parameters, 
the solar neutrino mixing angle $\theta_{12}$ is 
large though non-maximal, but still possesses a sizable error. The magnitude of 
$U_{e3}$ is not known, as is the deviation from maximal atmospheric neutrino 
mixing $\theta_{23} - \pi/4$, and the value of the $CP$ phase $\delta$. The 
latter three observables all have a variety of implications 
both in the low and high energy sector, but correspond to suppressed effects. 
In the course of this paper we 
will expand the mixing probabilities of high energy 
neutrinos in terms of the 
small parameters $|U_{e3}|$ and the deviation from maximal 
atmospheric neutrino mixing. The zeroth order expression is just a 
function of the solar neutrino mixing angle.   
The first order correction $\Delta$ 
due to non-zero $\theta_{13}$ and non-maximal $\theta_{23}$ 
is universal, i.e.,  
the same for all probabilities. It depends in a characteristic way on the 
parameters, which we study in detail. The dependence on $\theta_{23}$ is 
strongest. 
Since $\theta_{12}$ is non-maximal and non-zero, no probability is zero or one, 
therefore high energy neutrinos are guaranteed to change their flavor.
As neutrino flux ratios are functions of 
the mixing probabilities, they are most of the times given by a zeroth 
order expression and a first order correction. 
This allows for a comparably simple analytical understanding of the 
measurements in terms of their implications on the neutrino mixing parameters. 
If the zeroth order expression of a flux ratio 
is independent of $\theta_{12}$, then 
measuring the ratio means measuring the correction $\Delta$ without any 
uncertainty due to our imprecise knowledge of $\theta_{12}$. Hence, the 
ratio is a good probe for 
$U_{e3}$, $\theta_{23} - \pi/4$ and $\delta$. This happens typically 
for neutrinos from pion sources. If in turn 
the zeroth order expression depends on $\theta_{12}$, 
which happens for neutrinos from neutron and muon-damped sources, 
then the corresponding ratio is a good candidate 
to probe this angle\footnote{We will 
not discuss other interesting aspects of neutrino flux ratios, such as 
neutrino decay \cite{decay}, Pseudo-Dirac structure \cite{Pseudo}, 
magnetic moments \cite{magn} or breakdown of fundamental 
symmetries \cite{CPT}.}. In addition, we also find for these sources 
that there are certain 
neutrino flux ratios for which there is no first order correction. 
They are functions of $\theta_{12}$ and receive only quadratic corrections 
from the small parameters. Hence, they allow for a simple 
extraction of $\theta_{12}$ 
and underline the usefulness of the source for measuring this parameter.\\

The outline of this paper is as follows: 
first we will study some general aspects of 
the mixing probabilities in Section \ref{sec:nus}. 
We will then discuss several possible 
flux ratios in Section \ref{sec:fluobs}. 
The flavor compositions and the 
flux ratios are evaluated in terms of $\theta_{12}$ and the universal 
correction $\Delta$, furthermore their allowed ranges are presented. 
It is shown that typically flux ratios from pion sources are at zeroth 
order independent of $\theta_{12}$, while ratios for neutrinos from 
other sources are in leading order $\theta_{12}$-depending.  
We then present in Section \ref{sec:special} 
ratios for which there is no first order correction. 
Finally, we analyze in Section \ref{sec:AvsB} 
the dependence of the ratios in cases when $\theta_{13}$ is 
sizable but $\theta_{23}$ 
very close to maximal and when $\theta_{13}$ is negligible but $\theta_{23}$ 
deviates sizably from maximal. 
Section \ref{sec:concl} is devoted to the conclusions.

\section{\label{sec:nus}Neutrino Mixing}
In this Section we will first summarize our current understanding of 
neutrino mixing before we investigate the properties and allowed ranges 
of the mixing probabilities of high energy neutrinos.

\subsection{\label{sec:form}Neutrino Data and $\mu$--$\tau$ Symmetry}
Let us first set the stage for the discussion by shortly summarizing the 
neutrino data and the implied mixing schemes. 
The physics of neutrino mixing is described by the 
neutrino mass matrix 
\be
m_\nu = U \, m_\nu^{\rm diag} \, U^T ~,
\ee 
where $U$ is the leptonic mixing, or 
Pontecorvo-Maki-Nakagawa-Sakata (PMNS) \cite{PMNS}, 
matrix in the basis in which the charged lepton mass 
matrix is real and diagonal. The three neutrino masses 
are contained in $m_\nu^{\rm diag} = {\rm diag}(m_1, m_2, m_3)$. 
A useful parameterization for the unitary PMNS matrix $U$ is  
\bea 
\label{eq:Upara}
U = 
\left( \bad 
U_{e1} & U_{e2} & U_{e3} \\[0.2cm] 
U_{\mu 1} & U_{\mu 2} & U_{\mu 3} \\[0.2cm] 
U_{\tau 1} & U_{\tau 2} & U_{\tau 3} 
\ea
\right) \\[0.3cm] 
= \left( \bad 
c_{12} \, c_{13} & s_{12} \, c_{13} & s_{13} \, e^{-i \delta}  \\[0.2cm] 
-s_{12} \, c_{23} - c_{12} \, s_{23} \, s_{13} \, e^{i \delta} 
& c_{12} \, c_{23} - s_{12} \, s_{23} \, s_{13} \, e^{i \delta} 
& s_{23} \, c_{13}  \\[0.2cm] 
s_{12} \, s_{23} - c_{12} \, c_{23} \, s_{13} \, e^{i \delta} & 
- c_{12} \, s_{23} - s_{12} \, c_{23} \, s_{13} \, e^{i \delta} 
& c_{23} \, c_{13}  \\ 
               \ea   
\right) ~,
\eea 
where we have used the usual notations $c_{ij} = \cos\theta_{ij}$, 
$s_{ij} = \sin\theta_{ij}$ and introduced the Dirac $CP$-violating
phase $\delta$. The angles can lie anywhere between zero and $\pi/2$, 
whereas the phase is allowed to take values between zero and $2\pi$. 
Possible Majorana phases are neglected here 
since they do not play a role 
in the oscillation framework. In fact, the analysis presented in this paper 
does not depend on whether neutrinos are Dirac or Majorana particles. 
Various experiments and their analyzes  
revealed the following allowed 2, 3 and 4$\sigma$ 
ranges of the mixing angles \cite{thomas}:
\begin{eqnarray} \label{eq:data}
\sss &=& 0.30^{+0.06, \,0.10, \, 0.14}_{-0.04, \, 0.06,\, 0.08} 
~,\nonumber\\
\sin^2\theta_{23} &=& 0.50^{+0.13, \, 0.18,  \,0.21}
_{-0.12, \, 0.16,  \,0.19} ~,\\
\sch &<& 0.025~(0.041, \, 0.058)~.\nonumber
\end{eqnarray}
The present best-fit value for $\sch$ is zero, and there is as yet 
no information on any of the phases.\\

The remarkable best-fit values of $\theta_{13}=0$ and 
$\theta_{23} = \pi/4$ have let many authors to study $\mu$--$\tau$ 
symmetry \cite{mutau}. 
This exchange symmetry leads to $\theta_{13}=0$ and 
$\theta_{23} = \pi/4$ by enforcing the low energy mass 
matrix to have the form (for Dirac neutrinos this would concern 
$m_\nu \, m_\nu^\dagger$) 
\be
m_\nu = 
\left( 
\bad
A & B & B \\[0.2cm]
\cdot & D & E \\[0.2cm]
\cdot & \cdot & D
\ea
\right)~.
\ee
The solar neutrino mixing angle and the neutrino masses 
are not predicted by $\mu$--$\tau$ symmetry. The PMNS matrix reads 
\be
U = 
\left( 
\bad
\cos \theta_{12} & \sin \theta_{12} & 0 \\[0.12cm] \D 
-\frac{\sin \theta_{12}}{\sqrt{2}} 
& \D \frac{\cos \theta_{12}}{\sqrt{2}} & \D \sqrt{\frac 12} \\[0.12cm] \D 
\frac{\sin \theta_{12}}{\sqrt{2}} 
& \D -\frac{\cos \theta_{12}}{\sqrt{2}} & \D \sqrt{\frac 12}
\ea 
\right) \stackrel{\rm TBM}{\longrightarrow}
\left( 
\bad
\D \sqrt{\frac 23} & \D \sqrt{\frac 13}  & 0 \\[0.12cm]
\D -\sqrt{\frac 16} & \D \sqrt{\frac 13}  & \D \sqrt{\frac 12}\\[0.12cm]
\D \sqrt{\frac 16} & \D -\sqrt{\frac 13}  & \D  \sqrt{\frac 12}
\ea 
\right)~.
\ee
Here we have also given the form of $U$ in case of 
tri-bimaximal mixing (TBM) \cite{TBM}, 
which is defined by $\mu$--$\tau$ symmetry with $\sin^2 \theta_{12} = 1/3$. 
It is a particularly simple mixing scheme which is very close to the 
current best-fit points. 
The analytical part of our analysis will rely on the fact that 
neutrino mixing can be accurately described by
$\mu$--$\tau$ symmetry, which implies that one can expand all relevant 
formulae in terms of the small parameters 
$\theta_{13}$ and $\pi/4 - \theta_{23}$.

\subsection{\label{sec:gen}Mixing Probabilities: General Considerations}

In this Section we will consider neutrino mixing probabilities 
of high energy neutrinos. We will show that 
there are only two independent probabilities (which is also 
true for the general case \cite{chef}), provide 
explicit expressions for them and give their simple forms in case 
of approximate $\mu$--$\tau$ symmetry and tri-bimaximal mixing. The  
universal first order correction to the probabilities is 
analyzed with respect to its dependence on the 
mixing angles and $CP$ phase. Then we obtain the allowed ranges of the 
probabilities.\\

Information on the mixing angles (and the phase $\delta$) is 
obtained of course by neutrino oscillations. The probability of flavor 
conversion is in general a function of the three mixing angles, the $CP$ phase 
$\delta$ and the mass-squared differences times baseline $L$ 
divided by energy $E$ \cite{chef}. 
In the context of astrophysical neutrinos, the latter do not play a role, 
since the neutrinos travel many oscillation lengths from source 
to detector, leading to\footnote{Since the oscillation 
probabilities do not depend anymore on the mass-squared differences,  
in principle the extraction of the mixing angles and the $CP$ phase is easier 
than in the general case. The possibility to probe the sign of the 
atmospheric $\Delta m^2$ is however lost.} 
\be \label{eq:Pab}
P_{\alpha \beta} \equiv P(\nu_\alpha \ra \nu_\beta) = 
\delta_{\alpha \beta} - 2\,{\cal R} \sum\limits_{i>j} \, 
U_{\alpha j} \, U_{\alpha i}^\ast \,  U_{\beta j}^\ast \, U_{\beta i} = 
\sum |U_{\alpha i}|^2 \, |U_{\beta i}|^2 ~, 
\ee
where ${\cal R}$ denotes the real part. 
As obvious from Eq.~(\ref{eq:Pab}), the probability is $CP$ conserving:  
$P(\nu_\alpha \ra \nu_\beta) = P(\bar \nu_\alpha \ra \bar \nu_\beta)$. 
Due to $CPT$ invariance, it follows that 
$P(\nu_\alpha \ra \nu_\beta) = P(\nu_\beta \ra \nu_\alpha)$. This is 
also obvious from Eq.~(\ref{eq:Pab}). Recall that there are no 
matter effects which could spoil these relations, though recently 
some cases have been investigated in which they might play a 
role \cite{matter?}. 
Note that Eq.~(\ref{eq:Pab}) displays no oscillatory behavior as a function 
of $L/E$ anymore. 
Unitarity provides the ``sum-rules''  
$\sum_\alpha P_{\alpha \beta} = \sum_\beta P_{\alpha \beta} = 1$.
Two other useful relations are  
\bea \label{eq:t23repl}
P_{e \tau} = P_{e \mu}(\theta_{23} \ra \theta_{23} + \pi/2 \mbox{ or } 
\theta_{23} \ra \theta_{23} + 3\pi/2)  ~,\\[0.2cm]
P_{\tau \tau} = P_{\mu \mu}(\theta_{23} \ra \theta_{23} + \pi/2\mbox{ or } 
\theta_{23} \ra \theta_{23} + 3\pi/2)~.
\eea
With these properties and relations it is easy to show that 
 -- as in the general case \cite{chef} -- there 
are only two independent mixing probabilities: for instance, 
it suffices to consider only $P_{e\mu}$ and $P_{\mu\mu}$. 
There are five other possible choices of two probabilities, 
namely $(P_{e\mu},P_{\tau\tau})$, $(P_{e\tau},P_{\mu\mu})$, 
$(P_{e\tau},P_{\tau\tau})$, $(P_{\mu\mu},P_{\mu\tau})$ and 
$(P_{\tau\tau},P_{\mu\tau})$. 
Note that $P_{ee}$ should not be one of them and that the pair 
should not be related via Eq.~(\ref{eq:t23repl}). 
The remaining probabilities for the pair $P_{e\mu}$ and $P_{\mu\mu}$ 
can be obtained by Eq.~(\ref{eq:t23repl}) and 
\bea \label{eq:sumrules}
P_{ee} = 1 - P_{e\mu} - P_{e\tau} ~,\\[0.2cm]
P_{\mu \tau} = 1 - P_{e \mu} - P_{\mu \mu} ~,\\[0.2cm]
P_{\tau \tau} = 1 -  P_{e \tau} - P_{\mu \tau} 
= P_{ee} + 2 \,  P_{e \mu} + P_{\mu\mu} - 1 = 
P_{e\mu} - P_{e\tau} + P_{\mu\mu}~.
\eea
The explicit expressions for $P_{e\mu}$ and $P_{\mu\mu}$ are 
delegated to the Appendix, we give here only the survival probability 
for electron neutrinos since it has the simplest 
structure of all:   
\be
P_{ee} =  1-2 \, c_{13}^2 \left(c_{12}^2 \, s_{12}^2 \, c_{13}^2
    +\, s_{13}^2\right)
~.
\ee 
Note that $P_{ee}$ depends mainly on $\theta_{12}$ and only very weakly 
(quadratically) on $\theta_{13}$. This will be useful later on.\\

In the limit of exact $\mu$--$\tau$ symmetry the flavors 
$\mu$ and $\tau$ are equivalent, and it will hold that 
$P_{\alpha \mu} = P_{\alpha \tau}$ for all $\alpha = e, \mu, \tau$. 
Assuming breaking of $\mu$--$\tau$ symmetry, 
one can expand the probabilities 
in terms of the small parameters $|U_{e3}|$ 
and $\epsilon = \pi/4 - \theta_{23}$. 
Introducing 
the symmetric matrix\footnote{This matrix is symmetric because 
of the $T$ conserving nature of the flavor transitions. In the 
general case $P_{\alpha \beta} \neq P_{\beta \alpha}$.} 
of probabilities $P$ we find that 
\bea \label{eq:probs}
P \equiv \left( 
\bad 
P_{ee} & P_{e \mu} & P_{e \tau} \\
\cdot & P_{\mu\mu} & P_{\mu\tau} \\
\cdot & \cdot & P_{\tau\tau} 
\ea
\right) 
\\[0.3cm]
\hspace{-0.61cm}
\simeq 
\left( 
\bad
(1 - 2 \, c_{12}^2 \, s_{12}^2) \, (1 - 2 \, |U_{e3}|^2) 
& c_{12}^2 \, s_{12}^2 + \Delta & 
c_{12}^2 \, s_{12}^2 - \Delta \\[0.2cm]
\cdot & \frac 12 \,  \left(1 - c_{12}^2 \, s_{12}^2\right) - \Delta & 
\frac 12 \,  (1 - c_{12}^2 \, s_{12}^2) 
\\[0.2cm]
\cdot & \cdot & \frac 12 \, (1 - c_{12}^2 \, s_{12}^2) + \Delta 
\ea
\right)~.
\eea
We have defined here the universal correction parameter 
\be \label{eq:Delta}
\Delta \equiv \frac 14 \, \cos \delta \, \sin 4 \theta_{12} \, |U_{e3}| 
+ 2 \, s_{12}^2 \, c_{12}^2 \, \epsilon 
+ {\cal O}(|U_{e3}|^2, \, \epsilon^2, \, |U_{e3}|\, \epsilon)~, 
\ee
where $\sin 4 \theta_{12} = 2 \, \sin 2  \theta_{12} \, \cos 2 \theta_{12} 
= 4 \, c_{12} \, s_{12} \, (c_{12}^2 - s_{12}^2)$. Unitarity (i.e., 
$\sum_\alpha P_{\alpha \beta} = \sum_\beta P_{\alpha \beta} = 1$) 
holds to first order in Eq.~(\ref{eq:probs}). 
We will encounter $\Delta$ throughout this paper and its importance has been 
noted first in Ref.~\cite{xing2}. 
Note that $\Delta$ can vanish not 
only in the trivial case of $\theta_{23} - \pi/4 = |U_{e3}| = 0$, 
but also when $|U_{e3}|/\epsilon \simeq - \tan 2 \theta_{12} /c_\delta$. 
In general, $\Delta$ has the range
\be \label{eq:Delta_range}
-0.09~(-0.11, \, -0.14) \le \Delta \le 0.08~(0.11, \, 0.13)~,
\ee
when the oscillation 
parameters are varied within their $2~(3, \, 4)\sigma$ ranges. 
The probabilities 
$P_{ee}$ and $P_{\mu\tau}$ receive corrections 
only at second order in the small parameters, where the lengthy contribution to 
$P_{\mu\tau}$ is 
$- (\frac 12 + 2 \, c_{2\delta} \, c_{12}^2 \, s_{12}^2) \, |U_{e3}|^2 
+ \frac 12 c_\delta \, \sin 4 \theta_{12} \, |U_{e3}| \, \epsilon 
-2 \, (1 - c_{12}^2 \, s_{12}^2) \, \epsilon^2$.\\  

\begin{figure}[htb]
\begin{center}
\begin{tabular}{cc}
\epsfig{file=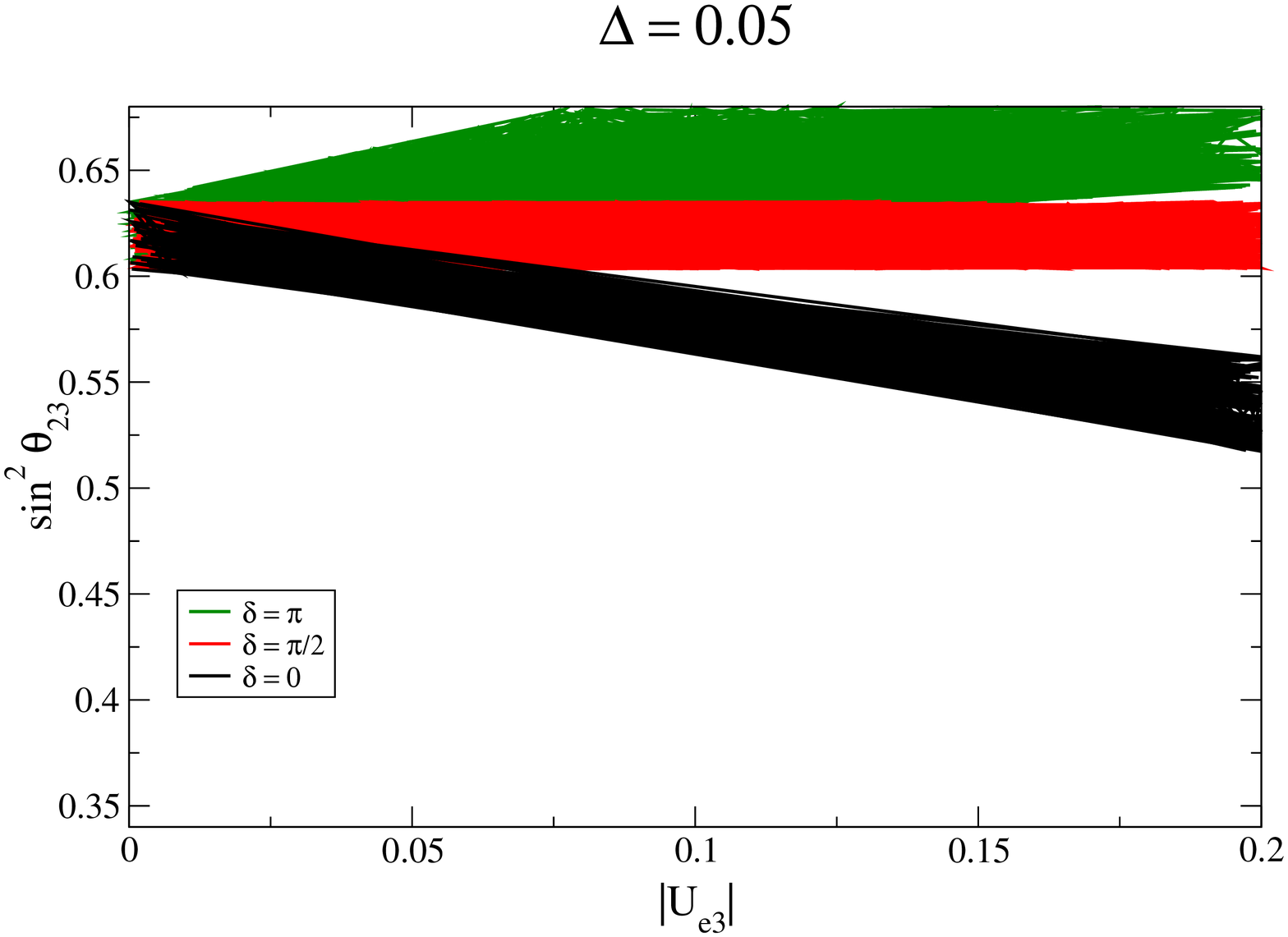,width=8cm,height=6.7cm} & 
\epsfig{file=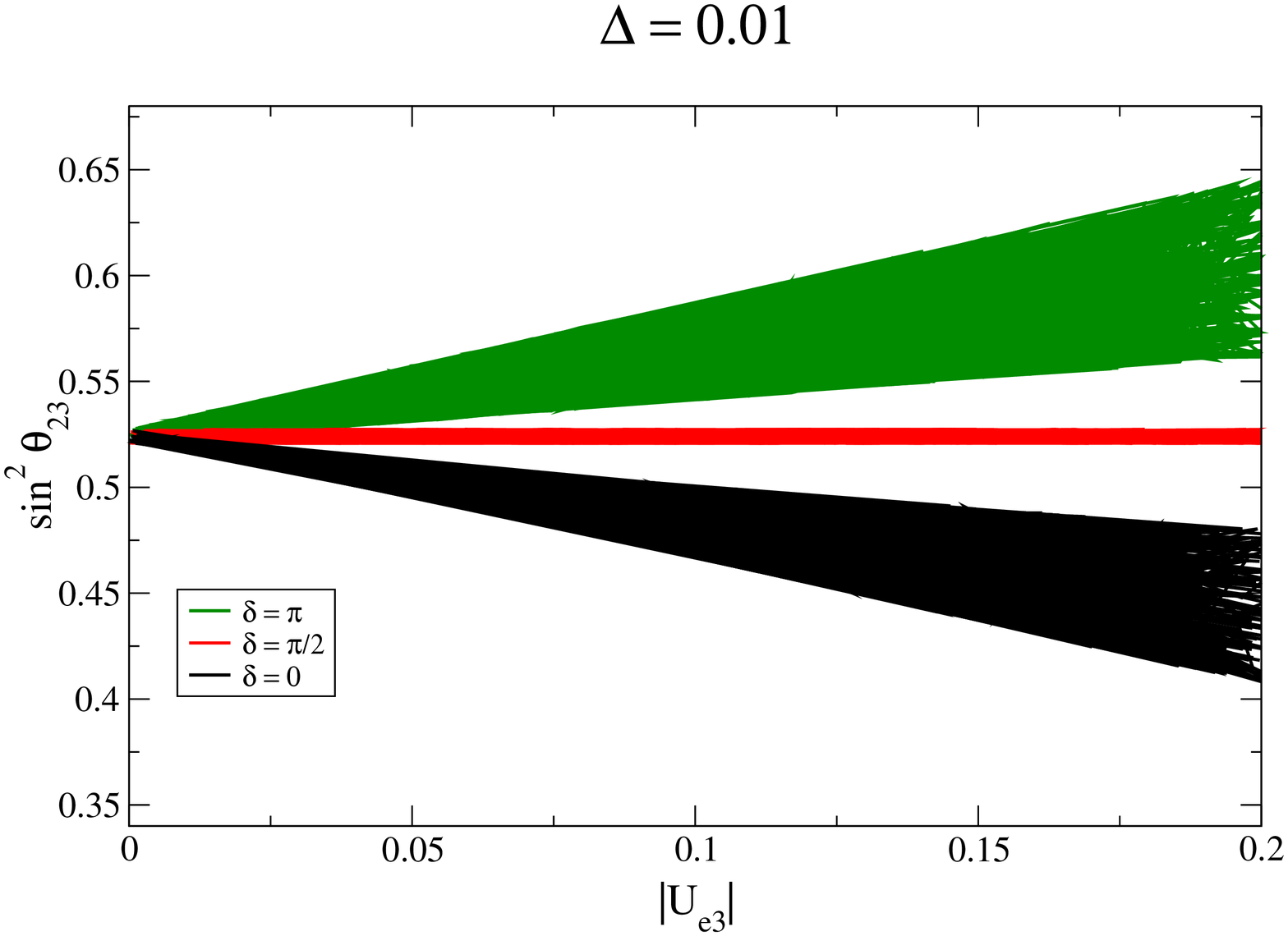,width=8cm,height=6.7cm} \\ 
\epsfig{file=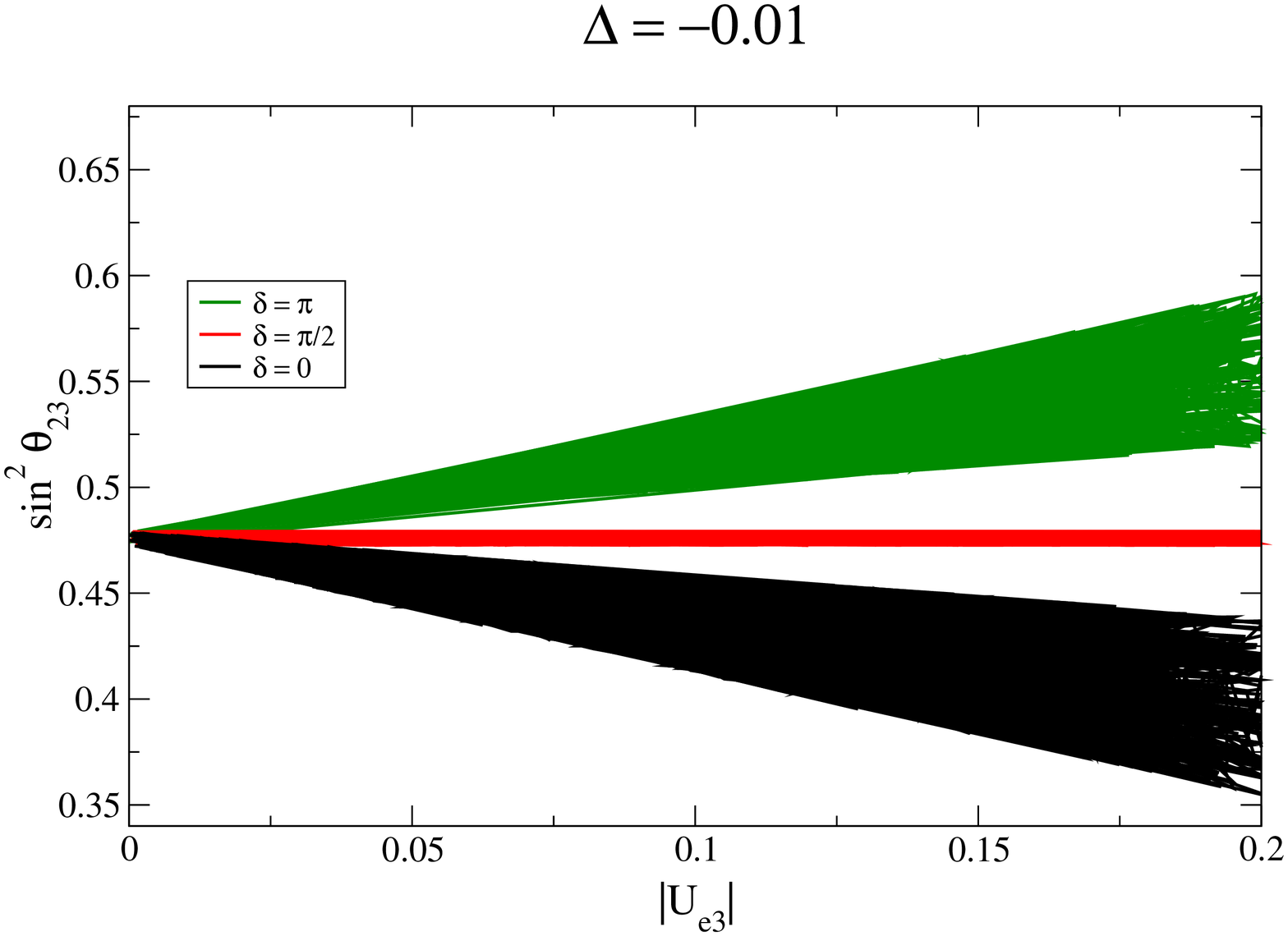,width=8cm,height=6.7cm} &  
\epsfig{file=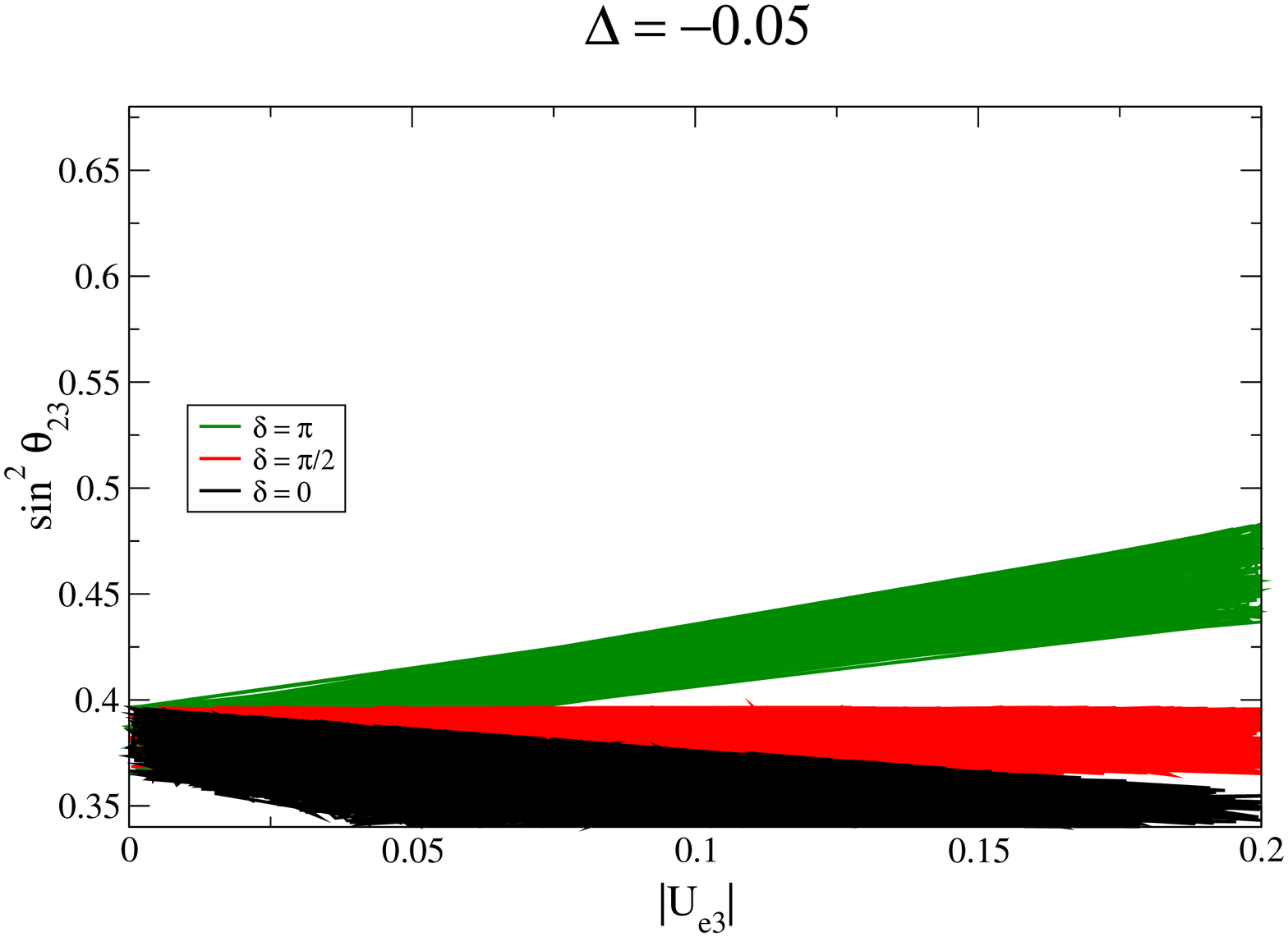,width=8cm,height=6.7cm}  \end{tabular}
\epsfig{file=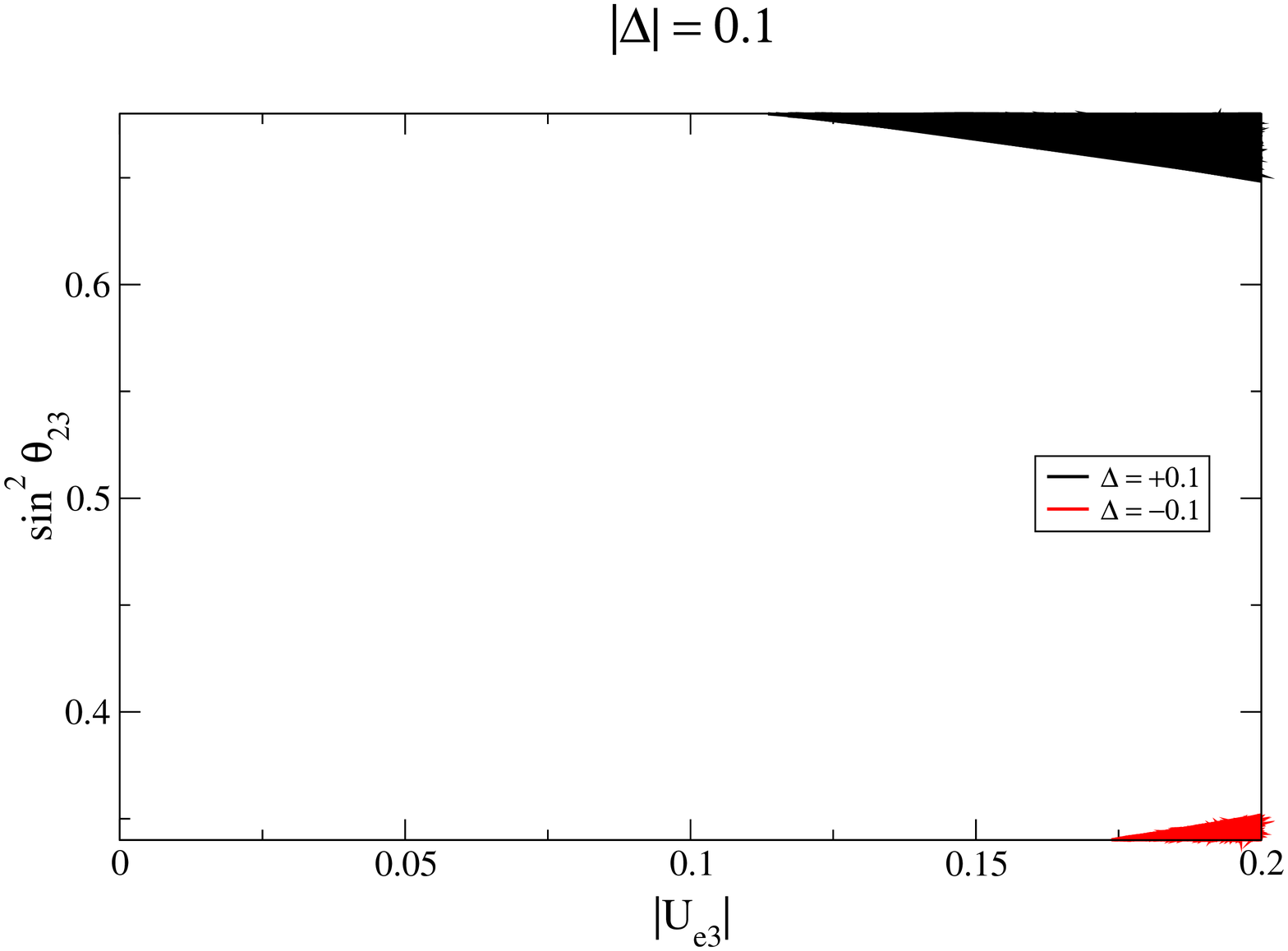,width=8cm,height=6.7cm}
\end{center}
\caption{\label{fig:Delta0}The ranges of $\theta_{13}$ and $\theta_{23}$ 
giving different values of 
$\Delta =  \frac 14 \, \cos \delta \, \sin 4 \theta_{12} \, |U_{e3}| 
+ 2 \, s_{12}^2 \, c_{12}^2 \, \epsilon$, where 
$\theta_{23} = \pi/4 - \epsilon$. The four upper plots are for 
different values of $\delta = \pi$ (upper or green range), 
$\delta = \pi/2$ (middle or red range) and $\delta = 0$ 
(lower or black range). The lowest plot has $\delta$ as a free parameter.}
\end{figure}

The message of Eq.~(\ref{eq:probs}) is that there is at first order a 
universal correction $\Delta$ to the neutrino mixing probabilities. 
To indicate here the importance of this fact (studied in mode detail in 
the remaining Sections), note that 
the neutrino fluxes arriving 
at terrestrial detectors will be functions of the mixing  
probabilities and therefore also functions of $\Delta$. Consequently, the flux 
ratios will depend on $\Delta$ (except for cases treated in Section 
\ref{sec:special}). 
A detailed study of $\Delta$ (see also \cite{xing2}) 
is therefore very useful. 
We plot in Fig.~\ref{fig:Delta0} for different 
values of $\Delta$ the ranges of 
$\theta_{13}$ and $\theta_{23}$ which yield this value. 
The mixing angle $\theta_{12}$ is varied within 
its current $3\sigma$ range. We have chosen 
for $|\Delta| = 0.01$, 0.05 three extreme values of the phases. 
For $\delta = \pi/2$ there is no dependence on $|U_{e3}|$ and if 
the phase is kept as a free parameter then 
the whole area between the upper line for $\delta = \pi$ and the lower line for 
$\delta = 0$ is covered. Hence, if $\theta_{13}$, $\theta_{23}$ 
and $\theta_{12}$ are known with sufficient precision, then 
information on the $CP$ phase $\delta$ can be obtained \cite{KS,WW}. 
The lowest plot in Fig.~\ref{fig:Delta0} is for $|\Delta| = 0.1$, in which case 
not all phase values are allowed (see below). 
Large positive $\Delta \simeq 0.05$ implies 
$\sin^2 \theta_{23} \ge 1/2$. Basically the same holds true for slightly 
smaller $\Delta \simeq 0.01$, unless $\delta \simeq 0$ and $|U_{e3}|$ is sizable. 
For sizable and negative $\Delta \simeq -0.05$ one has 
$\sin^2 \theta_{23} \le 1/2$.  
The same is true for $\Delta \simeq -0.01$ unless $\delta = \pi/2$ and 
$|U_{e3}|$ is sizable. Hence, the octant, i.e., $\sin^2 \theta_{23}$ above or below 
1/2, reflects for small $U_{e3}$ and no extreme values 
of $\delta$ in positive or negative $\Delta$. 
Therefore, in this kind of measurements there might be no octant degeneracy, 
from which the 
interpretation of the results of future long baseline neutrino 
experiments may suffer \cite{8fold}. 
The dependence on the atmospheric mixing angle is in general 
stronger than on $|U_{e3}|$. This occurs because in Eq.~(\ref{eq:Delta}) 
the contribution proportional 
to $|U_{e3}|$ is suppressed by a factor $\frac 14 \, 
\sin 4 \theta_{12}$, which is always approximately two to three times smaller 
than $2 \, c_{12}^2 \, s_{12}^2$ with which $\epsilon$ is multiplied. 
We will discuss the cases in which one of the 
$\mu$--$\tau$ breaking parameters is negligible and the other one sizable 
in Section \ref{sec:AvsB}.

\begin{figure}[htb]
\begin{center}
\begin{tabular}{cc}
\epsfig{file=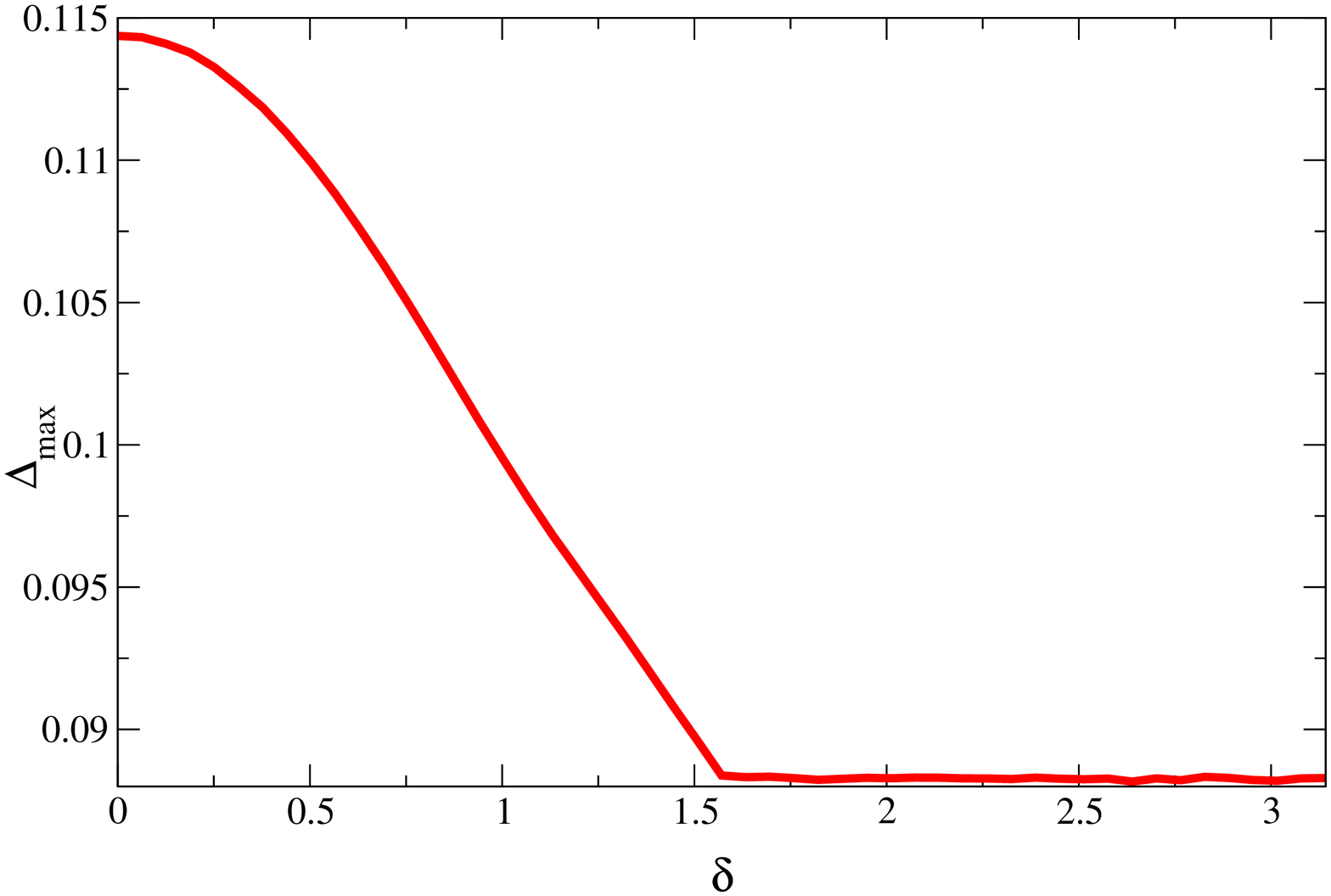,width=8cm,height=5.4cm} & 
\epsfig{file=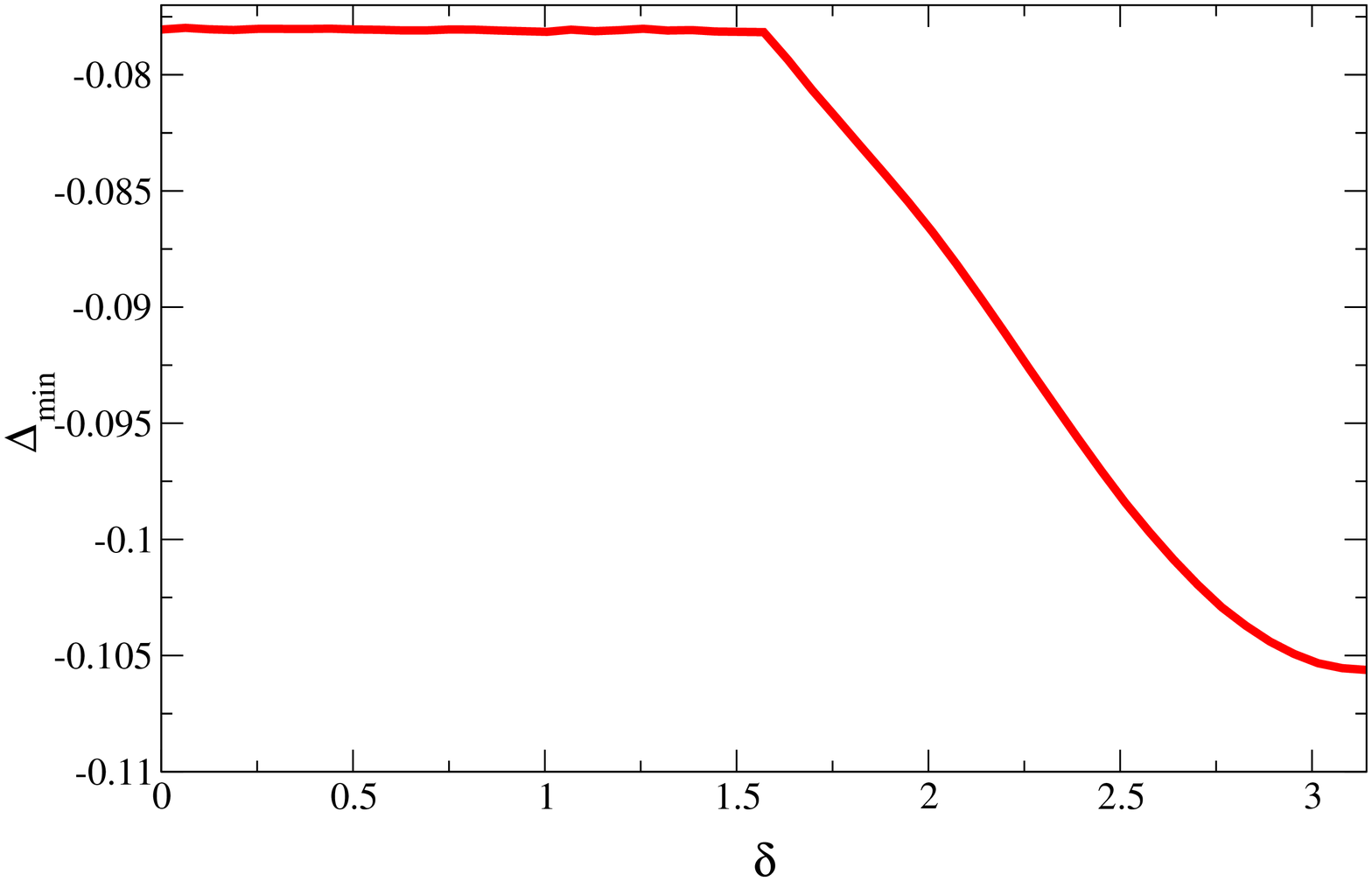,width=8cm,height=5.4cm} \\ 
\epsfig{file=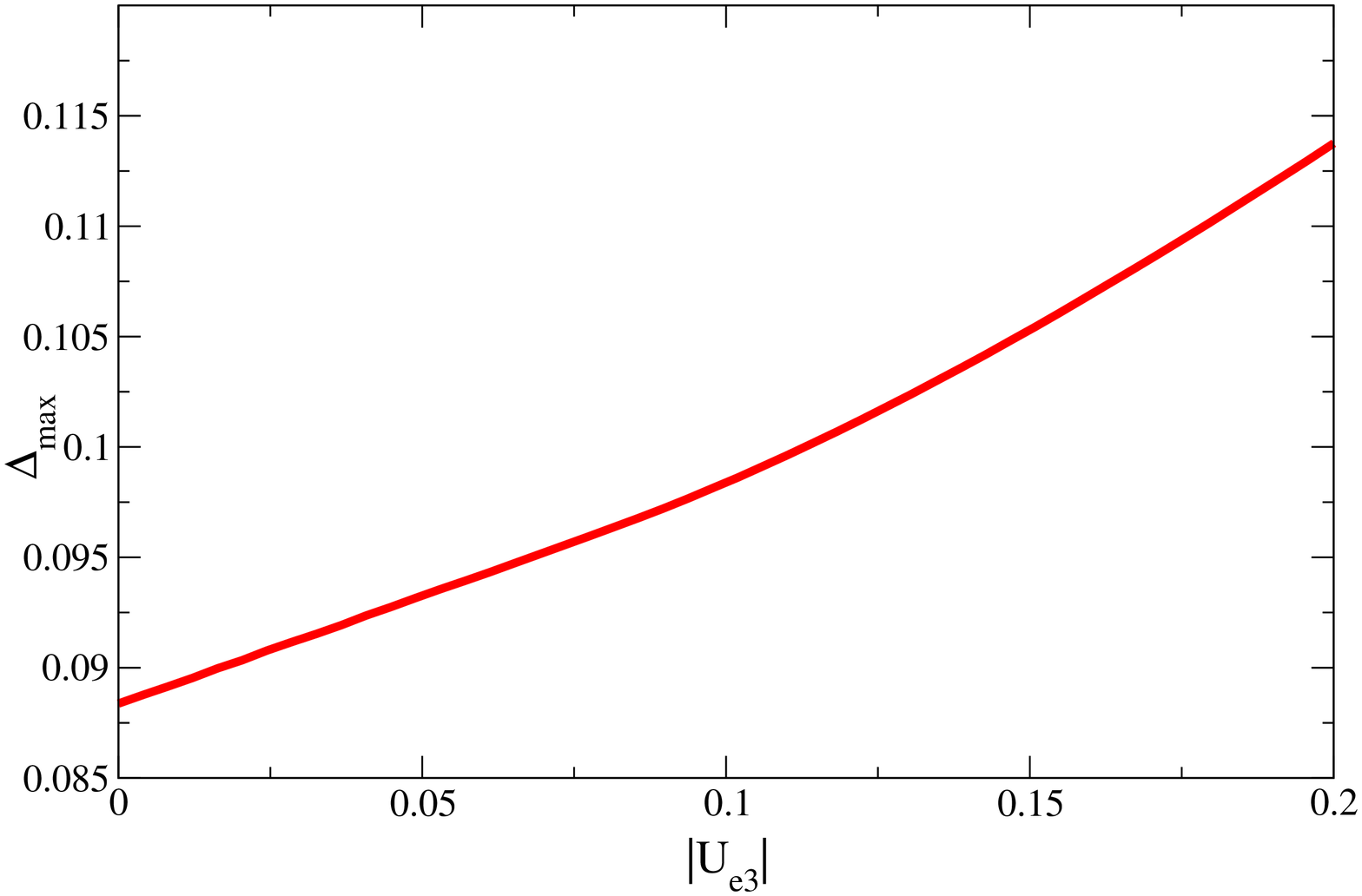,width=8cm,height=5.4cm} &  
\epsfig{file=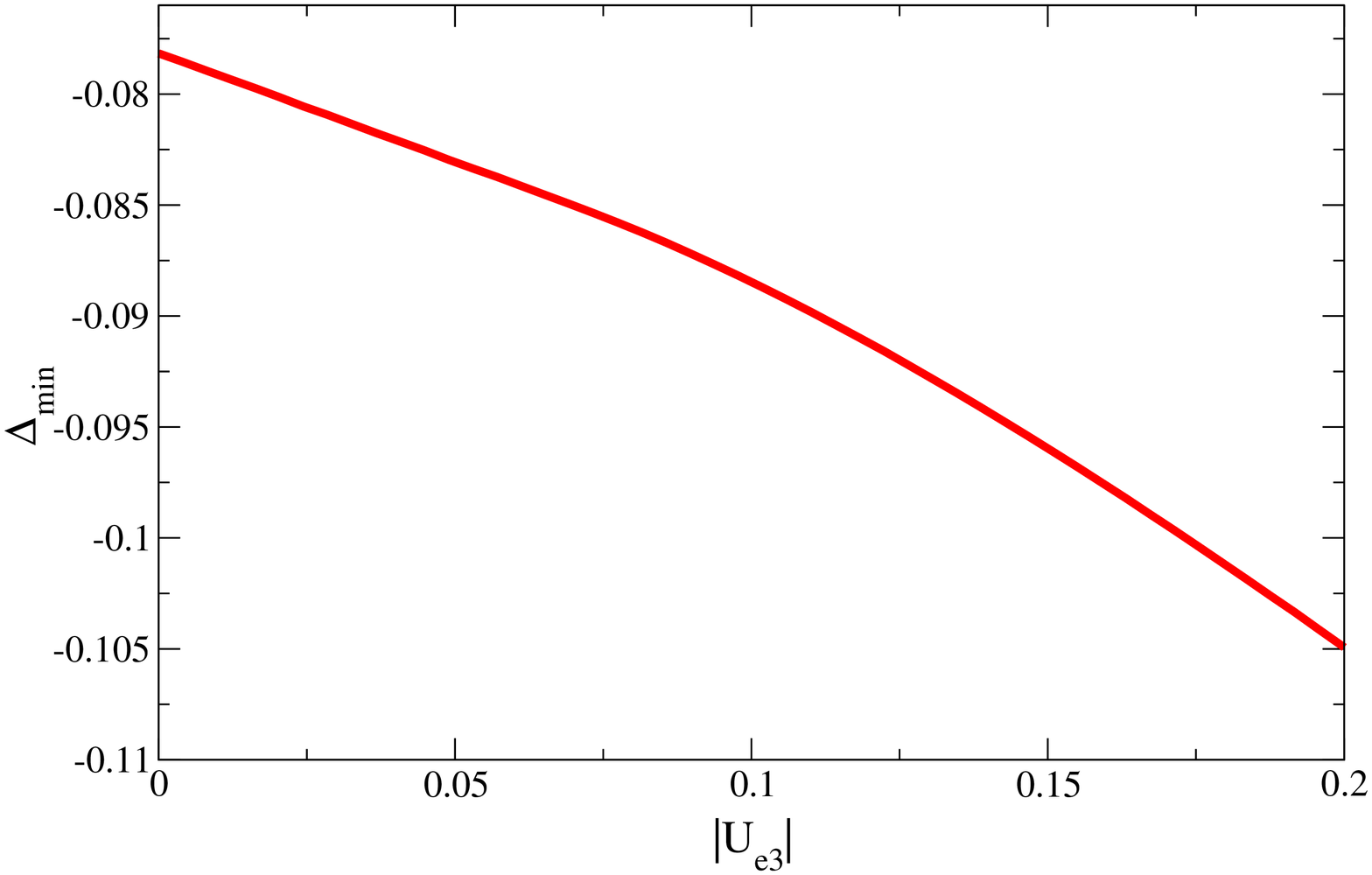,width=8cm,height=5.4cm} \\
\epsfig{file=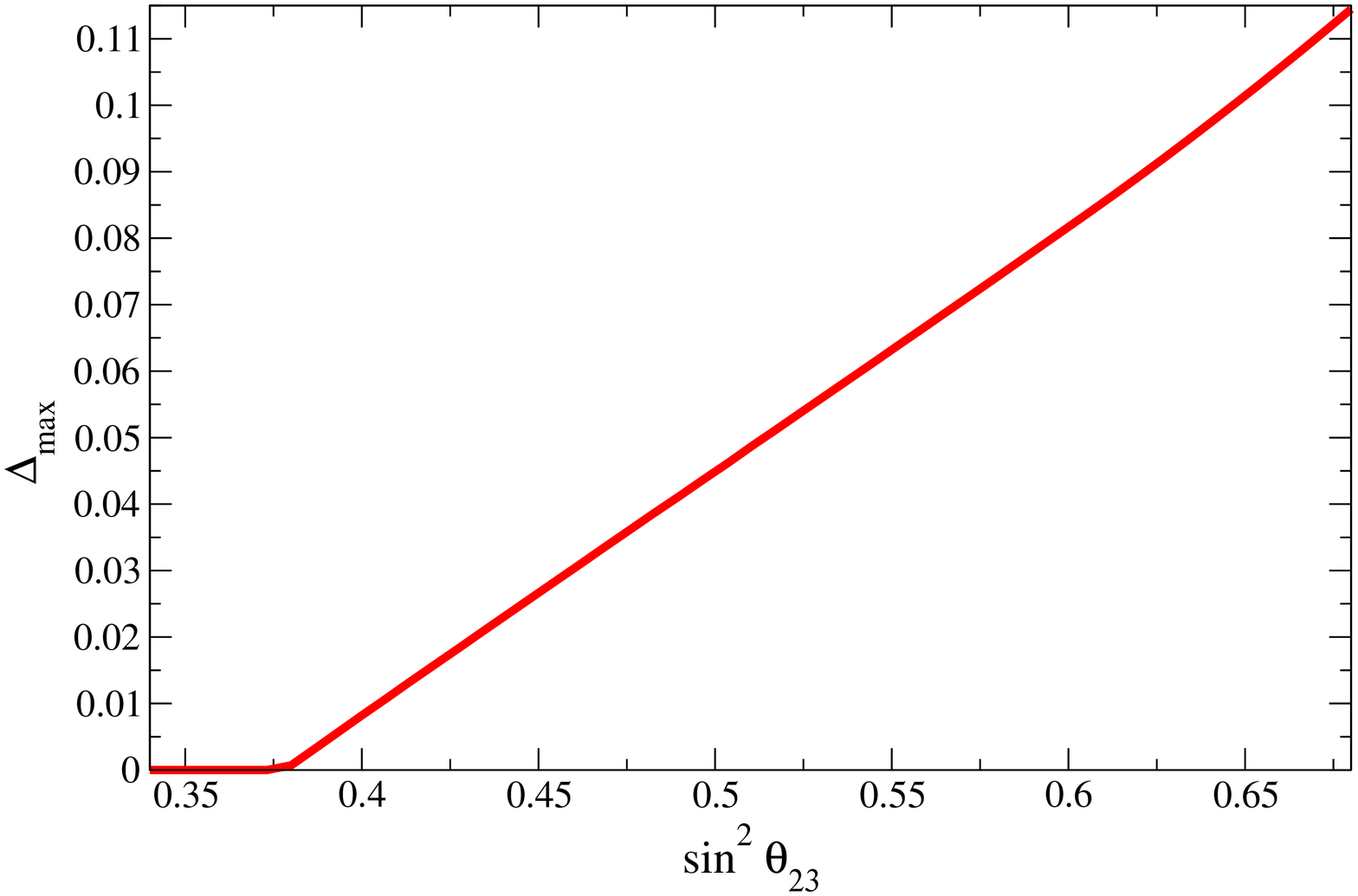,width=8cm,height=5.4cm} & 
\epsfig{file=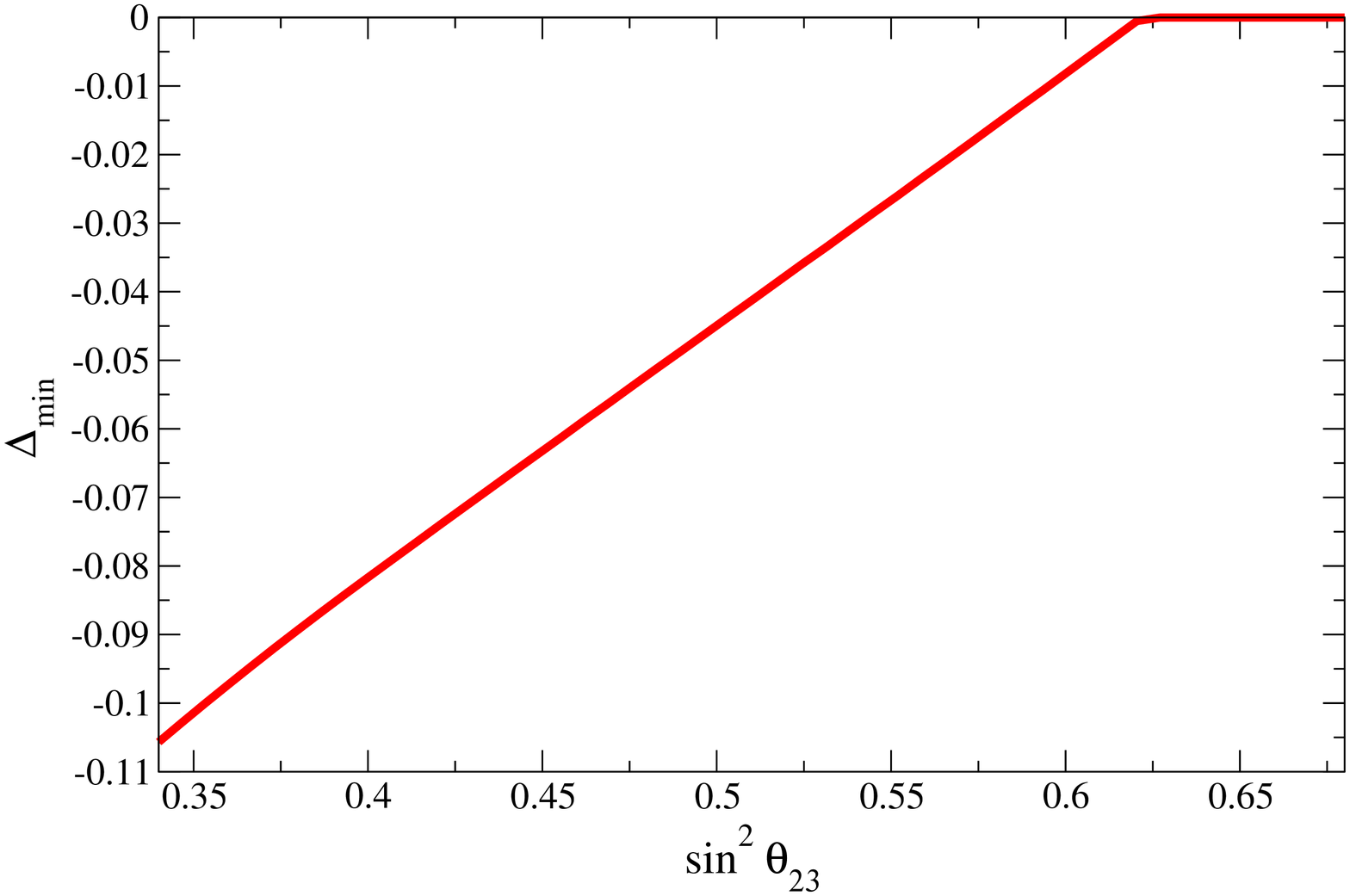,width=8cm,height=5.4cm} \\
\epsfig{file=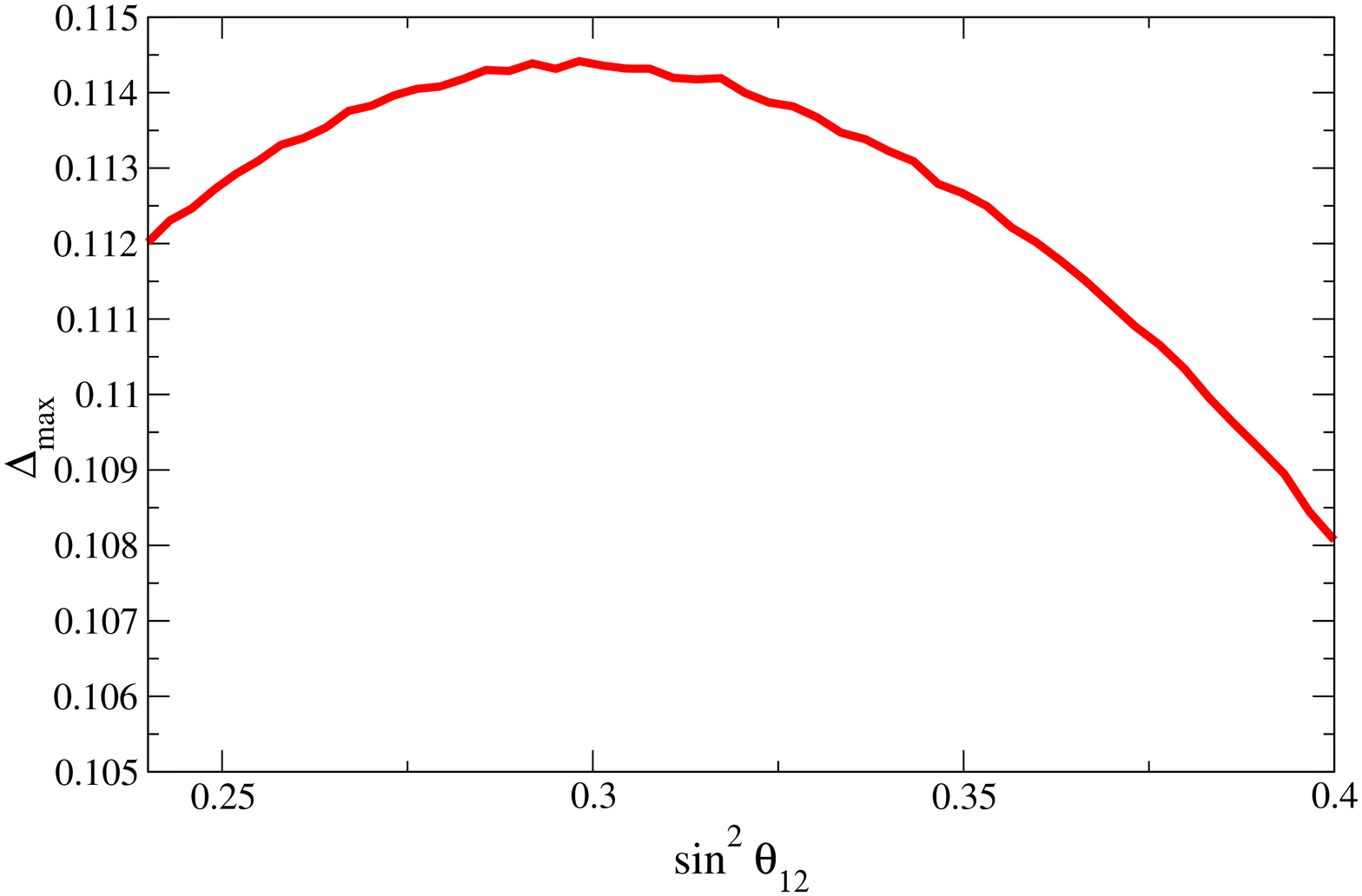,width=8cm,height=5.4cm} & 
\epsfig{file=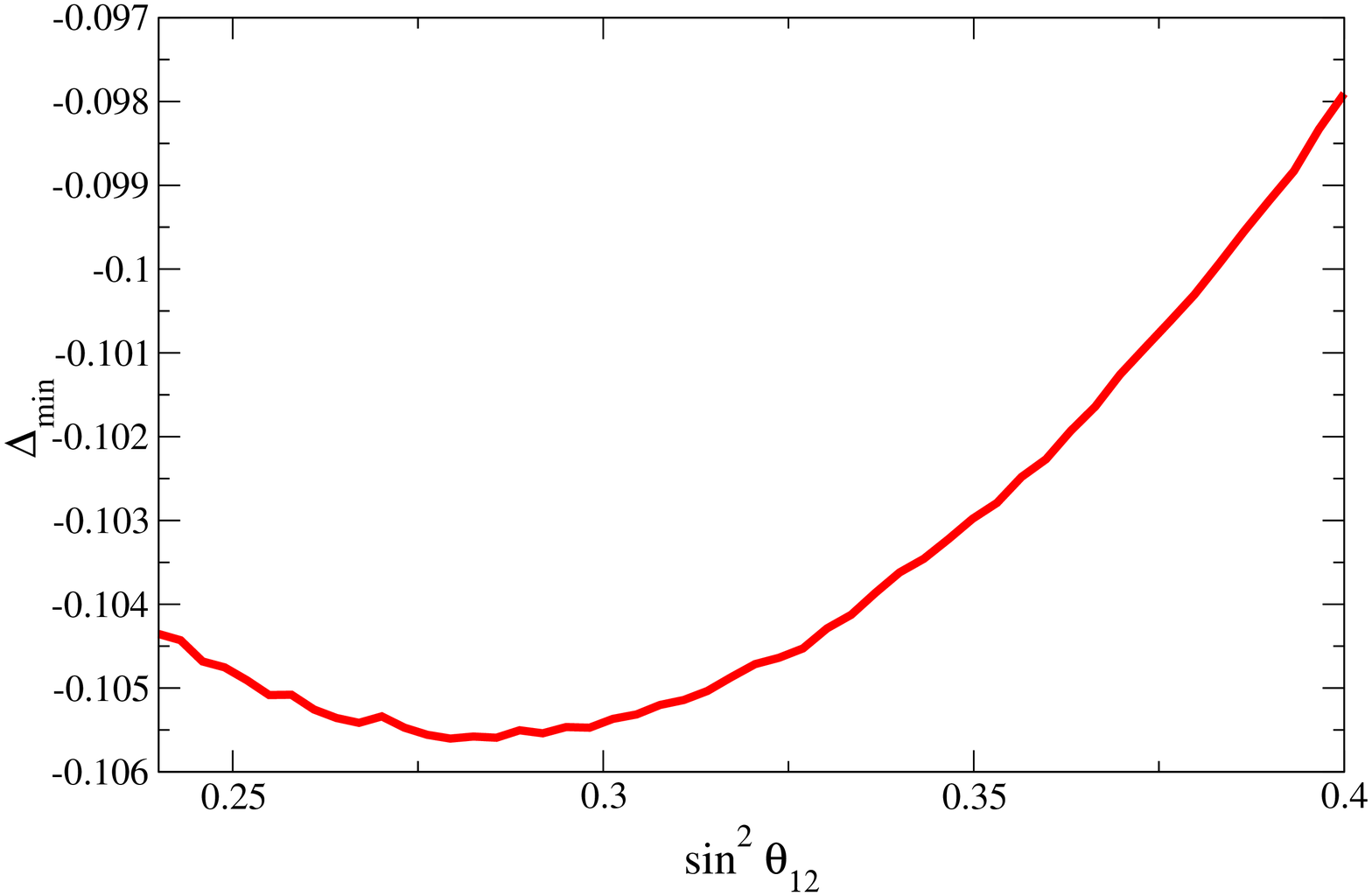,width=8cm,height=5.4cm}
\end{tabular}
\end{center}
\caption{\label{fig:Delta01}The largest and smallest possible 
value of $\Delta$ as a function 
of the mixing parameters $\theta_{12, 13, 23}$ and $\delta$, when the 
other three  
are inside their $3\sigma$ ranges.}
\end{figure}

We have seen that for large values of $|\Delta|$ not all phase 
values are allowed. This is illustrated further in Fig.~\ref{fig:Delta01}, 
where we give the largest and smallest possible value of 
$\Delta$ as a function of each of the mixing parameters 
$\theta_{12, 13, 23}$ and $\delta$, 
when the other three are varied within their current $3\sigma$ ranges. 
Both positive and negative $\Delta$ 
are considered. Again one can see that the dependence on 
the atmospheric mixing angle is stronger than on $|U_{e3}|$. 
One feature worth explaining is the horizontal line 
for $\Delta_{\rm max}$ against $\delta$. If $\delta \ge \pi/2$, then 
$\cos \delta \le 0$ and $\Delta_{\rm max} 
= 2 \, c_{12}^2 \, s_{12}^2 \, \epsilon$, a value reached when $|U_{e3}| = 0$. 
We stress that the dependence on $\theta_{12}$ is rather weak 
(see also \cite{xing2}):  
the maximum $\Delta$ as a function of $s_{12}^2$ changes only 
by $\simeq 5\%$, a much smaller effect than for the other cases.\\ 

A measurement of a flux ratio will be a measurement of 
$\Delta$ and -- depending on 
its magnitude -- the allowed ranges of the mixing parameters can be 
read off Eq.~(\ref{eq:Delta}) or Fig.~\ref{fig:Delta0}. Additional information 
on the angles, no matter if this means a better limit or even a precise 
measurement, will further constrain the possible values. 
Let us point out already at this stage of our investigation that 
there are basically two classes of flux ratios: 
those whose zeroth order expression, i.e., the ratio 
for zero $\theta_{13}$ and $\theta_{23}=\pi/4$, depends on $\theta_{12}$ 
and those whose zeroth order expression does not depend on $\theta_{12}$. 
If the zeroth order expression is independent of $\theta_{12}$, then a 
measurement of the ratio will be a clean measurement of $\Delta$. 
As we will see in the next Section, this occurs typically for neutrinos 
from pion sources. 
If the zeroth order ratio depends on $\theta_{12}$ 
there will be additional uncertainty due to our imprecise 
knowledge of $\theta_{12}$ and the sensitivity to $\Delta$ (or to 
$\theta_{13, 23}$ and $\delta$) is low. 
If $\theta_{13}$ and $\theta_{23}-\pi/4$ are close to zero, then 
the ratios will be a useful probe of $\theta_{12}$ if the zeroth order ratio 
is a function of this angle.\\

Now we turn to numerical values and ranges of the probabilities. 
The probabilities take a particularly simple form for 
tri-bimaximal mixing ($\mu$--$\tau$ symmetry with $\sin^2 \theta_{12} = 1/3$): 
\be
P_{\rm TBM}  = 
\left( 
\bad
\frac 59 & \frac 29 & \frac 29 \\[0.2cm]
\cdot & \frac{7}{18} & \frac{7}{18} \\[0.2cm]
\cdot & \cdot & \frac{7}{18} 
\ea
\right)~.
\ee 
In case of $\sin^2 \theta_{12} = 1/3$ the correction parameter 
$\Delta$ takes the form $\sqrt{2}/9 \, \cos \delta \, |U_{e3}| + 4/9 \, \epsilon$.
Let us obtain now the ranges of the 
mixing probabilities $P_{\alpha \beta}$. To this end, 
we vary the mixing angles in 
their allowed ranges given in Eq.~(\ref{eq:data}). 
Note that due to the fact that the probabilities are $CP$ 
conserving, they depend only on $\cos \delta$. Consequently 
the phase $\delta$ needs to 
be varied only between zero and $\pi$. 
The result is 
\be \label{eq:ranges}
P  = 
\left\{
\baz 
\left( 
\bad 
0.51 \div 0.62 & 0.13 \div 0.31 & 0.13 \div 0.32 \\
\cdot & 0.34 \div 0.50 & 0.33 \div 0.40 \\
\cdot & \cdot & 0.33 \div 0.49
\ea
\right) & (\mbox{at }2\sigma)~,\\[0.3cm]
\left( 
\bad 
0.48 \div 0.64 & 0.11 \div 0.34 & 0.11 \div 0.35 \\
\cdot & 0.33 \div 0.54 & 0.30 \div 0.41 \\
\cdot & \cdot & 0.33 \div 0.52
\ea
\right) & (\mbox{at }3\sigma)~,\\[0.3cm]
\left( 
\bad 
0.45 \div 0.66 & 0.09 \div 0.36 & 0.09 \div 0.36 \\
\cdot & 0.33 \div 0.56 & 0.27 \div 0.41 \\
\cdot & \cdot & 0.33 \div 0.55
\ea
\right) &  (\mbox{at }4\sigma)~.
\ea
\right. 
\ee
No survival probability is below $\simeq 33 \%$ or above $\simeq 66 \%$, whereas 
no transition probability is below $\simeq 10 \%$ or above $\simeq 40 \%$. 
Note that in the general oscillation framework the probabilities 
depend also on energy and baseline and in 
addition on the $\Delta m^2$. Consequently, 
several situations can occur in which a certain $P_{\alpha \beta}$ vanishes or 
becomes equal to one \cite{chef}. This is impossible for the 
high energy neutrinos under consideration, a 
property caused in particular by the non-maximal and 
non-zero value of the solar neutrino mixing angle. Therefore, neutrinos 
from astrophysical sources are guaranteed to change their flavor.

\section{\label{sec:fluobs}Possible Fluxes and Ratios}
In this Section we will first discuss various neutrino sources, all of which 
posses a characteristic flux composition. We analyze this composition 
in terms of the solar neutrino mixing angle and the universal 
correction parameter $\Delta$. Then 
we consider various flux ratios, obtain their zeroth order form and their 
correction, before studying the allowed ranges of the ratios.

\subsection{\label{sec:deco}Neutrino Sources and their Flux Composition}
A cosmic neutrino source can be classified for our purposes by the 
initial flux it generates (the flavor composition $F$). 
For neutrinos from pion decay one has 
\bea \label{eq:pi}
F_\pi^{p\gamma} = (\Phi_{e}:\Phi_{\bar e}:\Phi_{\mu}:\Phi_{\bar \mu}:
\Phi_{\tau}:\Phi_{\bar \tau}) = (1:0:1:1:0:0) ~,\\[0.2cm]
F_\pi^{pp} = (\Phi_{e}:\Phi_{\bar e}:\Phi_{\mu}:\Phi_{\bar \mu}:
\Phi_{\tau}:\Phi_{\bar \tau}) = (1:1:2:2:0:0) ~,\\[0.2cm]
\mbox{ or } F_\pi = (\Phi_{e + \bar e}:\Phi_{\mu +\bar \mu}:
\Phi_{\tau + \bar \tau}) = (1:2:0)~.
\eea
Here $\Phi_{\alpha}$ ($\Phi_{\bar\alpha}$) denotes the generated flux of 
neutrinos (anti-neutrinos) with flavor $\alpha = e, \mu, \tau$, 
whereas $\Phi_{\alpha + \bar \alpha}$ is their sum. 
We have given here two compositions for $F_\pi$, depending on whether 
the neutrinos are created by the $p\gamma$ or $pp$ mechanism. 
The next example are neutron beam sources: 
\bea \label{eq:neut}
F_n = (\Phi_{e}:\Phi_{\bar e}:\Phi_{\mu}:\Phi_{\bar \mu}:
\Phi_{\tau}:\Phi_{\bar \tau}) = (0:1:0:0:0:0) ~,\\[0.2cm]
\mbox{ or } F_n = (\Phi_{e + \bar e}:\Phi_{\mu +\bar \mu}:
\Phi_{\tau + \bar \tau}) = (1:0:0) ~.
\eea
We further have muon-damped sources, in which case 
\bea \label{eq:mu}
F_{\Slash{\mu}} = (\Phi_{e}:\Phi_{\bar e}:\Phi_{\mu}:\Phi_{\bar \mu}:
\Phi_{\tau}:\Phi_{\bar \tau}) = (0:0:1:1:0:0)~, \\[0.2cm]
\mbox{ or } F_{\Slash{\mu}} = (\Phi_{e + \bar e}:\Phi_{\mu +\bar \mu}:
\Phi_{\tau + \bar \tau}) = (0:1:0) ~.
\eea
We have given here the flux composition when the initial pions, for which the 
medium is optically thin, are created by the $pp$ mechanism. 
In case of the $p\gamma$ mechanism, $(0:0:1:1:0:0)$ is changed to 
$(0:0:1:0:0:0)$, i.e., there are no anti-neutrinos involved.\\ 

A measurement at a neutrino telescope will 
measure a flux $\Phi_\alpha^{\rm D}$, which is not the initial neutrino 
flux, but the sum of all initial fluxes $\Phi_\beta$ times the probability 
to oscillate into the flavor $\alpha$: 
\be \label{eq:phiD}
\Phi_\alpha^{\rm D} = \sum\limits_\beta P_{\beta \alpha} \, \Phi_\beta~.
\ee
An interesting exercise is to investigate how the composition has changed 
when the neutrinos reach the earth. While with exact 
$\mu$--$\tau$ symmetry it is well-known that 
an initial flux of $(1:2:0)$ is (independent of 
$\theta_{12}$) altered to $(1:1:1)$ \cite{120}, corrections due to 
$\theta_{13} \neq 0$ and $\theta_{23} \neq \pi/4$ can be sizable. 
Indeed, 
it holds that \cite{xing2}
\be \label{eq:pion_cor}
(1:2:0) \longrightarrow (1 + 2 \, \Delta:1 - \Delta: 1 - \Delta)~,
\ee
where the universal correction parameter $\Delta$ as 
defined in Eq.~(\ref{eq:Delta}) can be up to around $\pm 0.1$. 
If we consider 
the composition in terms of neutrinos and anti-neutrinos, then the 
detected fluxes depend on the source. For neutrinos from 
the $p \gamma$ mechanism one has 
\bea
(1:0:1:1:0:0) \longrightarrow \\[0.2cm]
\hspace{-.7cm}
( 1 - c_{12}^2 \, s_{12}^2 + \Delta : c_{12}^2 \, s_{12}^2 + \Delta 
: \frac 12 \, (1 + c_{12}^2 \, s_{12}^2) 
: \frac 12 \, (1 - c_{12}^2 \, s_{12}^2) - \Delta : 
\frac 12 \, (1 + c_{12}^2 \, s_{12}^2) - \Delta : 
\frac 12 \, (1 - c_{12}^2 \, s_{12}^2) ) \\[0.2cm]
\stackrel{\rm TBM}{\longrightarrow} (14:4:11:7:11:7)~ ,
\eea
with $c_{12} = \cos \theta_{12}$ and $s_{12} = \sin \theta_{12}$. 
We have also simplified the expression for tri-bimaximal (TBM) mixing. 
If the neutrinos are generated by the $pp$ mechanism, then 
\bea
(1:1:2:2:0:0) \longrightarrow
(1 + 2 \, \Delta : 1 + 2 \, \Delta: 1 - \Delta : 1 - \Delta : 1 - \Delta 
: 1 - \Delta)~,\\[0.2cm]
\stackrel{\rm TBM}{\longrightarrow} (1:1:1:1:1:1)
\eea
where the leading term is independent of $\theta_{12}$. 
If we sum up neutrinos and anti-neutrinos, then 
the corrections due to breaking of 
$\mu$--$\tau$ symmetry add up to the one from Eq.~(\ref{eq:pion_cor}). 
For neutrinos from a neutron source: 
\be
(1:0:0) \longrightarrow 
(1 - 2 \, c_{12}^2 \, s_{12}^2 : c_{12}^2 \, s_{12}^2 + \Delta : 
 c_{12}^2 \, s_{12}^2 - \Delta)
\stackrel{\rm TBM}{\longrightarrow} (5:2:2)~,
\ee
or written in terms of neutrino and anti-neutrino flavor:
\bea
(0:1:0:0:0:0) \longrightarrow
(0 : 1 - 2 \, c_{12}^2 \, s_{12}^2 : 0 : c_{12}^2 \, s_{12}^2 + \Delta : 0 : 
c_{12}^2 \, s_{12}^2 - \Delta) \\[0.2cm]
\stackrel{\rm TBM}{\longrightarrow} (0:5:0:2:0:2)~. 
\eea
Finally, muon-damped sources lead to 
\bea
(0:1:0) \longrightarrow  
(c_{12}^2 \, s_{12}^2 + \Delta : 
\frac 12 \, (1 - c_{12}^2 \, s_{12}^2) - \Delta : 
\frac 12 \, (1 - c_{12}^2 \, s_{12}^2) )
\stackrel{\rm TBM}{\longrightarrow} (4:7:7)~,
\eea
or for neutrinos and anti-neutrinos: 
\bea \label{eq:m}
(0:0:1:1:0:0) \longrightarrow \\[0.2cm]
c_{12}^2 \, s_{12}^2 + \Delta : c_{12}^2 \, s_{12}^2 + \Delta : 
\frac 12 \, (1 - c_{12}^2 \, s_{12}^2) - \Delta : 
\frac 12 \, (1 - c_{12}^2 \, s_{12}^2) - \Delta : 
\frac 12 \, (1 - c_{12}^2 \, s_{12}^2) : 
\frac 12 \, (1 - c_{12}^2 \, s_{12}^2) ) \\[0.2cm]
\stackrel{\rm TBM}{\longrightarrow} (4:4:7:7:7:7)~. 
\eea
Recall that we only have discussed the $pp$ mechanism, for the $p\gamma$ case 
there are no anti-neutrinos involved, and the corresponding fluxes  
in Eq.~(\ref{eq:m}) have to be set to zero.
With the results obtained for the flux compositions, one can 
evaluate now the flux ratios, which is what will be done in the 
following Subsection.

\subsection{\label{sec:ratios}Flux Ratios}
The obvious and well-known conclusion from the previous Subsection is 
that the relative 
flux composition is characteristic for the source. 
If in an experiment one measures the flux ratios instead of the 
fluxes, the uncertainties associated in particular with the overall 
magnitude of the fluxes cancel. 
We will consider here 
the following possible observables: 
\bea \label{eq:obs}
R \equiv \frac{\D \Phi_{\mu + \bar\mu}^{\rm D}}
{\D \Phi_{e + \bar e}^{\rm D} + \Phi_{\tau + \bar \tau}^{\rm D}}~~,~~~~
S \equiv \frac{\D \Phi_{e + \bar e}^{\rm D}}
{\D \Phi_{\mu + \bar \mu}^{\rm D} + \Phi_{\tau + \bar \tau}^{\rm D}}~~,~~~~
T \equiv \frac{\D \Phi_{e + \bar e}^{\rm D}}
{\D \Phi_{\tau + \bar \tau}^{\rm D}}~~~~\\[0.3cm]
U \equiv \frac{\D \Phi_{\bar e}^{\rm D}}
{\D \Phi_{\mu + \bar \mu}^{\rm D}}~~,~~~~
V \equiv \frac{\D \Phi_{\bar e}^{\rm D}}
{\D \Phi_{\mu + \bar \mu}^{\rm D} + \Phi_{\tau + \bar \tau}^{\rm D}}~.
\eea
Note that only two of the three ratios 
$R, \, S$ and $T$ are independent. 
The quantities $U$ and $V$ are required 
for Glashow resonance related measurements. Perhaps more directly 
related to an actual measurement (though easily obtainable 
from $R, S, T$ and $U,V$) is the ratio of the muon neutrino, 
or anti-electron neutrino, flux divided by the total flux: 
\be \label{eq:obs1}
Q \equiv \frac{\D \Phi_{\mu + \bar\mu}^{\rm D}}
{\D \Phi_{e + \bar e}^{\rm D} + 
\Phi_{\mu + \bar\mu}^{\rm D} + \Phi_{\tau + \bar \tau}^{\rm D}} 
~\mbox{ and }~
W \equiv \frac{\D \Phi_{\bar e}^{\rm D}}
{\D \Phi_{e + \bar e}^{\rm D} + 
\Phi_{\mu + \bar \mu}^{\rm D} + \Phi_{\tau + \bar \tau}^{\rm D}}~.
\ee
\begin{table}[t]\hspace{-1.3cm}
\begin{tabular}{|c|c|c|c|c|}\hline 
           & \multicolumn{4}{c|}{Source} \\ \hline 
Observable & \multicolumn{2}{c|}{pion} & neutron & muon-damped 
\\ \hline \hline 
$Q\equiv \frac{\D \Phi_{\mu + \bar\mu}^{\rm D}}
{\D \Phi_{e + \bar e}^{\rm D} + \Phi_{\mu + \bar\mu}^{\rm D} + 
\Phi_{\tau + \bar \tau}^{\rm D}}$ 
&  \multicolumn{2}{c|}{$\frac{\D 1}{\D 3} (P_{e \mu} + 2 \, P_{\mu\mu}) $} 
& $ P_{e \mu}$ 
& $ P_{\mu\mu}$ \\ \hline
$R\equiv \frac{\D \Phi_{\mu + \bar\mu}^{\rm D}}
{\D \Phi_{e + \bar e}^{\rm D} + \Phi_{\tau + \bar \tau}^{\rm D}}$ 
&  \multicolumn{2}{c|}{$\frac{\D P_{e\mu} + 2 \, P_{\mu\mu}}
{\D 1 - P_{e\mu} + 2\, (1 - P_{\mu\mu})}$} 
& $\frac{\D P_{e\mu}}{\D 1 - P_{e\mu}} $ 
& $\frac{\D P_{\mu\mu}}{\D 1 - P_{\mu\mu}}$ \\ \hline
$S\equiv \frac{\D \Phi_{e + \bar e}^{\rm D}}
{\D \Phi_{\mu + \bar \mu}^{\rm D} + \Phi_{\tau + \bar \tau}^{\rm D}}$ 
& \multicolumn{2}{c|}{$\frac{\D P_{ee} + 2 \, P_{e\mu}}
{\D 1 - P_{ee} + 2 \, (1 - P_{e \mu})}$} & 
$\frac{\D P_{ee}}{\D 1 - P_{ee}}$ 
& $\frac{\D P_{e\mu }}{\D 1 - P_{e\mu}}$\\ \hline
$T\equiv \frac{\D \Phi_{e + \bar e}^{\rm D}}
{\D \Phi_{\tau + \bar \tau}^{\rm D}}$ 
& \multicolumn{2}{c|}{$\frac{\D 1 - P_{e\tau} +  P_{e\mu}}
{\D P_{e\tau} + 2 \, P_{\mu \tau}}= \frac{\D P_{ee} + 2 \, P_{e\mu}}
{\D 3 \, (1 - P_{e\mu}) - P_{ee} - 2 \, P_{\mu\mu}}$} & 
$\frac{\D P_{ee}}{\D P_{e\tau}}$ & $\frac{\D P_{e\mu }}{\D P_{\mu\tau}}$\\ 
\hline 
& pion $(p\gamma)$ & pion $(pp)$ & &  \\ \hline 
$W \equiv \frac{\D \Phi_{\bar e}^{\rm D}}
{\D \Phi_{e + \bar e}^{\rm D} + 
\Phi_{\mu + \bar \mu}^{\rm D} + \Phi_{\tau + \bar \tau}^{\rm D}}$ &  
$ \frac{\D 1}{\D 3} \, P_{e \mu}$  
& $ \frac{\D 1}{\D 6} \, (P_{ee} + 2 \, P_{e \mu})$  
& $ P_{ee} $ & $  \frac{\D 1}{\D 2} \, P_{e\mu}$ \\ \hline
$U \equiv \frac{\D \Phi_{\bar e}^{\rm D}}
{\D \Phi_{\mu + \bar \mu}^{\rm D}}$ 
& $\frac{\D P_{e\mu}}{\D P_{e\mu} + 2 \, P_{\mu\mu}}$ & 
$\frac{\D 1}{\D 2} \, 
\frac{\D P_{ee} + 2 \, P_{e \mu}}{\D P_{e \mu} + 2 \, P_{\mu\mu}}$ 
& $\frac{\D P_{ee}}{\D P_{e\mu}}$ 
& $\frac{\D P_{e\mu}}{\D 2 \, P_{\mu\mu}}$ \\ \hline
$V\equiv \frac{\D \Phi_{\bar e}^{\rm D}}
{\D \Phi_{\mu + \bar \mu}^{\rm D} + \Phi_{\tau + \bar \tau}^{\rm D}}$ &  
$\frac{\D P_{e \mu}}{\D (1 - P_{ee}) + 2 \, (1 - P_{e\mu})}$  
& $\frac{\D 1}{\D 2} \, \frac{\D P_{ee} + 2 \, P_{e\mu}}
{\D 1 - P_{ee} + 2 \, (1 - P_{e \mu})}$  
& $\frac{\D P_{ee}}{\D 1 - P_{ee}}$ & $ \frac{\D 1}{\D 2} \, 
\frac{\D P_{e\mu}}{\D 1 - P_{e\mu}}$ \\ \hline
\end{tabular}
\caption{\label{tab:obs}Result for the neutrino flux ratios in terms of 
the mixing probabilities.}
\end{table}
With the help of 
Eq.~(\ref{eq:phiD}), the flavor compositions in 
Eqs.~(\ref{eq:pi}, \ref{eq:neut}, \ref{eq:mu}) and the mixing  
probabilities in Eq.~(\ref{eq:Pab}) we can evaluate 
these possible observables in terms of the 
mixing probabilities $P_{\alpha \beta}$. 
The results are shown in Table \ref{tab:obs}. 
There are several relations between the 
observables, for instance $R_n = S_{\Slash{\mu}} = 2 \, V_{\Slash{\mu}}$, 
$S_n = V_n$ and $S_\pi = 2 \, V_\pi^{pp}$. Note that the ratios $W$ and $Q$ are in some 
cases direct measures of mixing probabilities. 
\begin{sidewaystable}
\vspace{-1cm}\hspace{-1cm}
\begin{tabular}{|c|c|c|c|c|}\hline 
           & \multicolumn{4}{c|}{Source} \\ \hline 
Observable & \multicolumn{2}{c|}{pion} & neutron 
& muon-damped \\ \hline \hline 
$Q = \frac{\Phi_{\mu + \bar \mu}^{\rm D}}
{ \Phi_{\rm total}^{\rm D}}$ 
& \multicolumn{2}{c|}{$ \frac{\D 1}{\D 3} (1 - \Delta)$} & 
$c^2 \, s^2 + \Delta \stackrel{\rm TBM}{\longrightarrow} \frac{\D 2}{\D 9}$ & 
$\frac{\D 1}{\D 2} \, (1 - c^2 \, s^2) - \Delta  
\stackrel{\rm TBM}{\longrightarrow} \frac{\D 7}{\D 18}$ \\ \hline  
$R = \frac{ \Phi_{\mu + \bar\mu}^{\rm D}}
{ \Phi_{e + \bar e}^{\rm D} + \Phi_{\tau + \bar \tau}^{\rm D}}$ 
&  \multicolumn{2}{c|}{$\frac {\D 1}{\D 2} - \frac{\D 3}{\D 4} \, \Delta$} & 
$\frac{\D c^2 \, s^2}{\D 1 - c^2 \, s^2} 
+ \frac{\D \Delta}{\D (1 - c^2 \, s^2)^2}
\stackrel{\rm TBM}{\longrightarrow} \frac{\D 2}{\D 7}$ 
& $\frac{\D 1 - c^2 \, s^2}{\D 1 + c^2 \, s^2} 
- 4 \, \frac{\D \Delta}{\D (1 +  c^2 \, s^2)^2} 
\stackrel{\rm TBM}{\longrightarrow} \frac {\D 7}{\D 11}$  \\ \hline
$S= \frac{ \Phi_{e + \bar e}^{\rm D}}
{ \Phi_{\mu + \bar \mu}^{\rm D} + \Phi_{\tau + \bar \tau}^{\rm D}} $ 
& \multicolumn{2}{c|}{$\frac {\D 1}{\D 2} + \frac{\D 3}{\D 2} \, \Delta$} & 
$  
\frac{\D 1 - 2 \, c^2 \, s^2}{\D 2 \, c^2 \, s^2} 
\stackrel{\rm TBM}{\longrightarrow} \frac{\D 5}{\D 4}$ 
& $\frac{\D c^2 \, s^2}{\D 1 - c^2 \, s^2} + 
\frac{\D \Delta}{\D (1 - c^2 \, s^2)^2} 
\stackrel{\rm TBM}{\longrightarrow} \frac {\D 2}{\D 7}$  \\ \hline
$T=\frac{ \Phi_{e + \bar e}^{\rm D}}
{ \Phi_{\tau + \bar \tau}^{\rm D}}$ 
& \multicolumn{2}{c|}{$1 + 3 \, \Delta$} & 
$ \frac{\D 1  - 2 \, c^2 \, s^2}{\D c^2 \, s^2} + 
\frac{\D 1 - 2 \, c^2 \, s^2}{\D c^4 \, s^4} \, \Delta
\stackrel{\rm TBM}{\longrightarrow} \frac{\D 5}{\D 2}$ 
& $ \frac{\D 2 \, c^2 \, s^2}{\D 1 - c^2 \, s^2} 
+ 2 \, \frac{\D \Delta}{\D 1 - c^2 \, s^2}
\stackrel{\rm TBM}{\longrightarrow} \frac {\D 4}{\D 7}$  \\ \hline
\hline 
& pion $(p\gamma)$ & pion $(pp)$ & &  \\ \hline 
$W = \frac{ \Phi_{\bar e}^{\rm D}}
{ \Phi_{\rm total}^{\rm D}}$
& $ \frac{\D 1}{\D 3} \, (c^2 \, s^2 + \Delta) 
\stackrel{\rm TBM}{\longrightarrow} \frac {\D 2}{\D 27}$ 
& $ \frac{\D 1}{\D 6} \, (1 + 2 \, \Delta) 
\stackrel{\rm TBM}{\longrightarrow} \frac {\D 1}{\D 6}$ 
& $ (1 - 2 \, c^2 \, s^2 ) 
\stackrel{\rm TBM}{\longrightarrow} \frac{\D 5}{\D 9}$ 
& $ \frac{\D 1}{\D 2} \, (c^2 \, s^2 + \Delta)
\stackrel{\rm TBM}{\longrightarrow} \frac{\D 1}{\D 9}$ \\ \hline
$U =\frac{ \Phi_{\bar e}^{\rm D}}
{ \Phi_{\mu + \bar \mu}^{\rm D}} $ & $c^2 \, s^2 + (1 + c^2 \, s^2) \, \Delta 
\stackrel{\rm TBM}{\longrightarrow} 
\frac{\D 2}{\D 9}$  
& $\frac {\D 1}{\D 2} + \frac{\D 3}{\D 2} \, \Delta$ & 
$\frac{\D 1 - 2 \, c^2 \, s^2}{\D c^2 \, s^2} - 
\frac{\D 1 - 2 \, c^2 \, s^2}{\D c^4 \, s^4 } \, \Delta  
\stackrel{\rm TBM}{\longrightarrow} \frac{\D 5}{\D 2}$ 
& 
$\frac{\D c^2 \, s^2}{\D 1 - c^2 \, s^2} 
+ \frac{\D 1 + c^2 \, s^2}{\D (1 - c^2 \, s^2)^2} \, \Delta
\stackrel{\rm TBM}{\longrightarrow} \frac {\D 2}{\D 7}$ 
\\ \hline
$V=\frac{ \Phi_{\bar e}^{\rm D}}
{ \Phi_{\mu + \bar \mu}^{\rm D} + \Phi_{\tau + \bar \tau}^{\rm D}}$ &  
 $ \frac{\D 1}{\D 2}\, c^2 \, s^2 + \frac{\D 1}{\D 2} \, 
(1 + c^2 \, s^2) \, \Delta 
\stackrel{\rm TBM}{\longrightarrow} \frac{\D 1}{\D 9}$  
& $\frac{\D 1}{\D 4} + \frac{\D 3}{\D 4} \, \Delta$ & 
$\frac{\D 1 - 2 \, c^2 \, s^2}{\D 2 \, c^2 \, s^2} 
\stackrel{\rm TBM}{\longrightarrow} \frac{\D 5}{\D 4}$ 
& 
$\frac{\D 1}{\D 2} \, 
\frac{\D c^2 \, s^2}{\D 1 - c^2 \, s^2}  + \frac{\D 1}{\D 2} \, 
\frac{\D \Delta}{\D (1 - c^2 \, s^2)^2}
\stackrel{\rm TBM}{\longrightarrow} \frac{\D 1}{\D 7}$ 
\\ \hline
\end{tabular}
\caption{\label{tab:obs_spe}The neutrino flux ratios in case of 
almost exact $\mu$--$\tau$ symmetry and 
for exact $\mu$--$\tau$ symmetry with tri-bimaximal mixing. 
We have defined $c = \cos \theta_{12}$, $s = \sin \theta_{12}$ and 
$\Delta$ is defined as 
$\Delta \simeq \frac 14 \, \cos \delta \, \sin 4 \theta_{12} \, |U_{e3}| 
+ 2 \, s_{12}^2 \, c_{12}^2 \, \epsilon$, where 
$\theta_{23} = \pi/4 - \epsilon$. $\Phi_{\rm total}^{\rm D}$ is the 
flux of all incoming neutrinos and anti-neutrinos at the detector.
If $\Delta$ does not appear in a ratio, then there is a quadratic correction, see 
Section \ref{sec:special}.}
\end{sidewaystable}

\begin{table}[ht]
\begin{center}
\begin{tabular}{|c|c|c|c|c|c|c|c|}\hline 
& $R_\pi$ & $S_\pi$ & $T_\pi$ & $U_\pi^{p\gamma}$ & $U_\pi^{pp}$ & 
$V_\pi^{p\gamma}$ & $V_\pi^{pp}$ \\ \hline \hline
$2\sigma$ & $0.47 \div  0.62 $ & $0.39 \div 0.61$ 
& $0.83 \div 1.30$  & $0.11 \div 0.31$ & $0.37 \div  0.59$ 
& $0.06 \div  0.17$ & $0.19 \div 0.30$  \\ \hline
$3\sigma$ & $0.47 \div 0.66$ & $0.37 \div  0.64$ 
& $0.81 \div  1.42$ & $0.09 \div  0.33$  & 
$0.34 \div  0.59$ & $0.05 \div  0.18$  & $0.18 \div  0.32$  \\ \hline
$4\sigma$ & $0.47 \div 0.71$ & $0.35 \div 0.66$ & $0.81  \div 1.51$ 
& $0.07 \div  0.34$ & $0.32 \div 0.59 $ & $0.04 \div  0.19$  & 
$0.18 \div  0.33$ \\ \hline
\end{tabular}
\end{center}
\caption{\label{tab:pi}Allowed ranges of the observables from 
Eq.~(\ref{eq:obs}) in case of a pion source.}
\begin{center}
\begin{tabular}{|c|c|c|c|c|c|}\hline 
& $R_n$ & $S_n$ & $T_n$ & $U_n$ & $V_n$ \\ \hline \hline
$2\sigma$ & $0.14 \div  0.45$ & $1.05 \div 1.60$ 
& $1.60 \div 4.48$ & $1.64 \div  4.73$  & $1.05 \div 1.60$ \\ \hline
$3\sigma$ & $0.12 \div 0.52$ & $0.92 \div  1.74$ 
& $1.37 \div  5.53$  & $1.41 \div  5.87$ & $0.92 \div  1.74$ \\ \hline
$4\sigma$ & $0.10 \div 0.56$ & $0.83 \div 1.91$ & $1.24 \div 6.74 $
& $1.26 \div 7.22$ & $0.82 \div 1.91$ \\ \hline
\end{tabular}
\end{center}
\caption{\label{tab:n}Same as Table \ref{tab:pi} for a neutron source.}
\begin{center}
\begin{tabular}{|c|c|c|c|c|c|}\hline 
& $R_{\Slash{\mu}}$ & $S_{\Slash{\mu}}$ & $T_{\Slash{\mu}}$ 
& $U_{\Slash{\mu}}$ & $V_{\Slash{\mu}}$ \\ \hline \hline
$2\sigma$ & $0.50 \div 1.02$ & $0.14 \div  0.45$ 
& $0.34 \div   0.89$ & $0.13 \div  0.46$ & $0.07 \div 0.23$ \\ \hline
$3\sigma$ & $0.50 \div  1.16$ & $0.12 \div  0.52$ 
& $0.29 \div  1.06$ & $0.10 \div  0.51$ & $0.06 \div  0.26$ \\ \hline
$4\sigma$ & $0.50 \div 1.30$ & $0.10 \div 0.56$ & $0.26 \div 1.20 $ 
& $0.08 \div 0.53$ & $0.05 \div 0.28$ \\ \hline
\end{tabular}
\end{center}
\caption{\label{tab:m}Same as Table \ref{tab:pi} for a muon-damped 
source.}
\end{table}

\begin{table}[ht]
\begin{center}
\begin{tabular}{|c|c|c|c||c|c||c|c|}\hline 
& $Q_\pi$ & $W_\pi^{ p \gamma}$ & $W_\pi^{ p p}$ & 
$Q_n$ & $W_n$ & $Q_{\Slash{\mu}}$ & $W_{\Slash{\mu}}$ \\ \hline \hline 
$2 \sigma$ & $0.32 \div 0.38$ & $0.04 \div 0.10$ & $0.14 \div 0.19$ 
& $0.13 \div 0.31$ & $0.51 \div 0.62$ & $0.34 \div 0.50$ & $0.06 \div 0.16$ 
\\ \hline   
$3 \sigma$ & $0.32 \div 0.40$ & $0.03 \div 0.11$ & $0.14 \div 0.20$ 
& $0.11 \div 0.34$ & $0.48 \div 0.63$ & $0.33 \div 0.54$ & $0.05 \div 0.17$ 
\\ \hline   
$4 \sigma$ & $0.32 \div 0.42$ & $0.03 \div 0.12$ & $0.13 \div 0.20$ 
& $0.09 \div 0.36$ & $0.45 \div 0.66$ & $0.33 \div 0.56$ & $0.04 \div 0.18$ 
\\ \hline   
\end{tabular}
\end{center}
\caption{\label{tab:WQ}Allowed ranges of the observables from 
Eq.~(\ref{eq:obs1}).}
\end{table}

We can also obtain the expressions for the ratios in case of 
approximate $\mu$--$\tau$ symmetry and for 
tri-bimaximal mixing. 
The results are shown in Table \ref{tab:obs_spe}. 
Except for three ratios (see Section \ref{sec:special}) 
the ubiquitous parameter 
$\Delta$ appears as a correction. This allows for a comparably 
easy understanding 
of the results of future measurements of the ratios. 
In case of a pion source and 
exact $\mu$--$\tau$ symmetry, only the observables related to 
$p\gamma$ neutrinos, which can be distinguished from $pp$ 
neutrinos only through measurements of the Glashow resonance, 
show a dependence on the 
solar neutrino mixing angle. As alluded to before in Section \ref{sec:gen}, 
there are ratios whose zeroth order expressions depend on $\theta_{12}$ 
and those whose do not. The latter are good candidates to 
probe $\theta_{23}$, $\theta_{13}$ and $\delta$, because the measured deviation 
will be directly related to $\Delta$, 
which in turn weakly depends on $\theta_{12}$. From Table \ref{tab:obs_spe} 
it is obvious that pion sources are the only cases in which the zeroth 
order ratio is independent of $\theta_{12}$. If the Glashow resonance is taken 
advantage of, then $pp$-generated neutrinos are necessary 
to measure $\Delta$ directly. All other sources and 
ratios depend at zeroth order on $\theta_{12}$ and therefore 
are useful to probe this observable.\\

Using the whole allowed 
ranges of the mixing probabilities we 
can also calculate the currently allowed ranges of the ratios. 
The results are summarized in 
Tables \ref{tab:pi}, \ref{tab:n} and \ref{tab:m}. 
For the sake of completeness, we also give the ranges of $Q$ and $W$ from 
Eq.~(\ref{eq:obs1}) in Table \ref{tab:WQ}. 
Recall again that with a muon-damped source we 
consider the $pp$ mechanism. For  
$p\gamma$ neutrinos and for damped muons the ratios $R,S,T$ and $Q$ 
are identical to 
the expressions in Tables \ref{tab:obs} and \ref{tab:obs_spe} and to the 
ranges in Tables \ref{tab:pi}, \ref{tab:n}, \ref{tab:m}, \ref{tab:WQ}. 
However, since no anti-neutrinos are involved in such cases, $U$, $V$ and $W$ 
are zero. 

\begin{figure}[htb]
\begin{center}
\epsfig{file=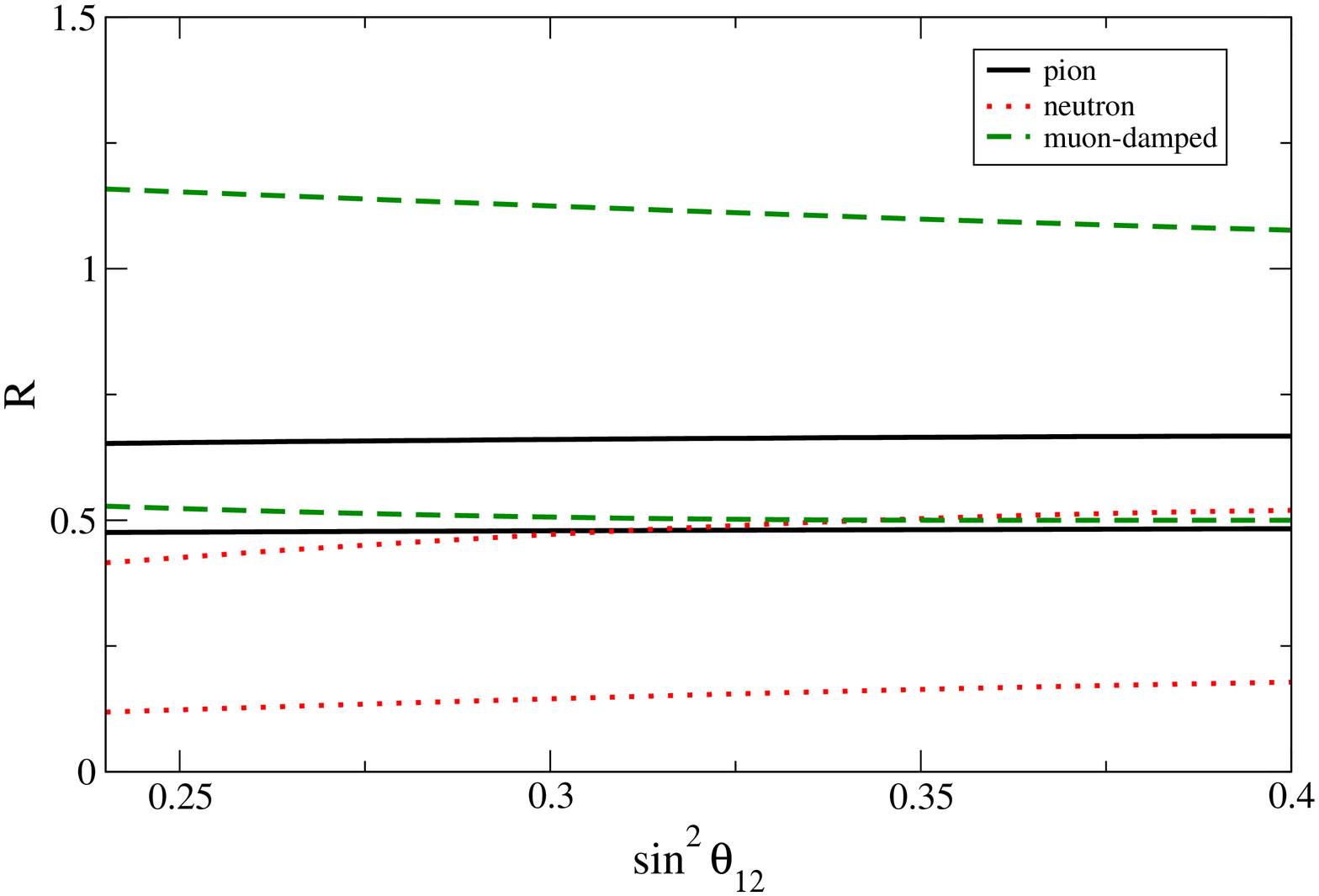,width=14cm,height=9.5cm} 
\epsfig{file=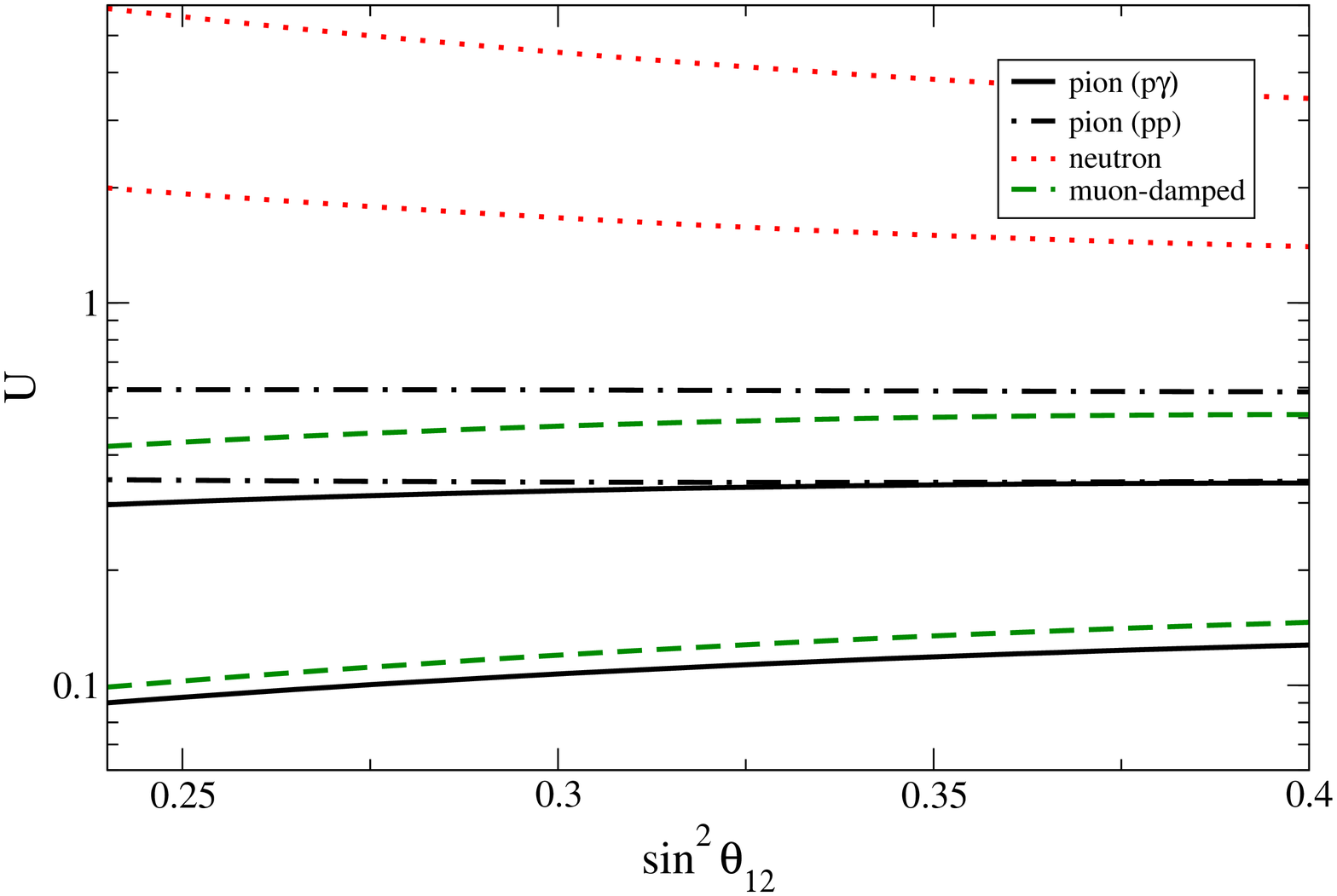,width=14cm,height=9.5cm} 
\end{center}
\caption{\label{fig:ratios12}The ratios $R = \Phi_{\mu + \bar \mu}^{\rm D}/(
\Phi_{e + \bar e}^{\rm D} + \Phi_{\tau + \bar \tau}^{\rm D})$ and 
$U = \Phi_{\bar e}^{\rm D}/\Phi_{\mu + \bar \mu}^{\rm D}$ as a function 
of $\sin^2 \theta_{12}$ for neutrinos from pion, neutron and muon-damped 
sources. 
}
\end{figure}

Even though we focus in this paper only on the functional dependence 
of the ratios, the question arises whether one can identify the neutrino source 
through the measurement of a ratio. 
While such an analysis is beyond the scope of the 
present paper (see, e.g., Refs.~\cite{energy,PdS,xing1,WW} for discussions), 
we give here a few comments. As seen from Tables \ref{tab:pi}, 
\ref{tab:n} and \ref{tab:m}, 
already for the 2$\sigma$ values of the oscillation 
parameters, the ranges of the ratios can overlap, 
in particular for pion and muon-damped sources. 
This is also illustrated in Fig.~\ref{fig:ratios12}, where  
we show the ratios $R$ and $U$ as a function of $\sin^2 \theta_{12}$ when the 
other oscillation parameters are allowed to vary within their current 
$3\sigma$ ranges. In this Figure and in all of the following ones we 
have been using the exact, lengthy mixing probabilities and not the 
approximate formulae from Table \ref{tab:obs_spe}. 
A value of $R \simeq 1/2$ seems to indicate from the upper plot 
of Fig.~\ref{fig:ratios12} that all three sources are allowed. 
Note however that, for instance, $R$ has  
quite different zeroth order values 
(for simplicity we use tri-bimaximal mixing here), namely 
$\frac 12$, $\frac 27$ and $\frac{7}{11}$ 
for pion, neutron and muon-damped sources, respectively. 
Indeed from Table \ref{tab:obs_spe} it is clear that 
$R_n$ and $R_{\Slash{\mu}}$ both around $\simeq 1/2$ would require large 
and positive $\Delta \sim 0.1$, 
whereas $R_\pi \simeq 1/2$ goes with small $\Delta$. 
In addition, neutron source fluxes have a large $\bar \nu_e$ content and 
therefore large $U$, as confirmed by Fig.~\ref{fig:ratios12}. 
Measuring smaller $U$ than expected for a neutron source leaves us 
only with pion or muon-damped sources. Since large and positive 
$\Delta$ is expected for the latter, there will be obvious neutrino 
oscillation phenomenology since almost maximal breaking of $\mu$--$\tau$ 
symmetry is required, see Figs.~\ref{fig:Delta0} and \ref{fig:Delta01}. 
Other examples easily read off Fig.~\ref{fig:ratios12} are the following: 
having $R$ above 
0.66 will mean a muon-damped source, $R$ well below 0.5 will rule out 
pion and muon-damped sources. Measuring $U$ below 0.5 rules out pions 
generated by the $p\gamma$ mechanism and also muon-damped sources. 
In general, ratios using the Glashow resonance have a smaller 
value for $p\gamma$ sources than for $pp$ sources, unless the 
oscillation parameters take extreme values. For instance, 
$U_\pi^{p\gamma}$ has the zeroth order value $\frac 29$, while 
$U_\pi^{pp} = \frac 12$. 
Another example is that a muon-damped source fed with pions from $p\gamma$ 
collisions contains no electron anti-neutrinos and would have $U=0$. 
All in all, in combination \cite{energy,PdS,xing1,WW} with 
issues like astrophysical observations, energy dependent 
differences of the various sources, there will be the 
possibility to distinguish 
the sources.

\section{\label{sec:special}Ratios almost only depending on $\theta_{12}$}

\begin{figure}[htb]
\begin{center}
\epsfig{file=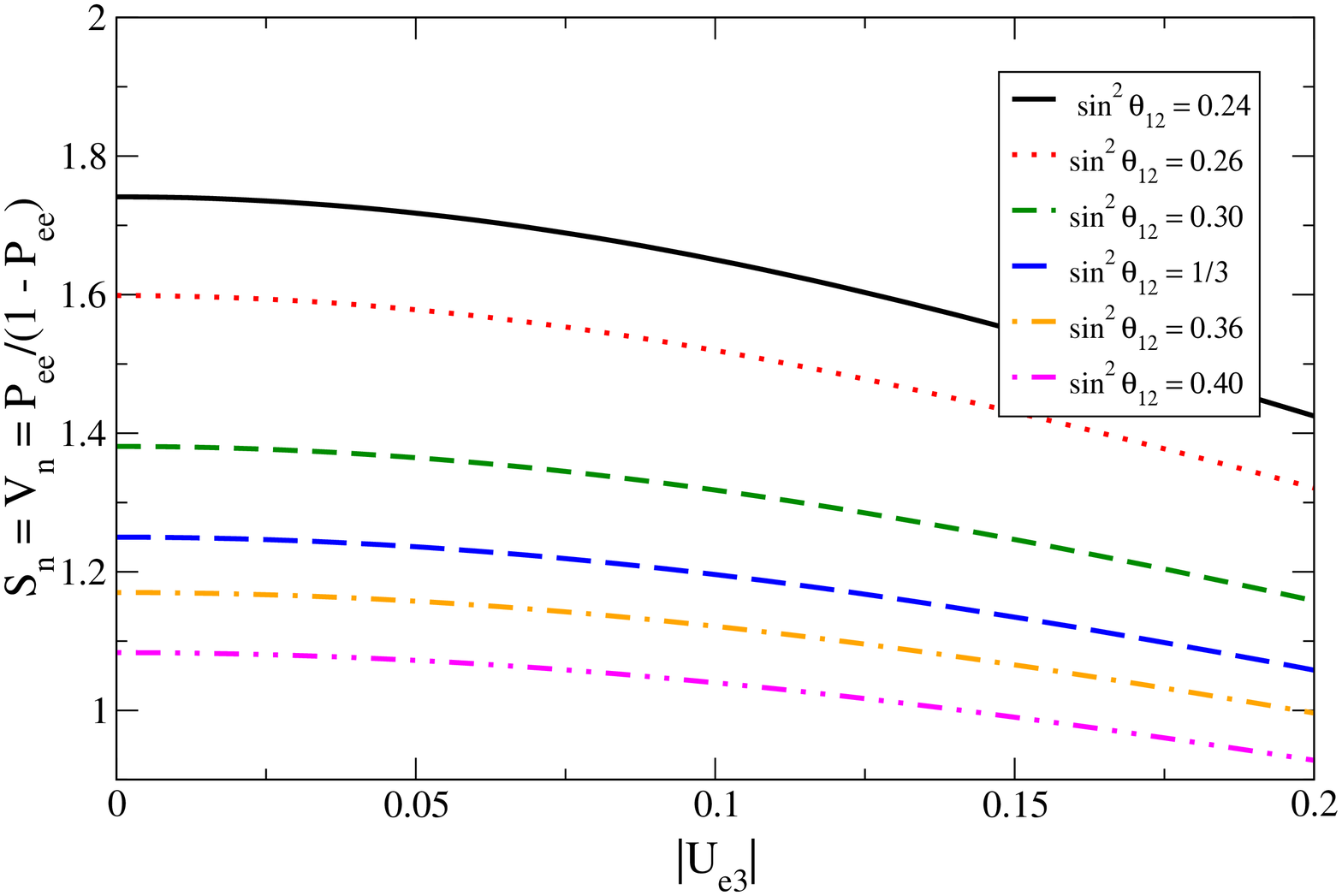,width=14cm,height=9cm}
\end{center}
\caption{\label{fig:Sneut}The ratios $S = \Phi_{e + \bar e}^{\rm D}/(
\Phi_{\mu + \bar \mu}^{\rm D} + \Phi_{\tau + \bar \tau}^{\rm D})$ and 
$V = \Phi_{\bar e}^{\rm D}/(\Phi_{\mu + \bar \mu}^{\rm D} 
+ \Phi_{\tau + \bar \tau}^{\rm D})$ for neutrinos from a neutron source 
as a function of $|U_{e3}|$ for different values of $\sin^2 \theta_{12}$. 
They do not depend on other parameters.}
%
\begin{center}
\epsfig{file=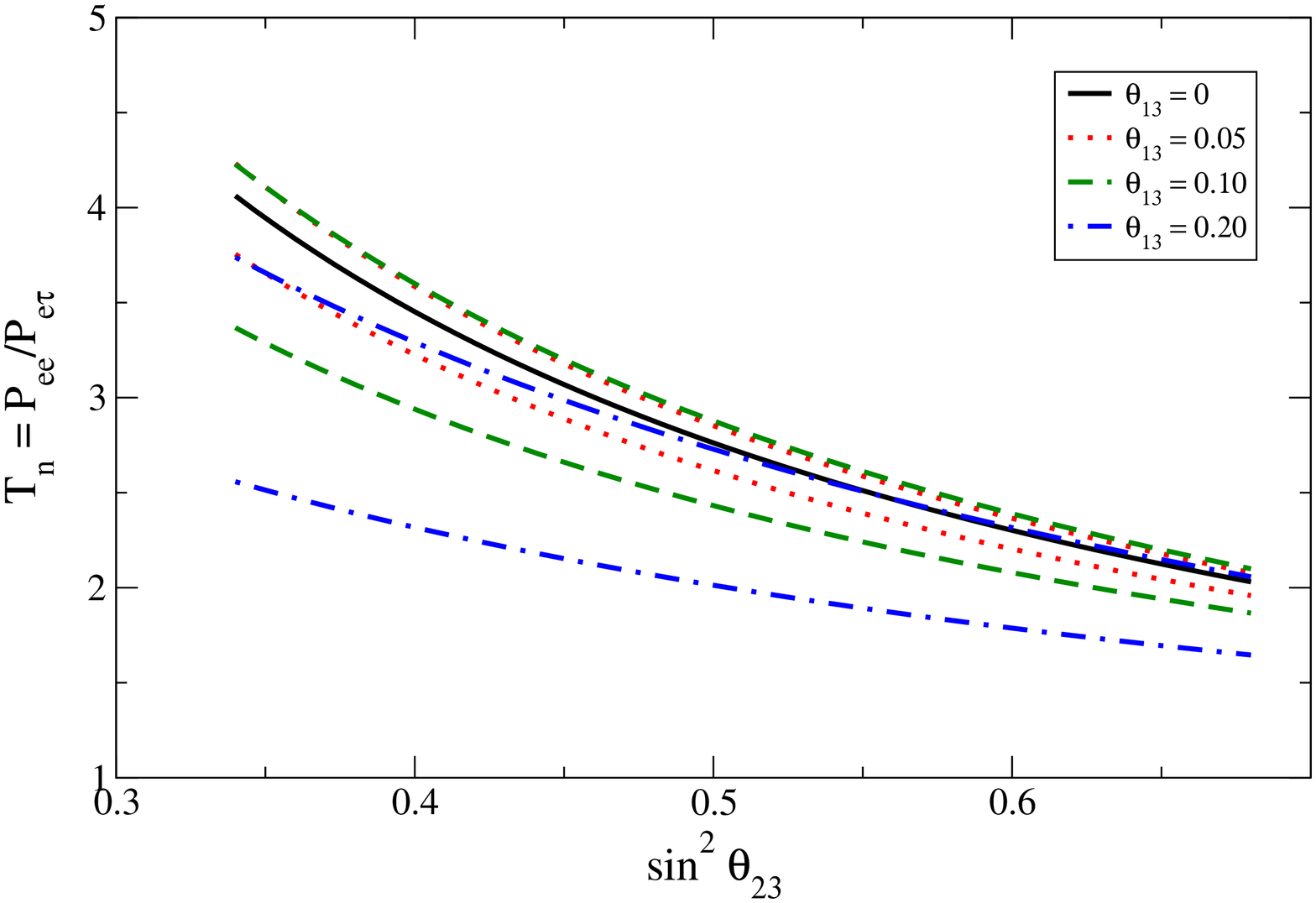,width=14cm,height=9cm}
\end{center}
\caption{\label{fig:Tneut}The ratio $T = \Phi_{e + \bar e}^{\rm D}/
\Phi_{\tau + \bar \tau}^{\rm D}$ for a neutrino flux from a 
neutron source for $\sin^2 \theta_{12} = 0.30$ 
as a function of $\sin^2 \theta_{23}$ for different values 
of $|U_{e3}|$. The $CP$ phase was chosen zero and $\pi$, thereby giving the 
two extreme values of $T$ for any non-zero $|U_{e3}|$. }
\end{figure}

We have seen in Section \ref{sec:gen} that the probabilities 
$P_{ee}$ and $P_{\mu\tau}$ are special since they receive corrections due 
to non-zero $\theta_{13}$ and $\cos 2 \theta_{23}$ only at second order. 
Moreover, the correction for $P_{ee}$ is a function only of $|U_{e3}|^2$. 
Constructing a ratio out of this probability 
will therefore be very interesting. 
Indeed, one can successfully construct such an observable, namely the 
ratios $S$ and $V$ in the case of a neutron source: 
\bea \label{eq:Sn}
S_n \equiv \frac{\D \Phi_{e + \bar e}^{\rm D}}
{\D \Phi_{\mu + \bar \mu}^{\rm D} + \Phi_{\tau + \bar \tau}^{\rm D}}
= V_n \equiv \frac{\D \Phi_{\bar e}^{\rm D}}
{\D \Phi_{\mu + \bar \mu}^{\rm D} + \Phi_{\tau + \bar \tau}^{\rm D}} 
 = \frac{\D P_{ee}}{\D P_{e\mu} + P_{e\tau}} 
= \frac{\D P_{ee}}{\D 1 - P_{ee}} \\[0.3cm]
\simeq 
\frac{\D 1 - 2 \, c_{12}^2 \, s_{12}^2}
{\D 2 \, c_{12}^2 \, s_{12}^2} - 
\frac{\D 1 - 2 \,c_{12}^2 \, s_{12}^2 }
{\D 2 \, c_{12}^4 \, s_{12}^4 } \, |U_{e3}|^2~,
\eea
which is independent of the atmospheric neutrino mixing angle and on the 
$CP$ phase $\delta$. With $\sin^2 \theta_{12} = 1/3$ one has 
$S_n = V_n \simeq 5/4 - 45/8 \, |U_{e3}|^2$.  
With the small dependence of this quantity on 
$|U_{e3}|$ it can serve as an interesting observable to probe 
the solar neutrino mixing angle. We show in Fig.~\ref{fig:Sneut} the 
dependence of $P_{ee}/(1 - P_{ee})$ on $|U_{e3}|$ for different values 
of $\sin^2 \theta_{12}$, which cover its currently allowed $3\sigma$ 
range. 
For all values of $|U_{e3}| \ls 0.1$ there is basically no 
dependence on it and the ratio varies strongly with $\sin^2 \theta_{12}$, from 
1.74 for $\sin^2 \theta_{12} = 0.24$ and 
1.08 for $\sin^2 \theta_{12} = 0.40$. 
We stress that the reason for the 
simple functional dependence of $S_n$ and $V_n$ is that in the denominator we 
have summed over muon and tau neutrinos. As a comparison, the ratio  
$T$, which is just the ratio of electron to 
tau neutrinos, is given for a neutron source by 
\be
T_n = \frac{\D \Phi_{e + \bar e}^{\rm D}}
{\D \Phi_{\tau + \bar \tau}^{\rm D}} = 
\frac{\D P_{ee}}{\D P_{e \tau}}
\simeq \frac{\D 1 - 2 \, c_{12}^2 \, s_{12}^2}
{\D c_{12}^2 \, s_{12}^2 } + 
\frac{\D 1 - 2 \, c_{12}^2 \, s_{12}^2}{\D c_{12}^4 \, s_{12}^4} \, \Delta 
~.
\ee
There can be sizable dependence on $|U_{e3}| \, \cos \delta$ and 
$\sin^2 \theta_{23}$ which is illustrated in 
Fig.~\ref{fig:Tneut}, where we show the ratio $T$ 
for a fixed value of $\sin^2 \theta_{12} = 0.30$. 
Another example for a simple ratio 
can also be identified in Tables \ref{tab:obs} and \ref{tab:obs_spe}. 
This is $W_n$, the flux of anti-electron neutrinos 
divided by the sum of all neutrinos for a neutron source. 
It is simply given by $P_{ee}$. Consequently, 
$W_n \simeq (1 - 2 \, c_{12}^2 \, s_{12}^2) \, (1 - 2 \, |U_{e3}|^2)$, 
which varies however less than $S_n$ with the solar neutrino mixing angle: 
it ranges from 0.64 for 
$\sin^2 \theta_{12} = 0.24$ to 0.48 for $\sin^2 \theta_{12} = 0.40$.

The other mixing probability receiving only quadratic corrections 
due to non-zero $|U_{e3}|$ and non-maximal $\theta_{23}$ is 
$P_{\mu\tau}$. It is also possible to construct a ratio depending solely on 
it: consider the ratio of tau neutrinos and the sum of electron and 
muon neutrinos\footnote{This flux can be obtained from the considered fluxes 
$R$ and $S$ by evaluating $(R/(1 + R) + S/(1 + S))^{-1} - 1$.}. 
If the neutrinos stem from a muon-damped source, then 
\be \label{eq:pmt}
\frac{\D \Phi_{\tau + \bar \tau}^{\rm D}}
{\D \Phi_{e + \bar e}^{\rm D} + \Phi_{\mu + \bar \mu}^{\rm D}} = 
\frac{P_{\mu\tau}}{P_{e \mu} + P_{\mu\mu}} = \frac{P_{\mu\tau}}{1 - P_{\mu\tau}} 
\simeq \frac{1 - c_{12}^2 \, s_{12}^2}{1 + c_{12}^2 \, s_{12}^2} ~,
\ee
plus corrections of second order in $|U_{e3}|$ and $\theta_{23} - \pi/4$. 
The range of the leading term is between 
0.69 for $\sin^2 \theta_{12} = 0.24$ and 
0.61 for $\sin^2 \theta_{12} = 0.40$, i.e., significantly smaller than the 
corresponding range for $S_n$ or $V_n$. Alternatively, 
in the denominator of Eq.~(\ref{eq:pmt}) one could include 
$\Phi_{\tau + \bar \tau}^{\rm D}$, in which case the ratio would be just 
$P_{\mu \tau}$. Its range is between 0.41 and 0.30 
if $\sin^2 \theta_{12}$ is varied between 0.24 and 0.40. 
 
Eventually, in an experiment one will presumably measure several 
ratios and reconstruct 
the neutrino parameters with global fits \cite{WW}. 
However, ratios which depend solely on 
one mixing angle would allow for a particularly illustrative, 
clear and clean measurement. In addition, the ratios presented 
here serve to illustrate that neutron and muon-damped sources are good 
candidates to probe $\theta_{12}$.\\

An issue related to the problem of identifying the neutrino source, 
as mentioned at the end of Section \ref{sec:ratios}, is the following: 
what if neutrinos from a neutron or a muon-damped source 
are polluted by neutrinos from usual pion sources? 
Denoting the relative flux from this 
source with $\eta$, one can show that 
\be
S_n = \frac{P_{ee} \, (1 + \eta) + 2 \, \eta \, P_{e\mu}}
{(1 - P_{ee}) \, (1 + \eta) + 2 \, \eta \, (1 -  P_{e\mu})} \simeq 
\frac{\D 1 - 2 \, c_{12}^2 \, s_{12}^2}
{\D 2 \, c_{12}^2 \, s_{12}^2} + \frac{3 \, c_{12}^2 \, s_{12}^2 - 1}
{2 \, c_{12}^4 \, s_{12}^4} \, \eta~,
\ee
plus higher order terms. The correction can be sizable, 
with $\sin^2 \theta_{12} = 1/3$ one has 
$S_n \simeq 5/4 - 27/8 \, \eta$. Knowing that the pollution is below 
5$\%$ would be necessary to have a correction to Eq.~(\ref{eq:Sn}) below 
10$\%$.

\section{\label{sec:AvsB}Comparison of Flux Ratios for 
almost maximal $\theta_{23}$ and for negligible $U_{e3}$}

Now we will study the sensitivity of the flux ratios if (A)
$\theta_{23}$ is close-to maximal and (B) if $|U_{e3}|$ is close to zero. 
One possibility for this situation to happen is through special 
breaking of $\mu$--$\tau$ symmetry, which usually always implies deviation from 
zero $U_{e3}$ and from maximal $\theta_{23}$ \cite{rabi,mutau}. 
In the symmetry limit and for a normal mass hierarchy we have the following 
mass matrix\footnote{This particular form of the mass matrix is motivated by an 
approximate flavor symmetry $U(1)_{L_e}$ times 
the $\mu$--$\tau$ exchange symmetry, 
where $\varepsilon$ denotes the strength of $U(1)_{L_e}$ 
breaking.} \cite{rabi}: 
\be
m_\nu = \frac{\sqrt{\dma}}{2} \, 
\left( 
\bad
a \, \varepsilon^2 & b \, \varepsilon & b \, \varepsilon \\[0.2cm]
\cdot & 1 + \varepsilon & -1 \\[0.2cm]
\cdot & \cdot & 1 + \varepsilon 
\ea
\right)~,
\ee
where $\varepsilon^2 = {\cal O}(\dms/\dma)$. 
The two interesting breaking scenarios correspond to different locations 
of the symmetry breaking term \cite{rabi}. 
The first case we call scenario A and it occurs if there is $\mu$--$\tau$ 
symmetry breaking in the electron sector of $m_\nu$, i.e., 
\be
m_\nu^{\rm A} = \frac{\sqrt{\dma}}{2} \, 
\left( 
\bad
a \, \varepsilon^2 & b \, \varepsilon & d \, \varepsilon \\[0.2cm]
\cdot & 1 + \varepsilon & -1 \\[0.2cm]
\cdot & \cdot & 1 + \varepsilon 
\ea
\right)~. 
\ee
Then we have that $|U_{e3}| 
\gg |\theta_{23} - \pi/4| 
$.
For breaking of the symmetry in the $\mu$--$\tau$ sector, i.e., 
\be
m_\nu^{\rm B} = \frac{\sqrt{\dma}}{2} \, 
\left( 
\bad
a \, \varepsilon^2 & b \, \varepsilon & b \, \varepsilon \\[0.2cm]
\cdot & 1 + \varepsilon & -1 \\[0.2cm]
\cdot & \cdot & 1 + d \, \varepsilon 
\ea
\right)~,
\ee
one finds that $|U_{e3}| 
\ll |\theta_{23} - \pi/4| 
$. 
We denote this case with scenario B. 
Of course, other scenarios are possible in which 
$|\theta_{23} - \pi/4| \gg \theta_{13}$ or $|\theta_{23} - \pi/4| \ll \theta_{13}$, 
and this particular example based on $\mu$--$\tau$ symmetry breaking serves just as an 
illustrative example.\\ 

\begin{figure}[htb]
\begin{center}
\epsfig{file=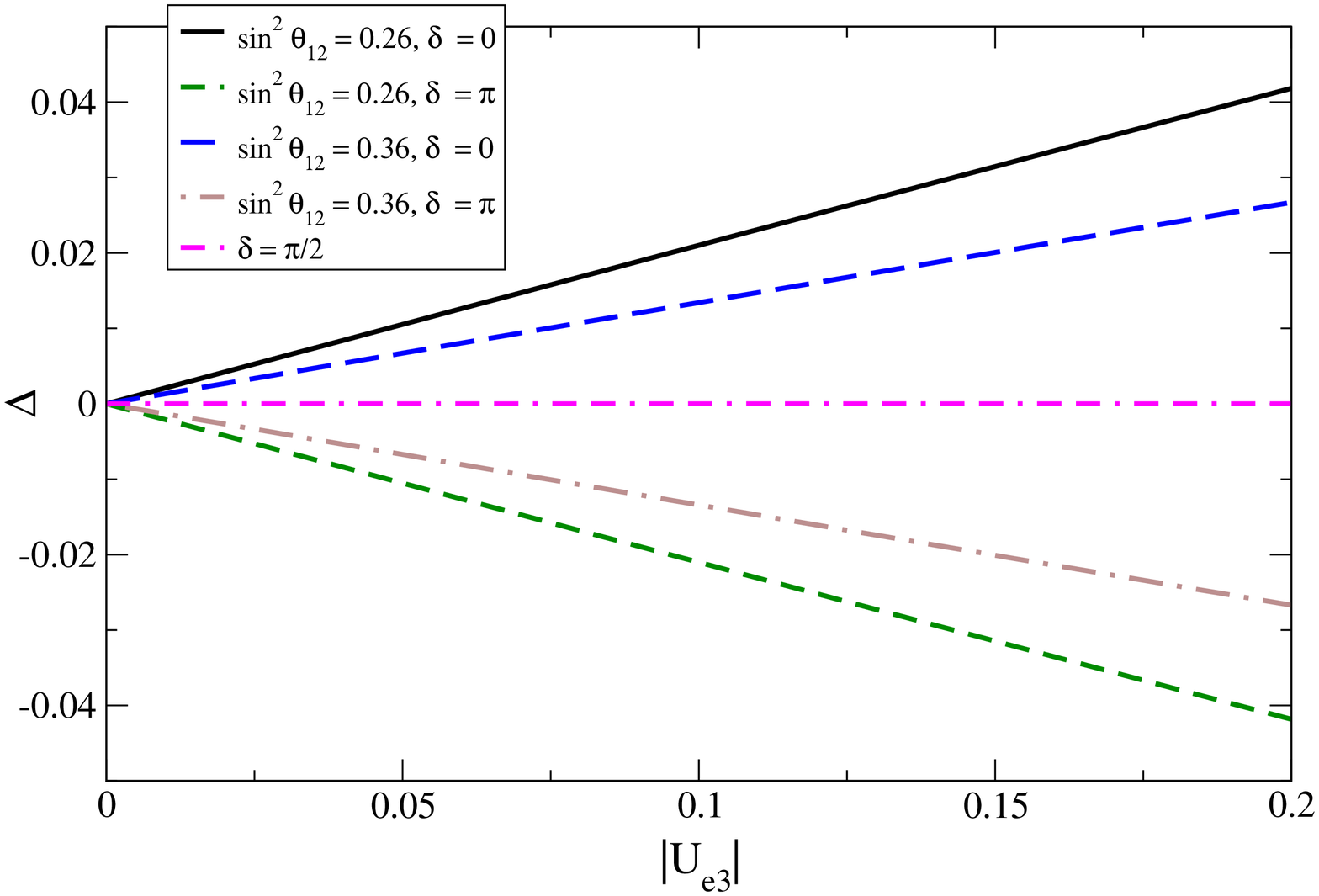,width=14cm,height=9.5cm}
\epsfig{file=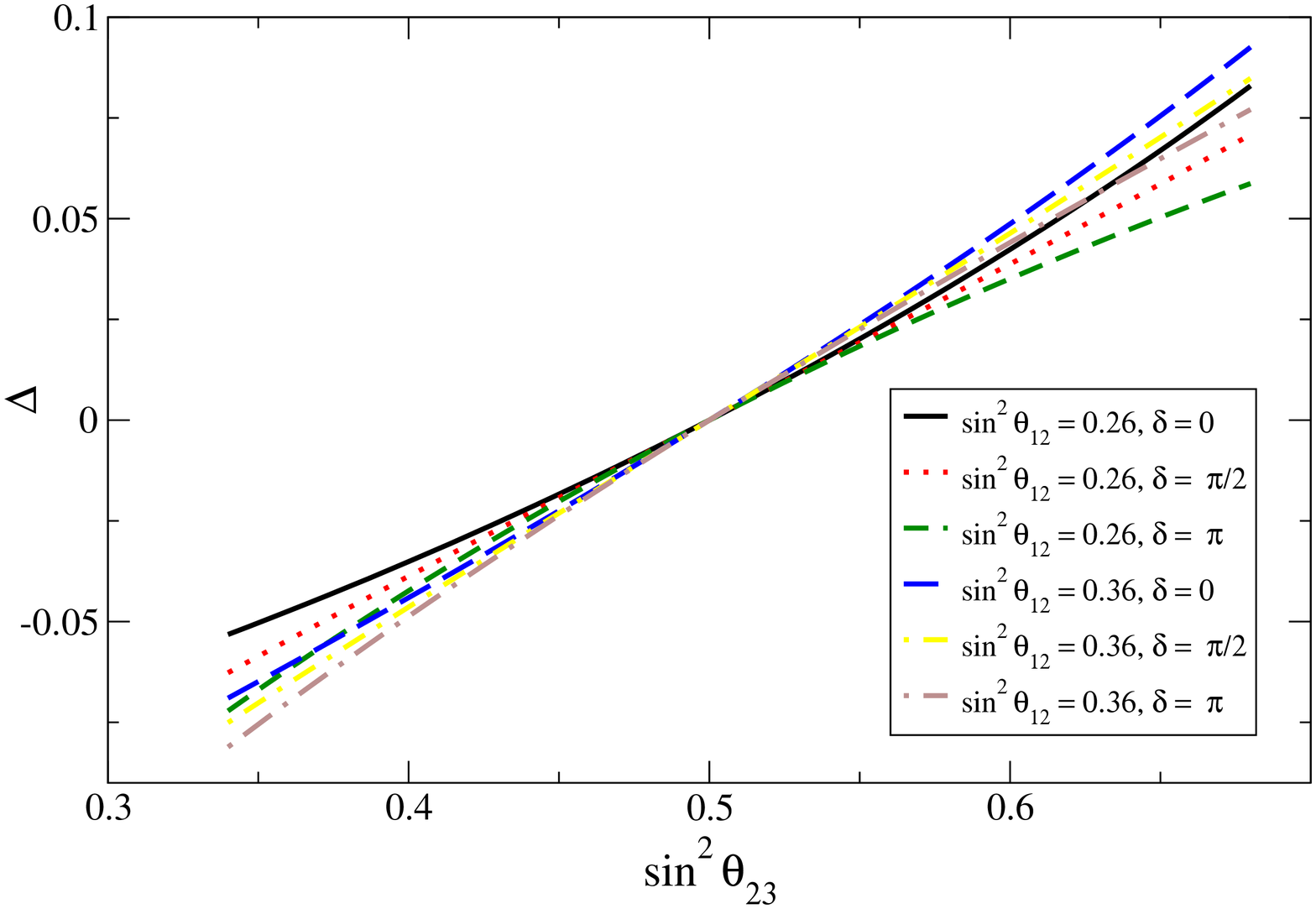,width=14cm,height=9.5cm}
\end{center}
\caption{\label{fig:Delta}The small parameter 
$\Delta$. The upper plot is for negligible $|\theta_{23} - \pi/4|$ 
and the lower plot for negligible $\theta_{13}$. }
\end{figure}

As shown in Sections \ref{sec:gen} and \ref{sec:fluobs}, the sensitivity on the 
neutrino parameters is governed by the universal correction parameter $\Delta$. 
We give the dependence of $\Delta$ on $|U_{e3}|$ 
and on $\sin^2 \theta_{23}$ in Fig.~\ref{fig:Delta} for 
scenario A (when 
$\theta_{23}$ is very close to $\pi/4$) and scenario B (when $|U_{e3}|$ 
is negligible). The two 
values of $\sin^2 \theta_{12}$ correspond to the current $2\sigma$ range. 
To take these scenarios into account, we varied in scenario A 
the parameter $|U_{e3}|$ and set atmospheric neutrino 
mixing to maximal. 
For scenario B we varied $\sin^2 \theta_{23}$ and fixed 
$|U_{e3}| = 1.7 \, (\theta_{23} - \pi/4)^2$.  
From the Figure we can draw the following lessons: 
in the case of negligible $|U_{e3}|$ the dependence on $\delta$ and $\theta_{12}$ 
is very weak. This is trivial for $\delta$,  
because it is always related to $|U_{e3}|$. Regarding $\theta_{12}$, the limited 
dependence of $\Delta$ on this parameter was noted in Section \ref{sec:gen}. 
In this special case of vanishing $U_{e3}$ it occurs because the allowed range of 
$s_{12}^2 \, c_{12}^2$ is smaller than the allowed range of $s_{12}^2$ alone. 
For instance, if $\sin^2 \theta_{12}$ ranges from 
0.26 to 0.36, then $s_{12}^2 \, c_{12}^2 = 0.19 \div 0.23$. 
This indicates that for small 
$|U_{e3}|$ the mixing probabilities and consequently the flux ratios 
can serve as a useful probe for the atmospheric neutrino 
mixing angle $\theta_{23}$. Its octant, 
i.e., $\sin^2 \theta_{23}$ above or below 
1/2, reflects always in positive or negative $\Delta$. 
If the zeroth order ratio 
is a function of $\theta_{12}$, 
then this ratio will be a particularly promising probe for this angle. 
In case the zeroth 
order ratio is just a number, the sensitivity for $\theta_{12}$ is low. 

We have seen in Section \ref{sec:gen} that $\Delta$ depends 
on $\theta_{23}$ stronger than on the other parameters. This can 
be underlined here: 
if $\theta_{23}$ is close to $\pi/4$, then $\Delta$ is typically 
smaller than for negligible $\theta_{13}$. Compare the situation 
$\epsilon = 0.1$ and $|U_{e3}| = 0$ with 
$|U_{e3}| = 0.1$ and $\epsilon = 0$, i.e., $\mu$--$\tau$ symmetry 
breaking of equal size (note that $\sin^2 \theta_{23} \simeq 1/2 - \epsilon$). 
$\Delta$ is around $-0.05$ for the first case, but smaller 
than 0.02 in absolute 
value for the second case. This feature has also been noted already in Section 
\ref{sec:gen}. It occurs because in Eq.~(\ref{eq:Delta}) 
the factor in front of $|U_{e3}|$ is roughly 
two times smaller than the term to which $\epsilon$ is proportional to.
A smaller correction means that the sensitivity to $\theta_{13}$ 
and $\delta$ in case of maximal $\theta_{23}$ 
will be smaller than the sensitivity to $\theta_{23}$ in case 
of negligible $U_{e3}$. 
However, a small correction implies that it will 
be easier to probe $\theta_{12}$ if the zeroth order ratio depends on it.\\

\begin{figure}[htb]
\begin{center}
\begin{tabular}{cc}
\epsfig{file=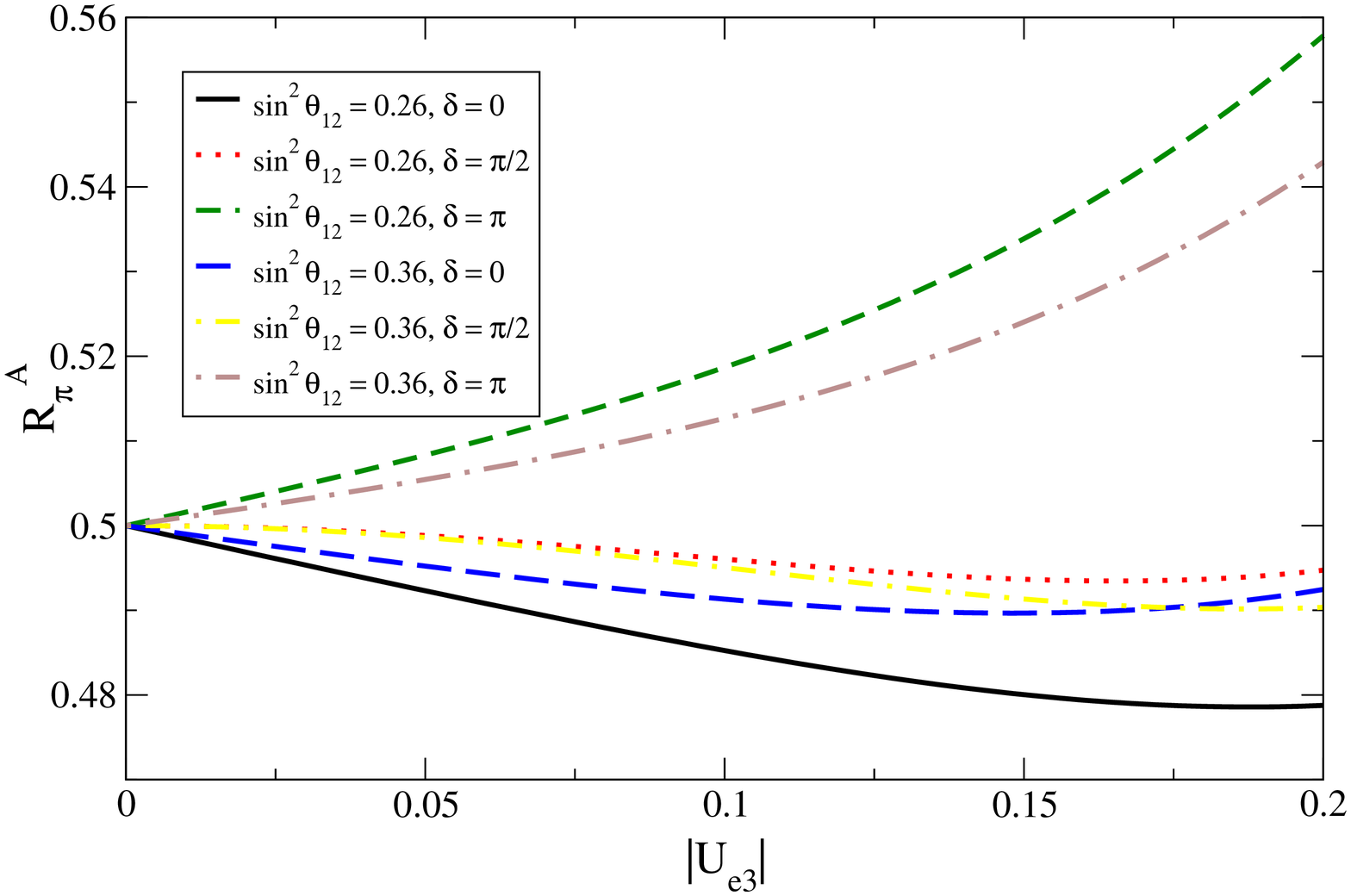,width=8cm,height=7cm} & 
\epsfig{file=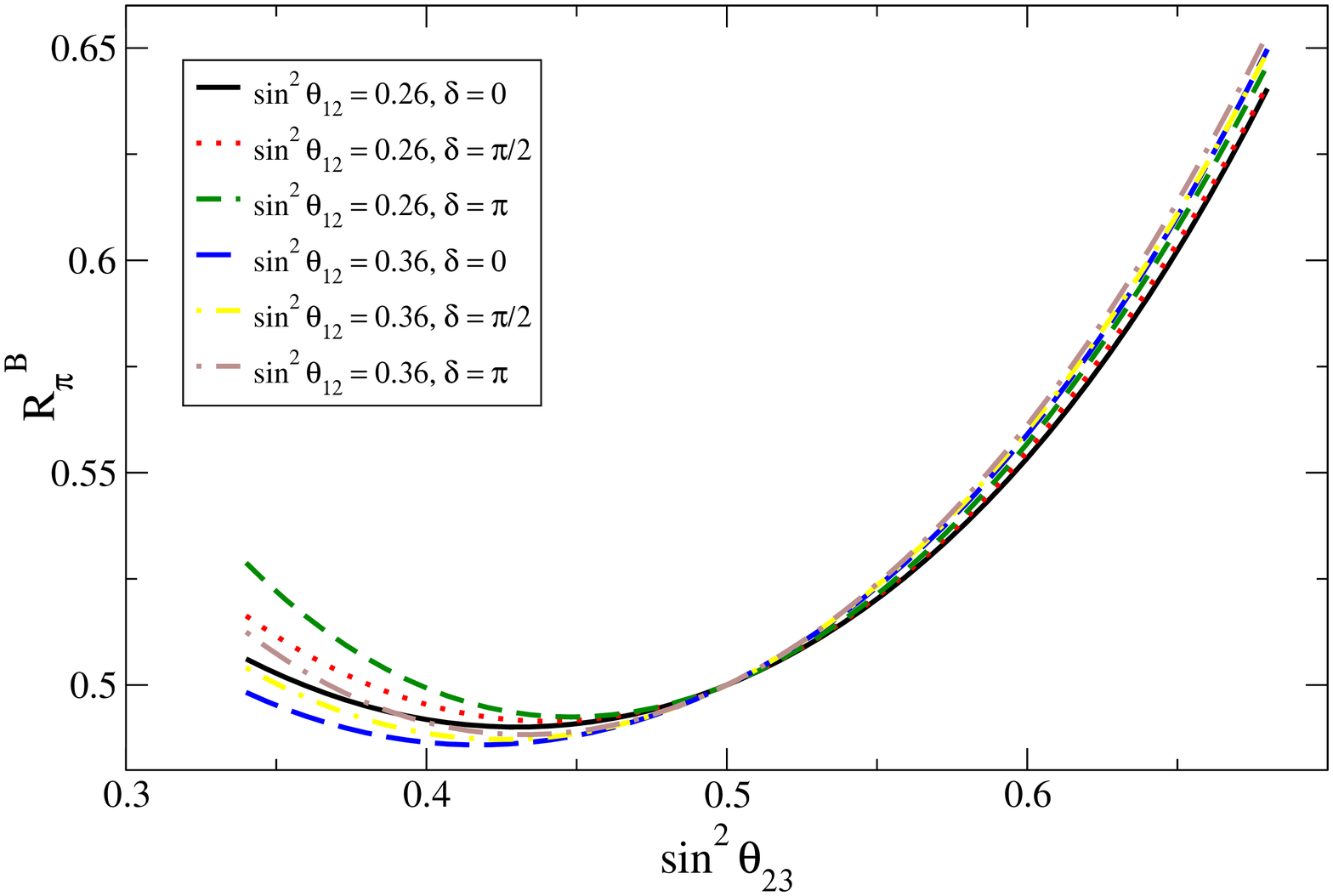,width=8cm,height=7cm} \\
\epsfig{file=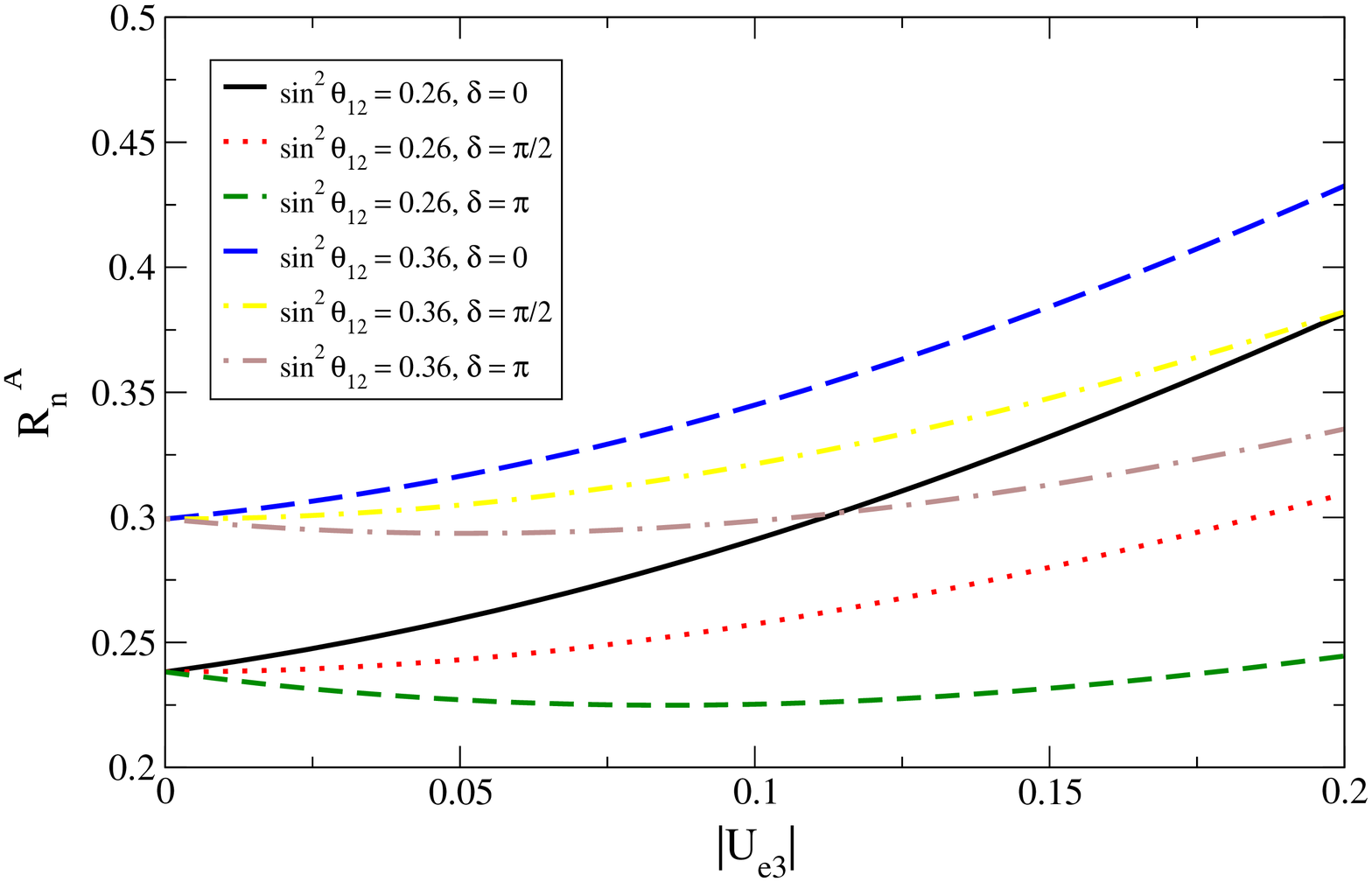,width=8cm,height=7cm} & 
\epsfig{file=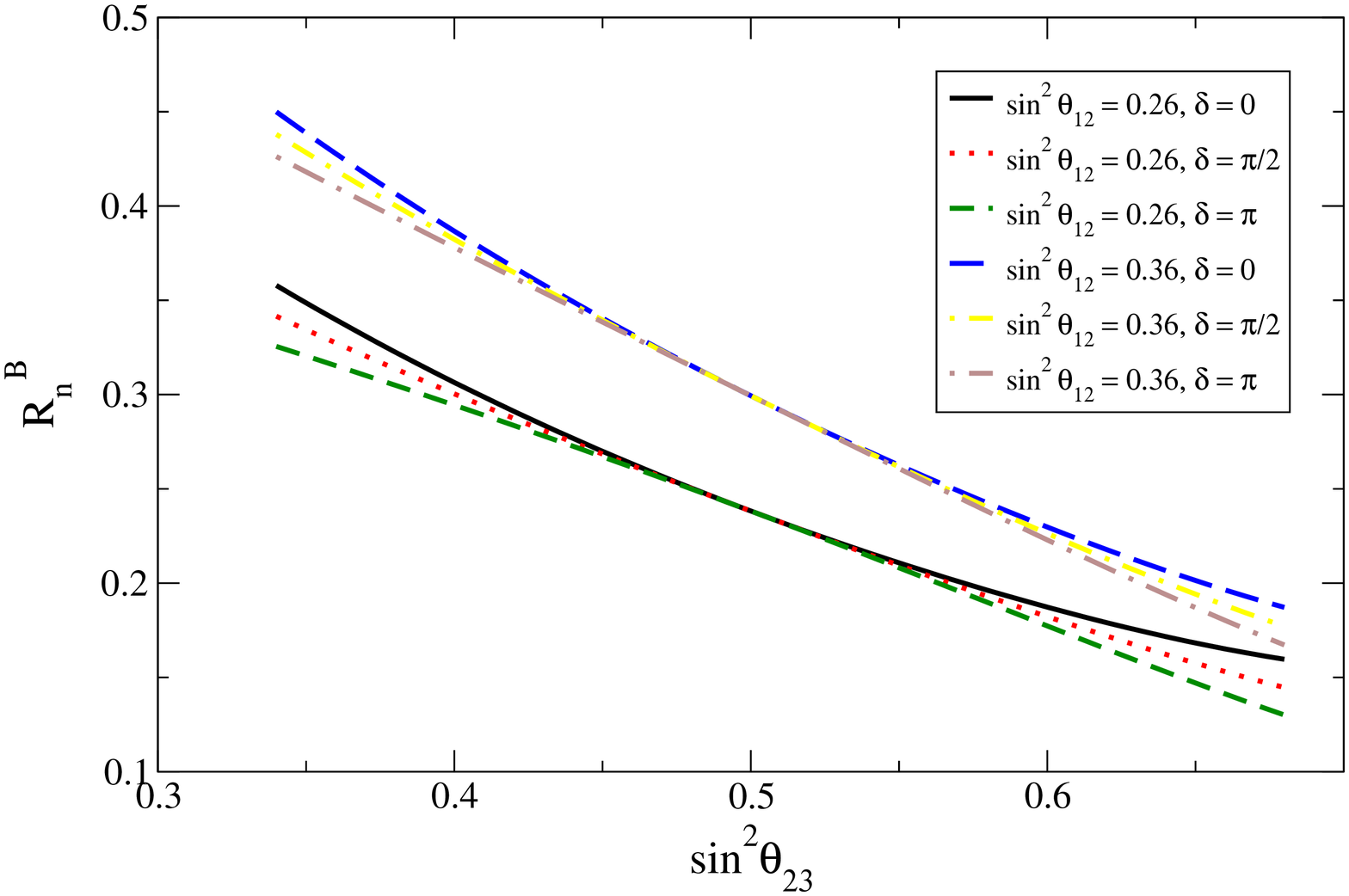,width=8cm,height=7cm} \\
\epsfig{file=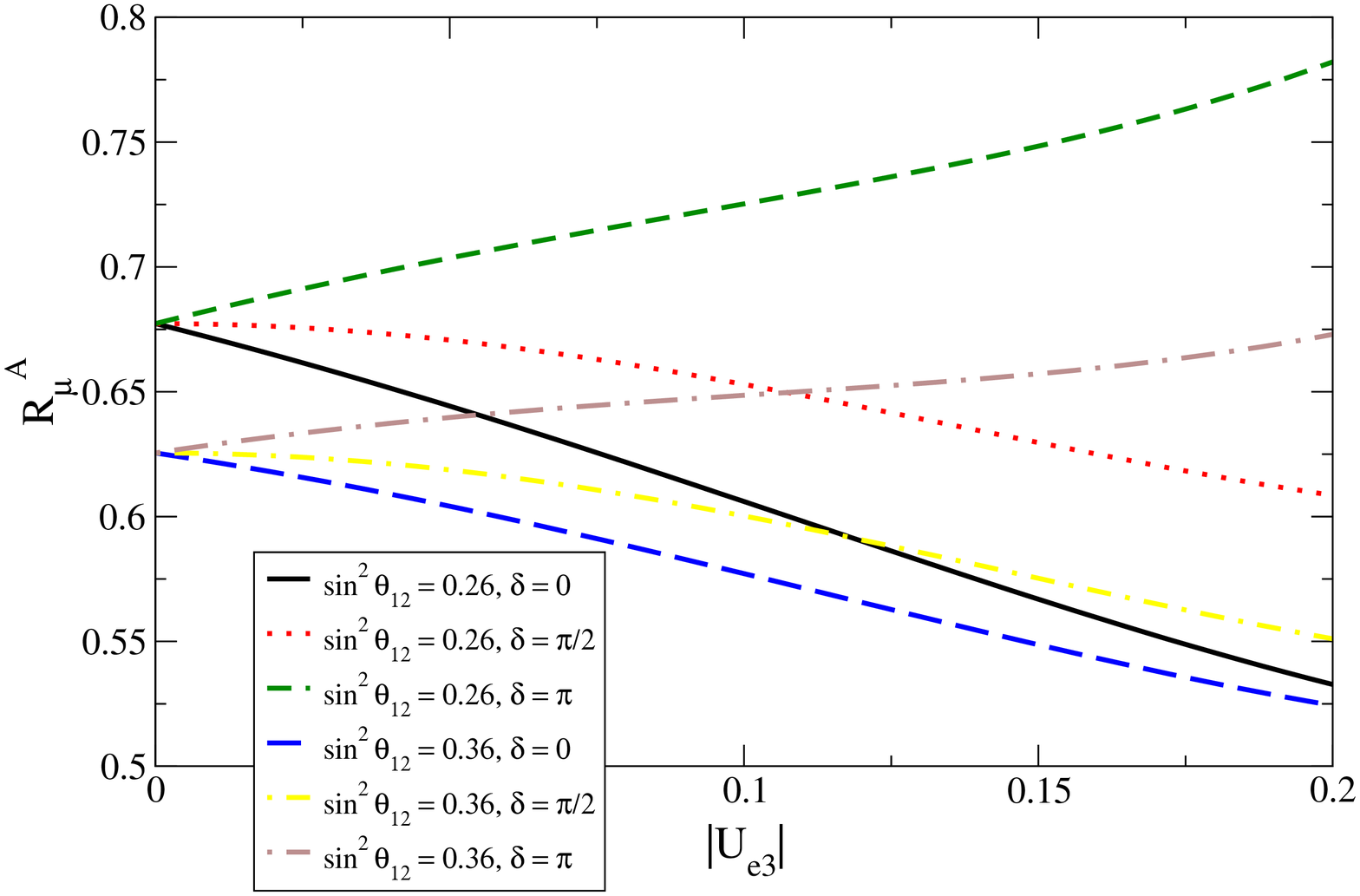,width=8cm,height=7cm} & 
\epsfig{file=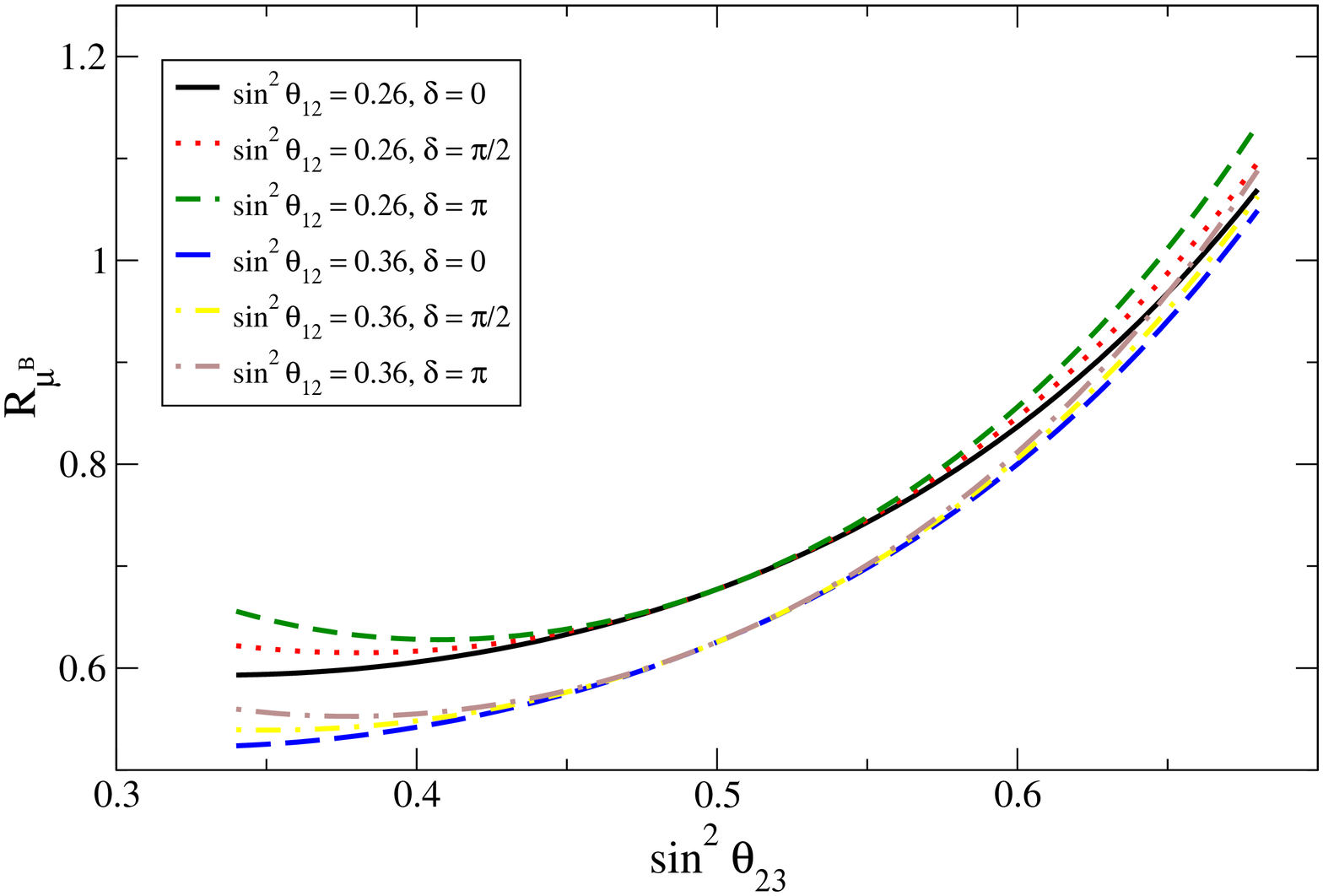,width=8cm,height=7cm} 
\end{tabular}
\end{center}
\caption{\label{fig:AvsB}
Comparison of $\mu$--$\tau$ breaking scenarios 
A (negligible $|\theta_{23} - \pi/4|$)
and B (negligible $\theta_{13}$) for the ratios  
$R = \Phi_{\mu + \bar\mu}^{\rm D}/(\Phi_{e + \bar e}^{\rm D} 
+ \Phi_{\tau + \bar \tau}^{\rm D})$ and 
neutrinos from pion, neutron and muon-damped sources.}
\end{figure}

\begin{figure}[htb]
\begin{center}
\begin{tabular}{cc}
\epsfig{file=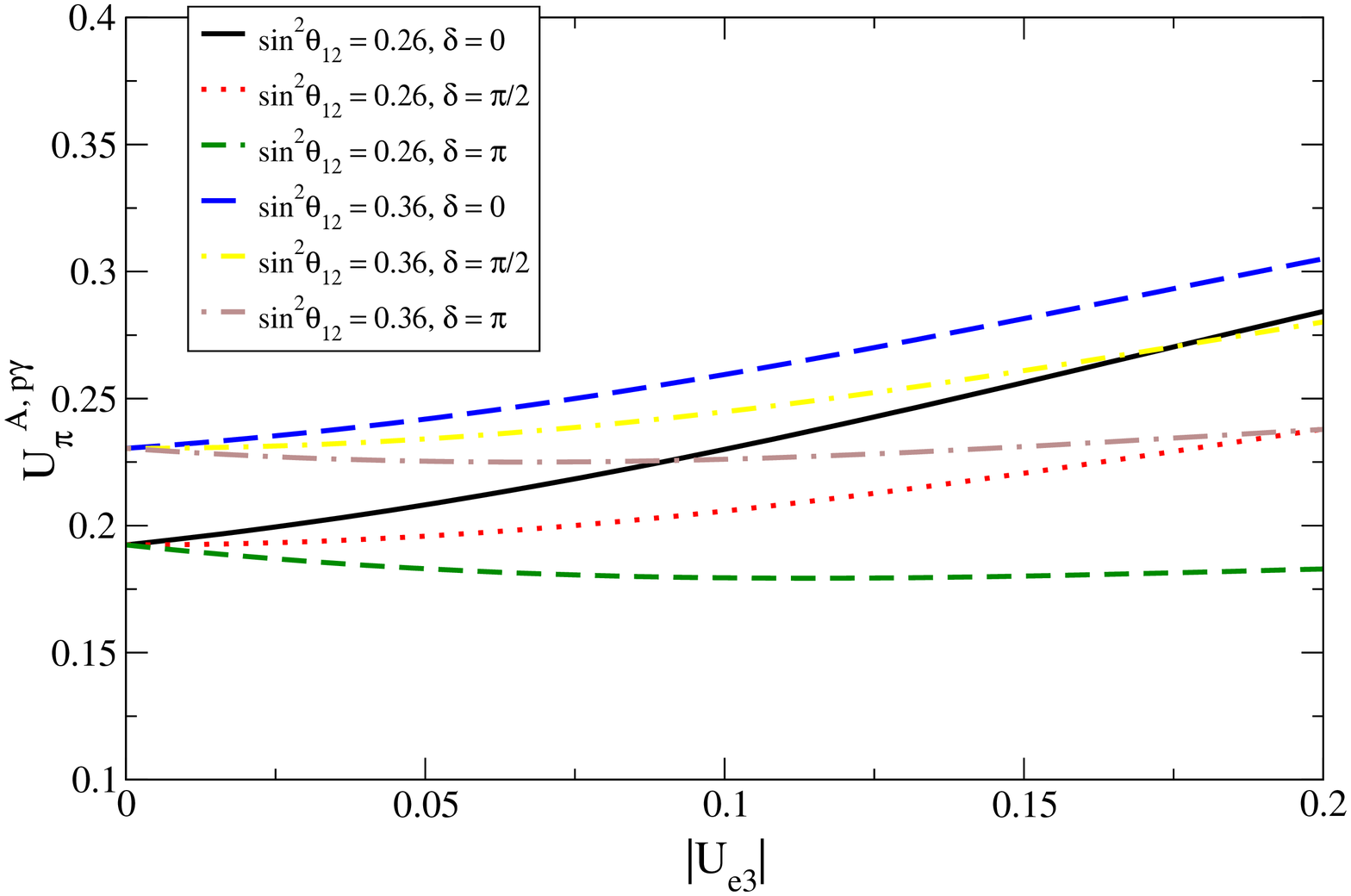,width=8cm,height=5.4cm} & 
\epsfig{file=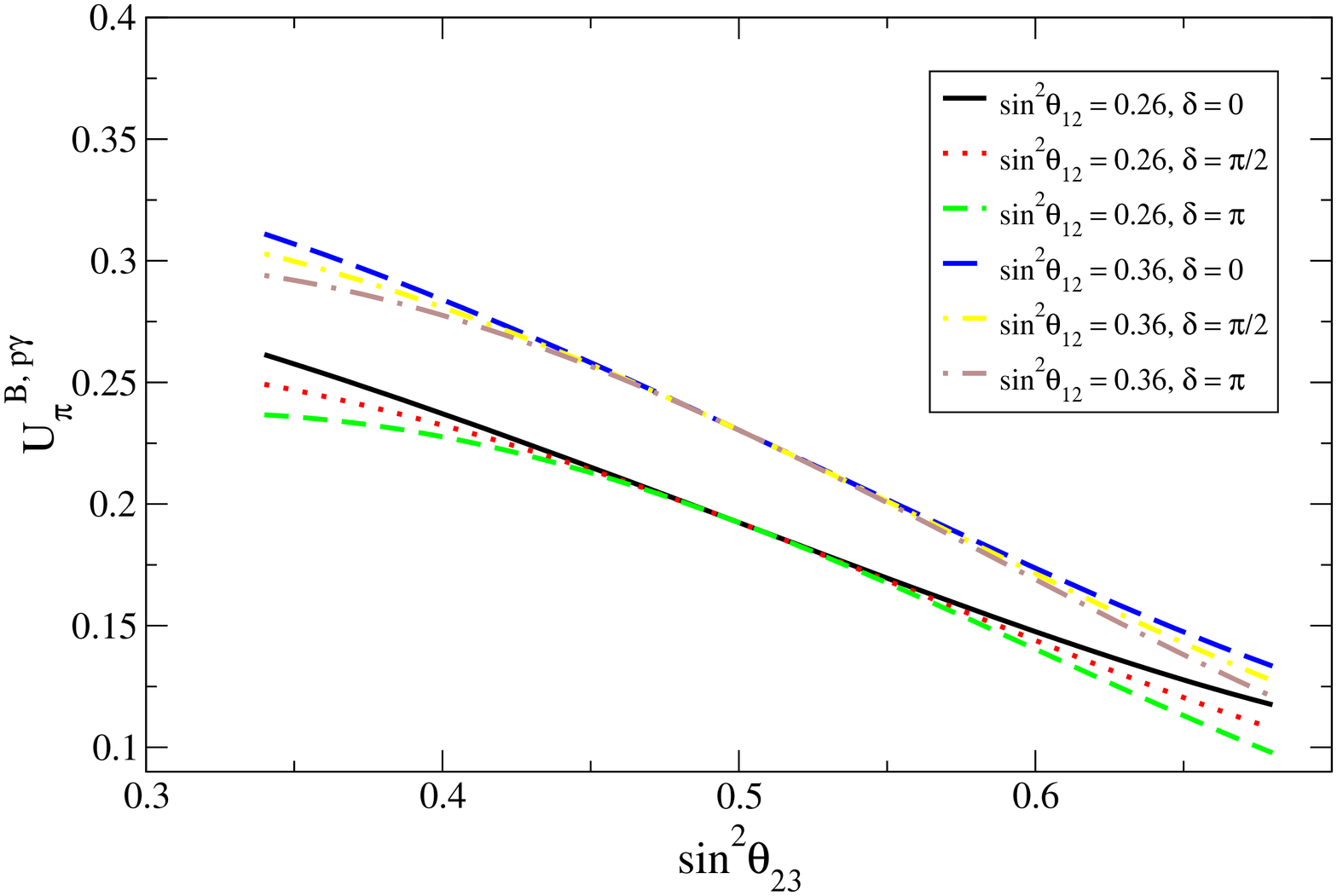,width=8cm,height=5.4cm} \\
\epsfig{file=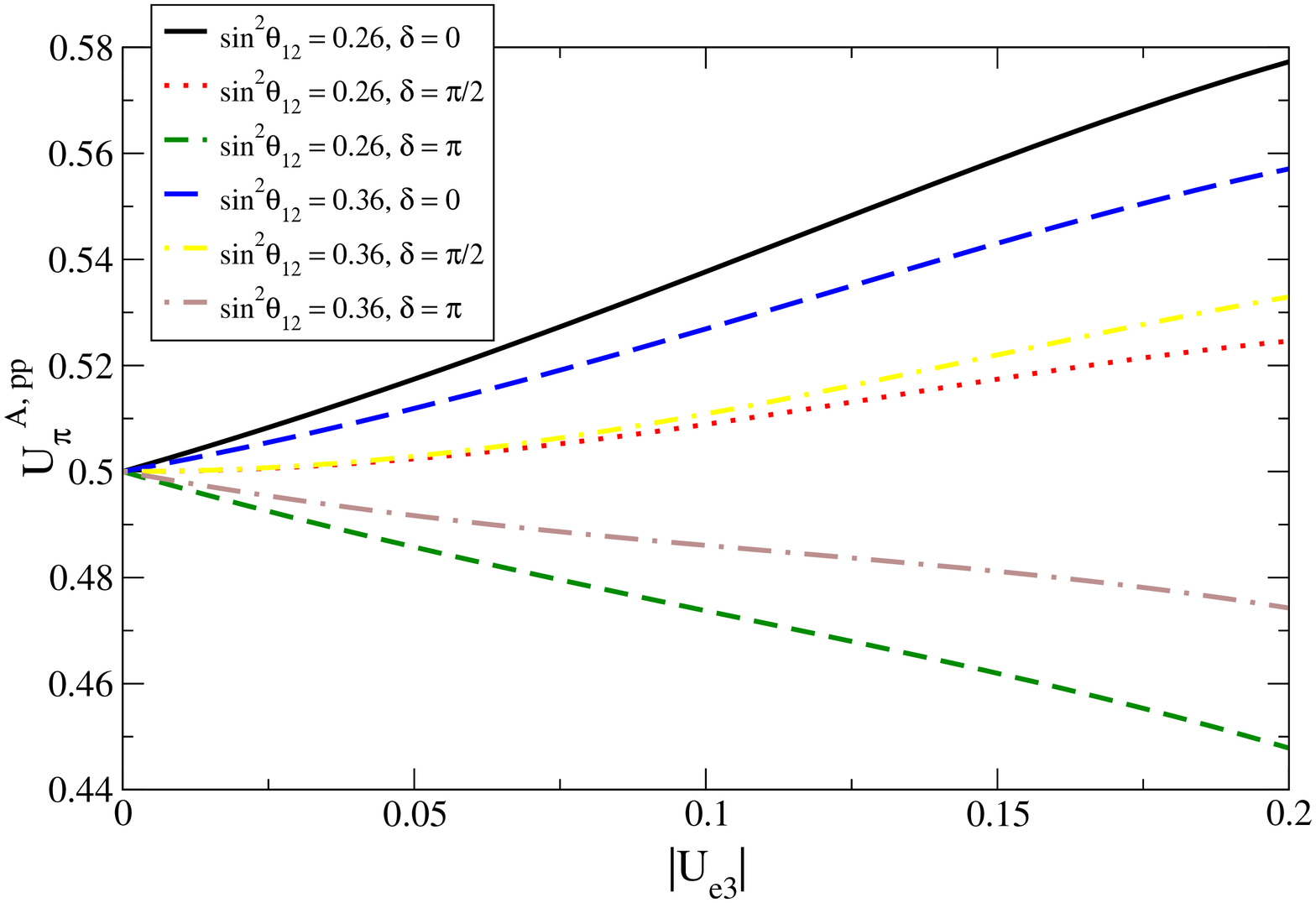,width=8cm,height=5.4cm} & 
\epsfig{file=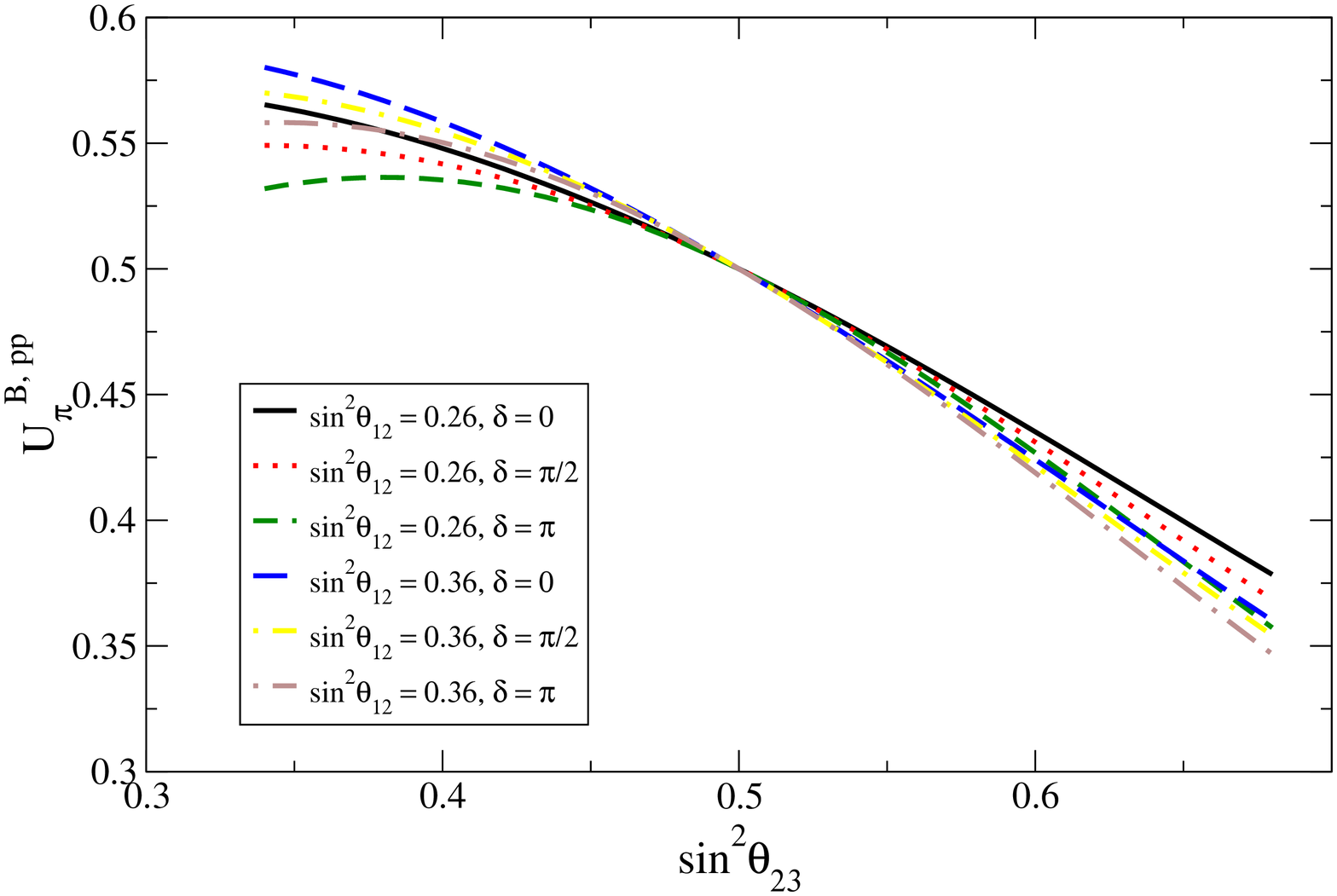,width=8cm,height=5.4cm} \\
\epsfig{file=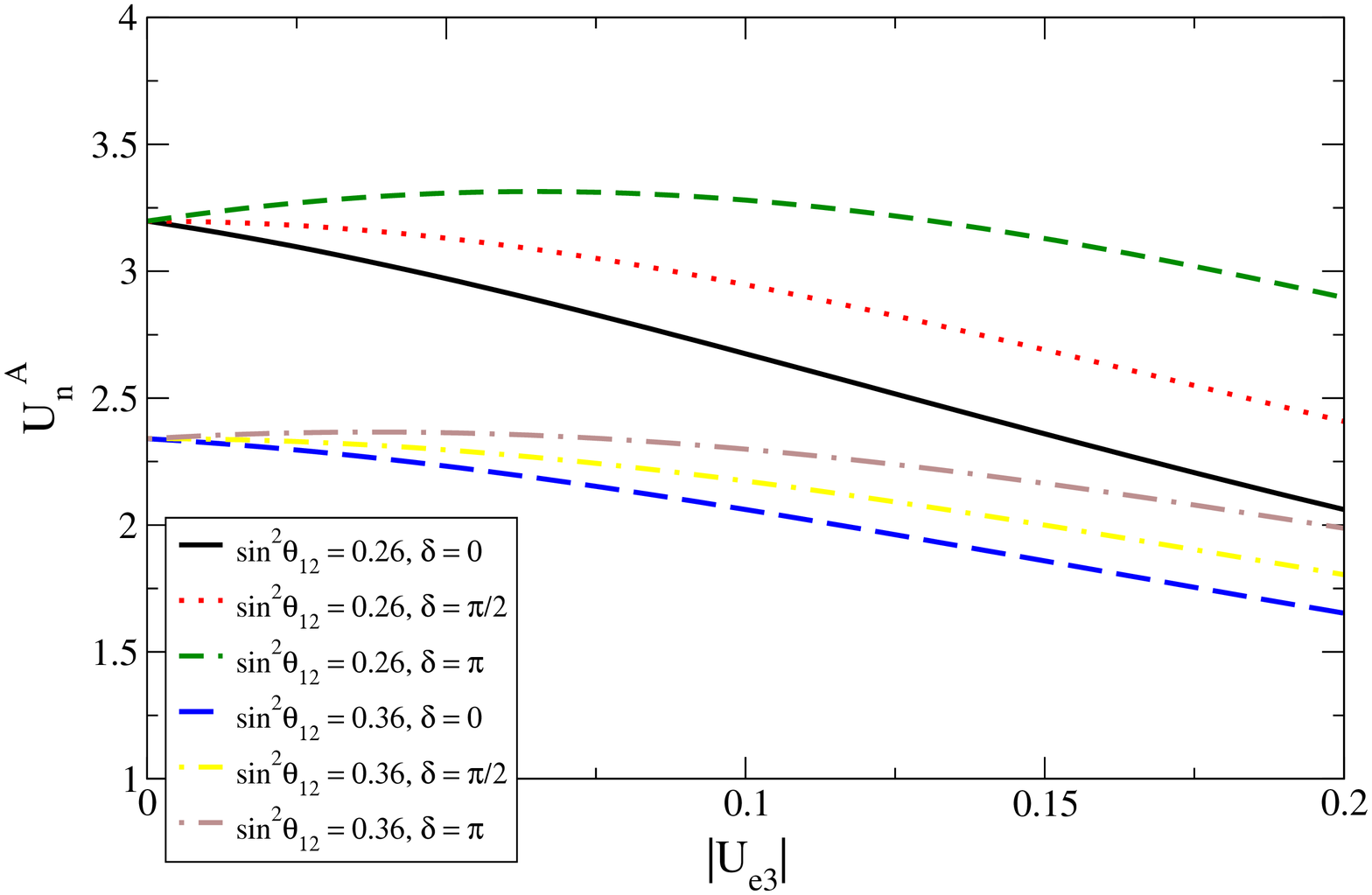,width=8cm,height=5.4cm} & 
\epsfig{file=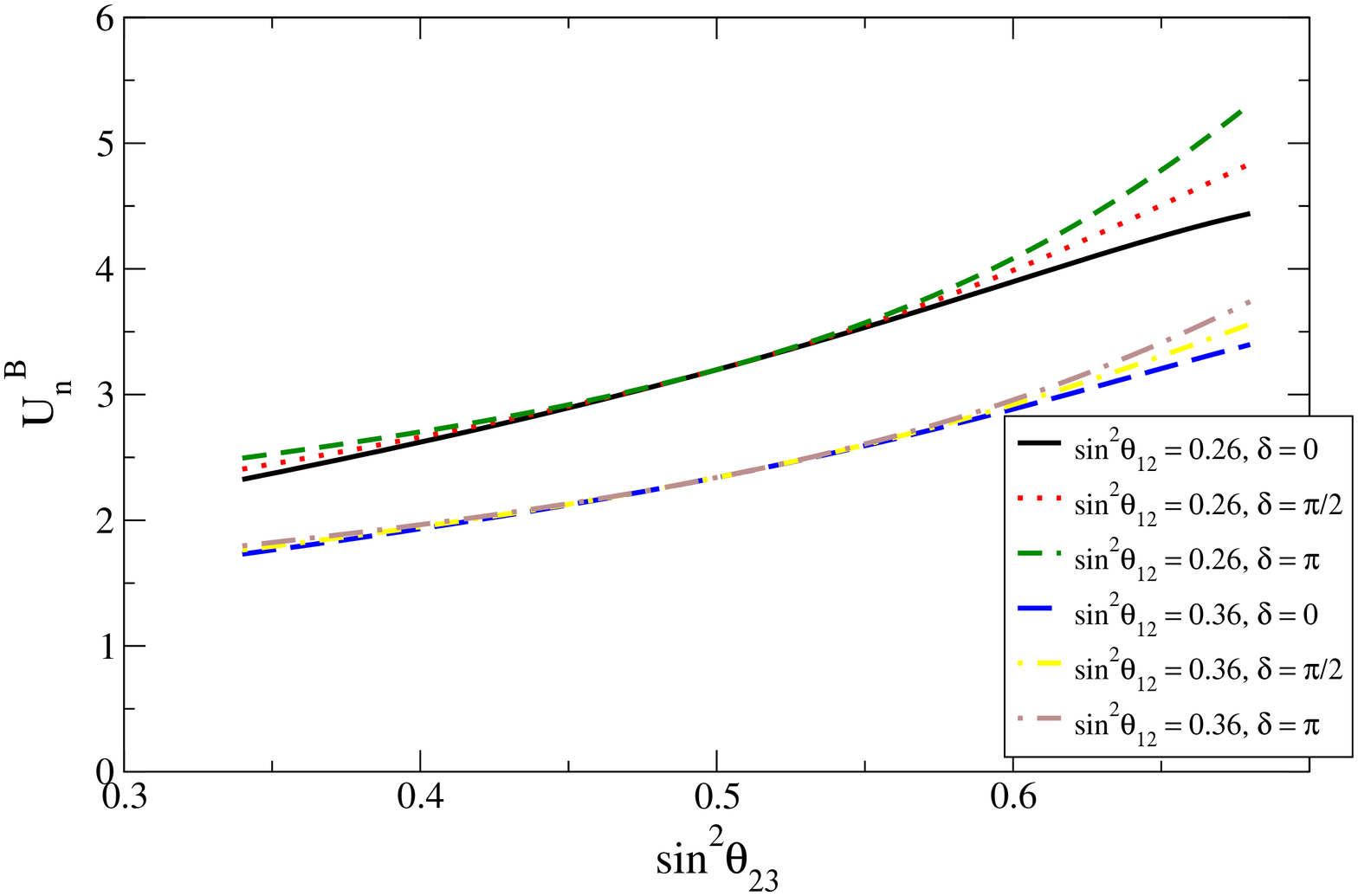,width=8cm,height=5.4cm} \\
\epsfig{file=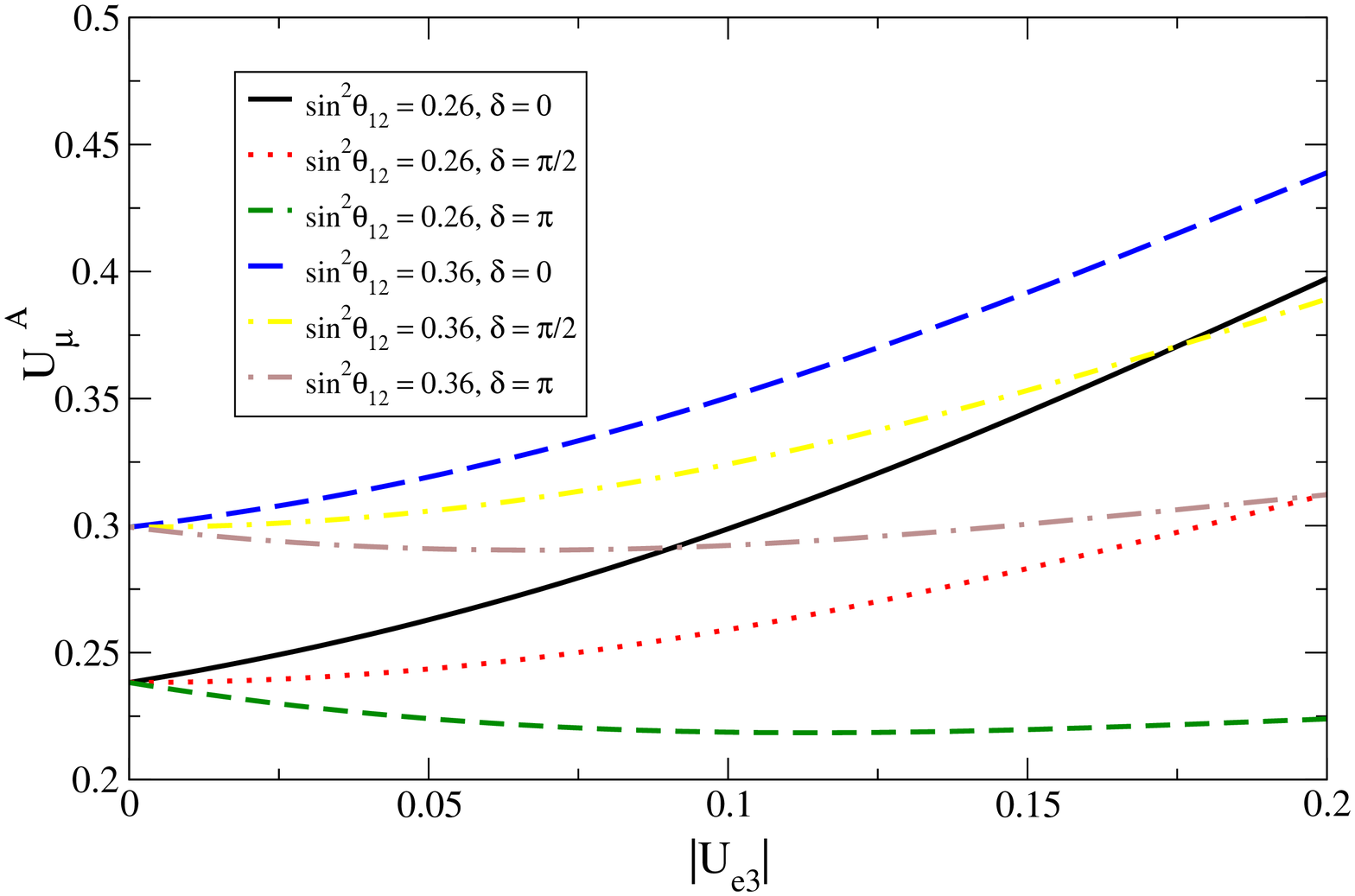,width=8cm,height=5.4cm} & 
\epsfig{file=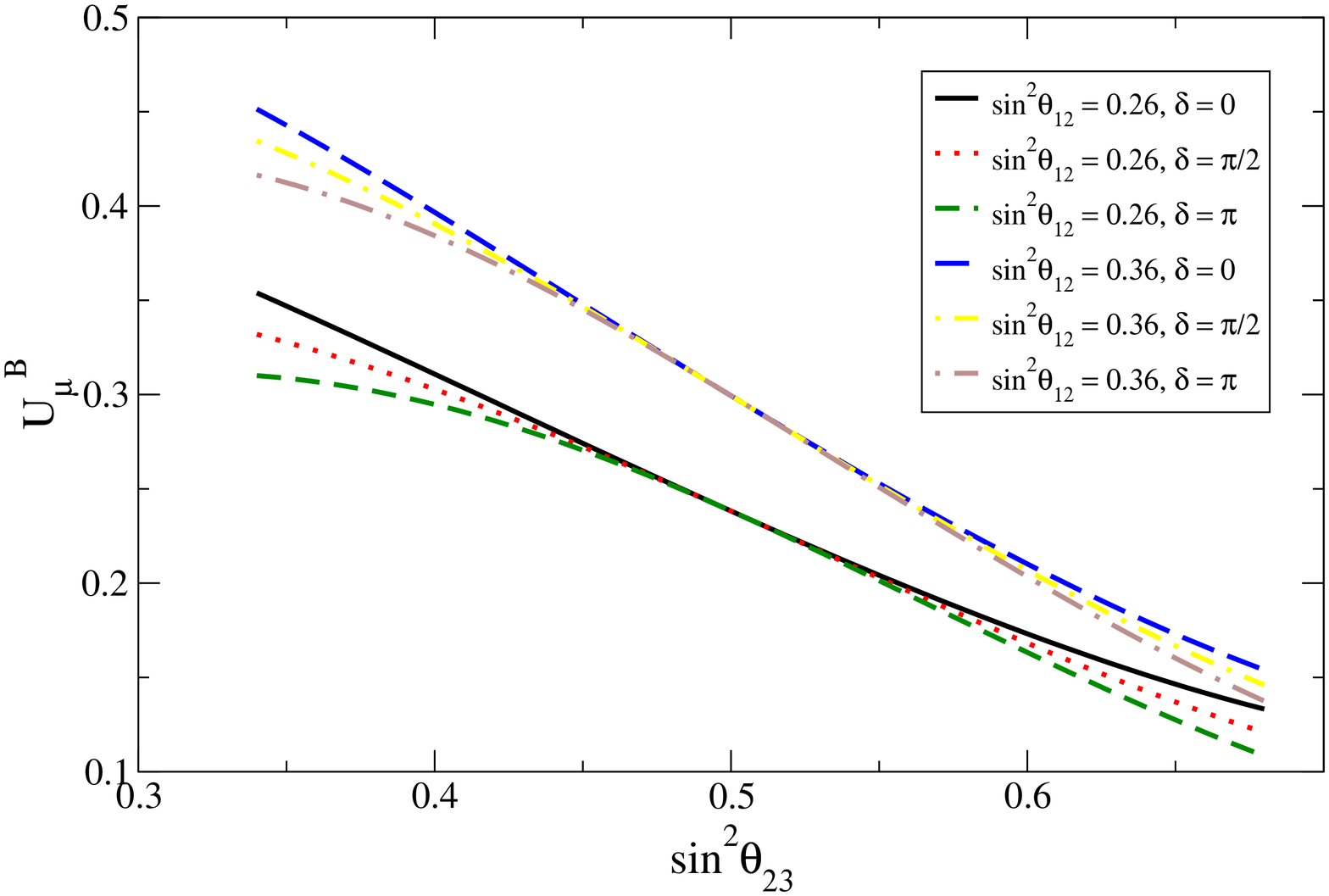,width=8cm,height=5.4cm} 
\end{tabular}
\end{center}
\caption{\label{fig:AvsB2}Same as previous figure for the ratio 
$U = \Phi_{\bar e}^{\rm D}/\Phi_{\mu + \bar \mu}^{\rm D}$.
}
\end{figure}

As an example for the difference of the ratios within these two scenarios, 
we show in Figs.~\ref{fig:AvsB} and \ref{fig:AvsB2} the ratios $R$ and $U$ 
for two representative values of $\sin^2 \theta_{12}$. 
For scenario A (B) we varied $|U_{e3}|$ ($\sin^2 \theta_{23}$) 
and fixed $\theta_{23} - \pi/4 = 1.7 \, |U_{e3}|^2$ 
($|U_{e3}| = 1.7 \, (\theta_{23} - \pi/4)^2$). 
Only $R_\pi$ and $U_\pi^{pp}$ are at zeroth order independent of $\theta_{12}$. 
If $\theta_{13}$ is negligible, then there is in this case hardly any dependence on 
$\theta_{12}$, but a rather strong dependence on $\theta_{23}$. 
Its octant, i.e., $\sin^2 \theta_{23}$ above or below 
1/2, reflects always in positive or negative $\Delta$. 
All other ratios will be good probes for the angle $\theta_{12}$. 
A criterion for this statement might be that the two branches 
for $\sin^2 \theta_{12} = 0.26$ and 0.36 are separated. If $\theta_{23}$ 
is close to $\pi/4$, then this is true even for very 
large values of $|U_{e3}|$. In particular for neutron sources the 
dependence on $\theta_{12}$ is strong. \\

Let us discuss one ratio in more detail, namely $R$ as plotted in 
Fig.~\ref{fig:AvsB}. 
In scenarios A and B it holds for neutrinos from a pion source  
\be
R_\pi^{\rm A} \simeq \frac 12 - 
\frac{3}{16} \, \cos \delta \, \sin 4 \theta_{12} \, |U_{e3}|~,~~
R_\pi^{\rm B} 
\simeq \frac{1}{2} - \frac 38 \, \sin^2 2 \theta_{12} \, \epsilon ~,
\ee
where $\epsilon = \pi/4 - \theta_{23} \simeq \sin^2 \theta_{23} - \frac 12$. 
Turning to neutrinos from muon-damped and neutron sources, we only note that for 
zero $\theta_{13}$ one can obtain simple exact expressions for $R$, namely 
\be
R_n^{\rm B} = 
\frac{2 \, c_{12}^2 \, s_{12}^2 \, c_{23}^2 }
{1 - 2 \, c_{12}^2 \, s_{12}^2 \, c_{23}^2} 
\simeq 
\frac{c_{12}^2 \, s_{12}^2}
{1 - c_{12}^2 \, s_{12}^2} + 
\frac{2 \, c_{12}^2 \, s_{12}^2}
{(1 - c_{12}^2 \, s_{12}^2)^2} \, \epsilon ~.
\ee
and 
\bea
R_{\Slash{\mu}}^{\rm B} = \frac{\D 1}{\D c_{23}^2} \, \frac{\D 1}
{\D 1 + c_{12}^2 \, s_{12}^2 
- (1 -c_{12}^2 \, s_{12}^2) \, \cos 2 \theta_{23}} 
- 1 \\[0.3cm]
\simeq  \frac{\D 1 - c_{12}^2 \, s_{12}^2}
{\D 1 + c_{12}^2 \, s_{12}^2 } - 8 \, 
\frac{\D c_{12}^2 \, s_{12}^2 }
{\D (1 + c_{12}^2 \, s_{12}^2)^2} \, \epsilon~.
\eea

\section{\label{sec:concl}Summary}

Flux ratios of high energy neutrinos are an alternative method to 
obtain information on the neutrino mixing angles and the $CP$ phase. 
Some issues related to the neutrino mixing phenomenology in this 
framework were investigated for the first time in this work. 
We have studied general aspects of the mixing probabilities, such as the 
number of independent probabilities (namely two) or its allowed ranges. 
Mainly due to non-zero and non-maximal solar neutrino mixing, no probability 
can take the trivial values zero or one, as can be 
seen in Eqs.~(\ref{eq:probs}, \ref{eq:ranges}). This is in contrast to the 
general oscillation framework. Hence, high energy neutrinos 
will certainly change their flavor when traveling from an astrophysical 
source to a terrestrial detector. 
At zeroth order, lepton mixing can be described by $\mu$--$\tau$ symmetry, 
and the first order correction $\Delta$ to the mixing probabilities 
due to non-zero $U_{e3}$ and non-maximal $\theta_{23}$ is universal.  
It has always the same dependence on $\theta_{13, 23}$ and $\delta$: 
\[
\Delta \equiv \frac 14 \, \sin 4 \theta_{12} \, 
\cos \delta \, |U_{e3}| + 
2 \, \cos^2 \theta_{12} \, \sin^2 \theta_{12} 
\, \left(\frac 12 - \sin^2 \theta_{23} \right) \simeq -0.1 \div 0.1~.
\]
We have used this to evaluate the relative compositions of the neutrino 
fluxes for three possible sources: pion decay, neutron decay, and muon-damping. 
For instance an initial flavor composition $(1:2:0)$ is modified by 
oscillations to $(1 + 2 \, \Delta : 1 - \Delta : 1 - \Delta)$. 
Various flux ratios were considered (Table \ref{tab:obs}), their allowed ranges 
were given (Tables \ref{tab:pi}, \ref{tab:n}, \ref{tab:m}, \ref{tab:WQ}) 
and their leading correction from $\mu$--$\tau$ symmetry 
breaking was obtained (Table \ref{tab:obs_spe}). 
The latter is almost always a function of $\Delta$. 
For the example of an initial flavor composition $(1:2:0)$ 
the measurable ratio of muon neutrinos to the sum of all neutrinos is 
$\frac 13 (1 - \Delta)$. The main point we wish to make in this paper is the 
following: 
a measurement of a ratio is a measurement of $\Delta$ 
-- and therefore of $\theta_{13}$, 
$\theta_{23}$ and $\delta$ -- if the zeroth order expression is known 
with sufficient precision. In particular, if the zeroth order expression  
does not depend on $\theta_{12}$, then there is no additional 
uncertainty on $\Delta$, which in turn weakly depends on $\theta_{12}$. 
This is true for flux ratios of neutrinos 
stemming from pion sources, except for Glashow resonance related measurements 
of $p\gamma$ neutrinos. 
Flux ratios from other sources depend 
in leading order on $\theta_{12}$ and are useful to probe this angle. 
Schematically, one may write
\[
\mbox{Ratio}(\pi) = f(\Delta)~~~,~~~
\mbox{Ratio}(\mbox{neutron, muon-damped}) 
= g(\theta_{12}) + h(\theta_{12}) \, \Delta~.
\]
The allowed 
ranges of $|U_{e3}|$ and $\sin^2 \theta_{23}$ 
leading to certain values of $\Delta$ are plotted in Fig.~\ref{fig:Delta0} and 
the maximal or minimal possible 
$\Delta$ is plotted in Fig.~\ref{fig:Delta01}. 
The dependence on $\sin^2 \theta_{23}$ is stronger than the dependence on 
$|U_{e3}|$. We have illustrated this feature in cases when 
$\theta_{13}$ is negligible and the deviation from maximal atmospheric mixing 
is sizable and vice versa. 

There can be certain situations for neutron and muon-damped sources 
in which there is no first order correction to a ratio. One interesting example  
is the ratio of electron to muon plus tau 
neutrinos in case of a neutron source, as given in Eq.~(\ref{eq:Sn}). 
It is a function of the electron neutrino survival probability 
only, and therefore a function of 
$\theta_{12}$ plus quadratic corrections of only $\sin^2 \theta_{13}$. 
This underlines the usefulness of the source for measuring $\theta_{12}$.

\vspace{0.5cm}
\begin{center}
{\bf Acknowledgments}
\end{center}
I thank Elisa Resconi for discussions and Kathrin Hochmuth for encouragement 
and careful reading of the manuscript. 
This work was supported by the ``Deutsche Forschungsgemeinschaft'' 
under project number RO--2516/3--1.

\renewcommand{\theequation}{A-\arabic{equation}}
\setcounter{equation}{0}
\begin{appendix}
\section{\label{sec:app}Appendix: Mixing Probabilities}

For the sake of completeness we give here the full expressions for the 
mixing probability of electron into muon neutrinos
\bea \D 
P_{e \mu} =   2 \, c_{13}^2 \, 
\left(c_{12}^2 \, s_{12}^2 \, c_{23}^2 
+ \left(c_{12}^4+ s_{12}^2 \right) \, s_{13}^2 \, s_{23}^2 
\right.
\\[0.2cm]
\left.
\D + c_{12} \, s_{12} \, 
    c_{23} \, s_{23} \, c_\delta \, (c_{12}-s_{12}) (c_{12}+ s_{12}) 
    \, s_{13} 
 \right)~,
\eea
and the survival probability of muon neutrinos: 
\bea \D 
P_{\mu \mu} = 1 - 2 \, c_{12}^4 \, c_{23}^2 \, s_{23}^2 \, s_{13}^2  \\[0.2cm]
 \D +2 \left(\left(s_{12}^2 \left(\left(s_{13}^4+\left(4
    \, c_\delta^2 - 1\right) \, s_{13}^2 +1 \right)
    \, s_{23}^2 - 1 \right)- c_{13}^2 \, s_{23}^2\right)
    \, c_{23}^2
\right.
\\[0.2cm]
\left. \D 
+ s_{13}^2 \, s_{23}^2 \left(c_{13}^2
    \, s_{12}^2-\left(c_{13}^2+ s_{12}^2\right)
    \, s_{23}^2\right)\right) \, c_{12}^2 \\[0.2cm]
 \D + s_{23} \left(-2
    \left( c_{23}^2 \, c_{13}^4+\left( c_{13}^2+ c_{23}^2
    \, s_{12}^2\right) \, s_{13}^2\right) \, s_{23} \, s_{12}^2
\right.
\\[0.2cm]
\left.
 \D -2 \, c_{12} \, s_{12} \, c_{23} \, c_\delta \, (c_{12}- s_{12})
    (c_{12}+ s_{12}) 
    \left(c_{13}^2+\left( s_{13}^2+1\right) ( c_{23}- s_{23})
    (c_{23}+ s_{23})\right) \, s_{13}\right)~.
\eea
With the help of Eqs.~(\ref{eq:t23repl}, \ref{eq:sumrules}) all other probabilities 
can be constructed out of these two formulae.  

\end{appendix}


\pagestyle{empty}

\end{document}